\documentclass[twocolumn]{aastex631}
\usepackage{hyperref}
\usepackage{graphicx}
\usepackage{float}
\usepackage{amsmath}
\usepackage{color}
\usepackage{silence}
\usepackage{multirow}
\usepackage{soul}
\usepackage{textgreek}

\newcommand{\Nc}{N$_{\mathrm{C}}$}
\newcommand{\aNc}{$\overline{N_{C}}$}
\newcommand{\afi}{$\overline{f_{i}}$}

\received{August 20, 2024}
\revised{November 6, 2024}
\accepted{November 27, 2024}

\shorttitle{pah modeling sensitivity analysis}
\shortauthors{Maragkoudakis \emph{et al.}}
\graphicspath{{./}{figures/}}

\begin{document}

\title{A sensitivity analysis of the modeling of Polycyclic Aromatic Hydrocarbon emission in galaxies}

\correspondingauthor{A.~Maragkoudakis}
\email{Alexandros.Maragkoudakis@nasa.gov}

\author[0000-0003-2552-3871]{A.~Maragkoudakis}
\affiliation{NASA Ames Research Center, MS 245-6, Moffett Field, CA 94035-1000, USA}
\affiliation{Oak Ridge Associated Universities, Oak Ridge, TN, USA}

\author[0000-0002-4836-217X]{C.~Boersma}
\affiliation{NASA Ames Research Center, MS 245-6, Moffett Field, CA 94035-1000, USA}

\author[0000-0002-8341-342X]{P.~Temi}
\affiliation{NASA Ames Research Center, MS 245-6, Moffett Field, CA 94035-1000, USA}

\author[0000-0002-1440-5362]{J.D.~Bregman}
\affiliation{NASA Ames Research Center, MS 245-6, Moffett Field, CA 94035-1000, USA}

\author[0000-0002-6049-4079]{L.J.~Allamandola}
\affiliation{NASA Ames Research Center, MS 245-6, Moffett Field, CA 94035-1000, USA}

\author[0000-0001-6035-3869]{V.J.~Esposito}
\affiliation{NASA Ames Research Center, MS 245-6, Moffett Field, CA 94035-1000, USA}
\affiliation{Oak Ridge Associated Universities, Oak Ridge, TN, USA}

\author[0000-0002-3141-0630]{A.~ Ricca}
\affiliation{NASA Ames Research Center, MS 245-6, Moffett Field, CA 94035-1000, USA}
\affiliation{Carl Sagan Center, SETI Institute, 189 Bernardo Ave., Mountain View, CA 94043, USA}

\author[0000-0002-2541-1602]{E.~Peeters}
\affiliation{Department of Physics and Astronomy, University of Western Ontario, London, ON, N6A 3K7, Canada}
\affiliation{Centre for Planetary Science and Exploration, University of Western Ontario, London, Ontario N6A 5B7, Canada}
\affiliation{Carl Sagan Center, SETI Institute, 189 Bernardo Ave., Mountain View, CA 94043, USA}

\begin{abstract}

We have conducted a sensitivity analysis on the mid-infrared spectral decomposition of galaxies and the modeling of the PAH emission spectrum with the NASA Ames PAH Infrared Spectroscopic Database (PAHdb) to assess the variance on the average galaxy PAH population properties under a grid of different modeling parameters. 
We find that the SL and SL+LL \textit{Spitzer}-IRS decomposition with \textsc{pahfit} provides consistent modeling and recovery of the 5--15 \micron{} PAH emission spectrum. 
For PAHdb modeling, application of a redshift to the calculated spectra to account for anharmonic effects introduces a $15\%$--$20\%$ variance on the derived parameters, while its absence improves the fits by $\sim$13\%. 
The 4.00-\textalpha{} release of PAHdb achieves the complete modeling of the 6--15 \micron{} PAH spectrum, including the full 6.2 \micron{} band, improving the average fitting uncertainty by a factor of 2.
The optimal PAHdb modeling configuration requires selection of pure PAHs without applying a redshift to the bands. 
Although quantitatively the PAHdb-derived parameters change under different modeling configurations or database versions, their variation follows a linear scaling, with previously reported trends remaining qualitatively valid. PAHdb modeling of JWST observations, and JWST observations smoothed and resampled to the \textit{Spitzer}-IRS resolution and dispersion have consistent PAHdb derived parameters. 
Decomposition with different codes, such as \textsc{pahfit} and \textsc{cafe}, produce PAH emission spectra with noticeable variation in the 11--15~\micron{} region, driving a $\sim$7\% difference in the neutral PAH fraction under PAHdb modeling. 
A new library of galaxy PAH emission templates is delivered to be utilized in galaxy SED modeling. 

\end{abstract}

\keywords{ISM: molecules --- ISM:lines and bands --- infrared: ISM --- galaxies: ISM}

\section{Introduction} \label{sec:intro}

The Aromatic Infrared Bands (AIBs) comprise the series of strong emission features found in the mid-Infrared (mid-IR) spectra of various astrophysical sources and are attributed to Polycyclic Aromatic Hydrocarbons (PAHs) and related species \citep{Leger1984, Allamandola1985, Tielens2008}. PAH emission has been used extensively to probe galactic and extragalactic environments \citep[e.g.,][]{Hony2001, Peeters2002, Smith07b, Galliano2008, Sandstrom2012, Boersma2018, Maragkoudakis2018a, Zang2022, Lai2022, Garcia-Bernete2022, Maragkoudakis2022, Chastenet2023, Egorov2023, Rigopoulou2024} due to the sensitivity of the PAH spectral characteristics to the astrophysical conditions of the emitting regions.  

The determination of PAH emission in astrophysical sources relies on the decomposition of their mid-IR spectrum, which often consists of several additional components such as stellar continuum emission from old stars, continuum emission from stochastically heated very small carbonaceous grains, absorption from silicate, and, in some cases, water-rich or interstellar mixed molecular ices, ro-vibrational and pure rotational H$_{2}$ lines, H$\,_{\textrm{I}}$ and He$\,_{\textrm{I}}$ recombination lines, and forbidden atomic lines. The spectral decomposition is then performed by building a model considering all components, accounting for dust extinction, and subsequently fitting the observed spectrum with the defined model (e.g., \textsc{pahfit} \citealp{Smith07b}, \textsc{cafe} \citealp{Marshall2007}). Another frequently employed method for isolated PAH band emission \citep[e.g.,][]{Hony2001, Peeters2002, Peeters2017} is based on the subtraction of a broadband or local spline continuum, that may or may not be tied to PAHs, defined on a given set of anchor points adjacent to the main PAH features. The PAH band fluxes are then estimated from the continuum-subtracted spectra.

Similarly, PAH characteristics such as their charge and size can be obtained either empirically using band strength ratios \citep[e.g.,][]{Galliano2008, Ricca2012, Croiset2016, Maragkoudakis2020, Knight2021, Draine2021, Maragkoudakis2023a, Maragkoudakis2023b}, or determined with a database fitting approach on the isolated PAH spectrum which results from the multi-component modeling of the mid-IR spectrum \citep[e.g.,][]{Cami2010, Boersma2013, Boersma2014a, Boersma2014b, Boersma2015, Boersma2018, Shannon2019, Maragkoudakis2022, Boersma2024}. In a database fitting approach an emission model is employed to convert either laboratory measured spectra or density functional theory (DFT)-computed PAH absorption spectra into emission spectra, which are subsequently used to fit the observed astronomical PAH spectrum.

Assumptions in the employed models, decomposition methods, but also the database content (in a database fitting approach), will have direct impact on the recovered properties of the different mid-IR spectral components, and the PAH characteristics. Such assumptions include the profile shape and width of the emission features, the adopted extinction curve in the attenuation model, the properties of the stellar populations (e.g., age, temperature, metallicity) which produce the stellar continuum or determine the radiation field strength, possible band shifts due to anharmonic effects, the relaxation process of PAHs after the absorption of UV photons which will determine their intensity, as well as the range that these parameters are permitted to vary within a fit. Additionally, the number of adopted (sub-)components in the decomposition (e.g., the number of black body components in the dust emission estimation, the number of dust features and their sub-components, and the number of PAH molecules used in a database fitting) will leverage the synthesized spectrum, and consequently the resulting determination of the components properties.   

In general for physical models, their assumptions and application are formed on the basis of our current knowledge on the involved processes and mechanisms. For instance, in the PAH (emission) model the band profiles are shaped by the intramolecular vibrational redistribution mechanism, where Lorentzian profiles provide adequate description. However, band profiles are also affected by intermode anharmonicities due to the coupling between modes, where PAHs can be considered to emit as oscillators randomly distributed around a mean, and as such Gaussian profiles can be more appropriate \citep[][]{Pech2002}. Regarding band widths and shifts, experiments on small PAHs have shown that the full-width-at-half maximum (FWHM) of PAH bands typically range between 10 and 30 cm$^{-1}$ \citep[][]{Peeters2004b} for PAH bands at wavelengths $<15$ \micron{}, and anharmonic effects usually shift band positions to lower frequencies, which can be accounted for by a uniform shift of all bands by 15 cm$^{-1}$ as a first approximation \citep[e.g.,][]{Cook1998, Pech2002, Boersma2013}, although \cite{Mackie2018} has argued that these shifts might not be required. The PAH emission model itself considers that PAHs will reach a maximum temperature after the absorption of an exciting photon, following the so-called thermal approximation \citep[e.g.,][]{Schutte1993}, where the emission from a given vibrational mode is calculated as the average emission of an oscillator at temperature $T$ over the cooling time, followed by a relaxation process which considers the emission process as a Poisson process \citep[][]{Bakes2001}. As these processes are not fully understood and/or are very difficult to model, models and their adopted parameters are also subjected to a certain degree of uncertainty. Therefore, the ``goodness" of a fit can have a less stringent, or more flexible, interpretation, as the application of an adopted model to the observations is dependent on (or limited by) our comprehensive understanding of the physical processes which synthesize the models. 

In \cite{Maragkoudakis2022}, hereafter Paper I, we studied the PAH component of over 900 \textit{Spitzer}-IRS galaxy spectra employing a database-fitting approach by utilizing the data and tools provided through the NASA Ames PAH IR Spectroscopic Database \citep[PAHdb;][]{Bauschlicher2010, Boersma2014b, Bauschlicher2018, Mattioda2020}. The adopted model components and their properties included radiation fields with a range of excitation energies, an emission model that takes the full emission cascade into account, Gaussian line profiles of a given FWHM, and a redshift application to mimic anharmonic effects. The pool of spectra was drawn from version 3.20 of PAHdb's library, containing 4233 quantum-chemically calculated absorption spectra of PAHs with various structures, charge states, sizes, and compositions.
In the current work, we conduct a sensitivity analysis, and examine the range of the derived PAH properties, i.e., the PAH charge, size, and composition breakdowns, as well as the PAH ionization fraction ($f_i$) and average number of carbon atoms (\aNc), when adopting different modeling parameters compared to Paper I, and quantify their impact on derived PAH properties and the astronomical interpretation. Although a vast parameter space can be explored, here we focus principally on the parameters employed in the database fitting, such as the adopted line profiles, emission models, molecule composition, altering one parameter at a time at a certain amount or direction.  

The structure of this paper is as follows. In section \ref{sec:sample} we define the sample of this work. Section \ref{sec:PAHFIT: SL+LL spectrum} examines the impact on the isolated/extracted PAH spectrum using \textsc{pahfit} when \textit{Spitzer}'s full 5--30 \micron{} spectral range is utilized to decompose a galaxy's spectrum, compared to 5--15 \micron{} in Paper I. Section \ref{sec:PAHdb} explores the variations in the PAHdb derived parameters when assuming line profiles with a FWHM of 10 cm$^{-1}$, a calculated temperature emission model, and omitting a 15 cm$^{-1}$ redshift (Sections \ref{sec:PAHdb-profiles}, \ref{sec:PAHdb-model}, and \ref{sec:PAHdb-redshift}, respectively). Section \ref{sec:v4.00} describes the impact of using the latest library of DFT-computed PAH spectra. Section \ref{sec:spitzer_vs_jwst} examines the results from observations of different spectral resolution, and Section \ref{sec:different_codes} explores the variance on the PAHdb derived parameters when the PAH emission spectrum is retrieved from different galaxy spectral decomposition codes (e.g. \textsc{cafe}). The results and a discussion of their implications are presented in Section \ref{sec:results}. A new and improved library of galaxy PAH spectral templates to be used in SED modeling is presented in Section \ref{sec:Templates}. Finally, a summary of our results and conclusions are given in Section \ref{sec:summary}.

\section{Sample} \label{sec:sample}

\begin{figure}
    \begin{center}     
    \includegraphics[scale=0.5]{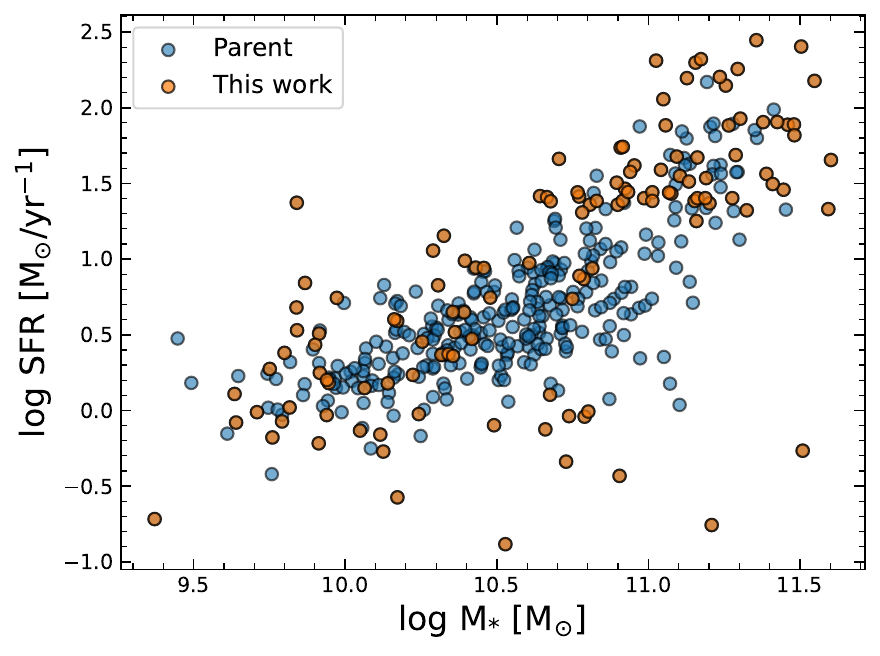}
    \includegraphics[scale=0.5]{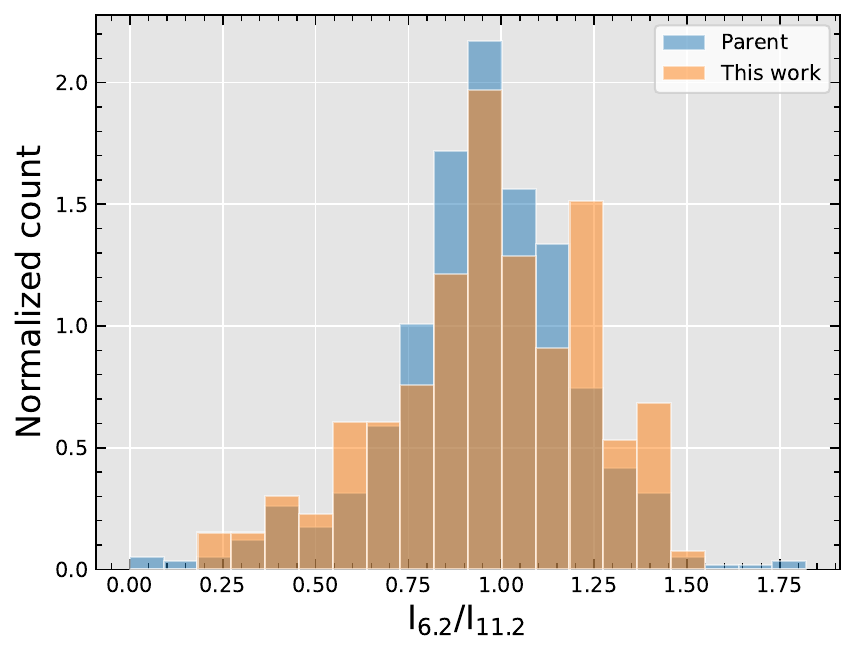}
    \includegraphics[scale=0.5]{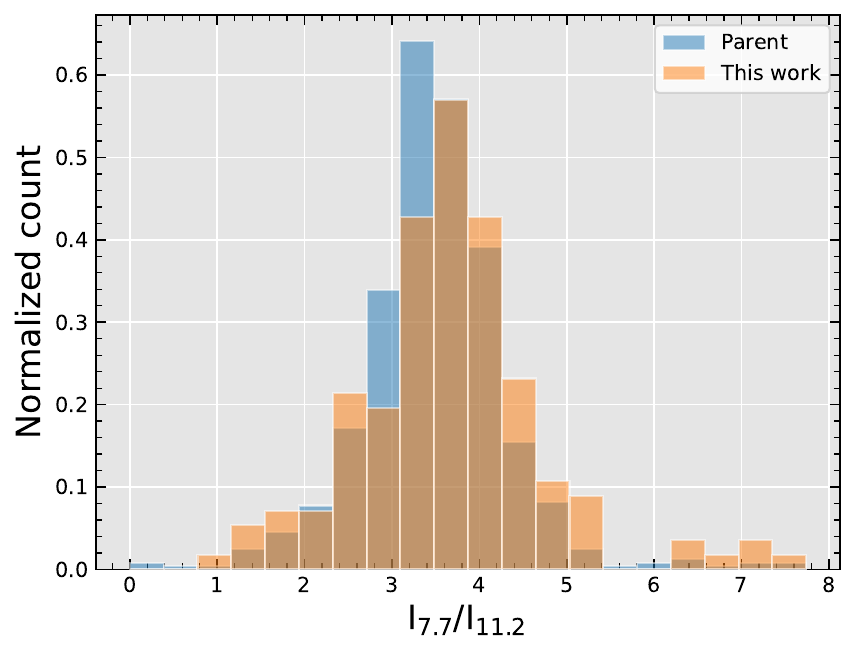}
    \caption{Top panel: Representation of the sample in this work (orange points) in the galaxy main-sequence plane (SFR -- M$_{*}$) with respect to the parent sample (blue points) defined in \cite{Maragkoudakis2022}. Middle and bottom panel: The I$_{6.2}$/I$_{11.2}$ and I$_{7.7}$/I$_{11.2}$ PAH intensity ratio distributions respectively, of the current (orange) and parent (blue) samples, showing a good representation of the parent distributions by the drawn subset.}
    \label{fig:sample}
    \end{center}
\end{figure}

The sample used in this work is a subset of the galaxies presented in Paper I, allowing a meticulous examination and visual inspection of the modeling results of individual galaxies, and was selected to be representative in the star-formation rate (SFR) and stellar mass (M$_{*}$) plane, i.e., across the galaxy main sequence \citep[e.g.,][]{Brinchmann2004, Daddi2007, Elbaz2007, Speagle2014, Maragkoudakis2017, Popesso2019}. We select up to 20 galaxies in each of the $7\times5$ SFR and M$_{*}$ 2D bins from the Q2 spectral quality class, as defined in Paper I, having the lowest spectral decomposition uncertainty obtained from \textsc{pahfit}\footnote{\href{https://github.com/PAHFIT/pahfit}{https://github.com/PAHFIT/pahfit}} ($\sigma_{\textrm{PAHFIT}}$) to ensure the highest quality PAH spectra are used for our sensitivity analysis study. The resulting sample consists of 147 galaxy spectra. The top panel of Figure \ref{fig:sample} shows the representation of the current sample in the galaxy main-sequence plane over-plotted on the parent sample. While not used as a requirement, this subset also samples well the I$_{6.2}$/I$_{11.2}$ and I$_{7.7}$/I$_{11.2}$\footnote{We note that the 11.2 \micron{} PAH feature is equivalent to} \textsc{pahfit}'s 11.3 \micron{} complex. PAH intensity ratio distributions of the parent sample (middle and bottom panel of Figure \ref{fig:sample}). 

\section{Analysis Methods} \label{sec:analysis}

In the following sections we describe and examine the methods and modeling parameters that have a direct impact on the characterization of the average PAH population properties, starting from the recovery of the PAH emission spectrum from the decomposition of the observational spectrum (Section \ref{sec:PAHFIT: SL+LL spectrum}), the adopted PAHdb modeling configuration (Sections \ref{sec:PAHdb}--\ref{sec:PAHdb-redshift}), different database version (Section \ref{sec:v4.00}), the modeling of observations with different spectral resolutions (Section \ref{sec:spitzer_vs_jwst}), and usage of different spectral decomposition codes (Section \ref{sec:different_codes}). 

\subsection{Spectral decomposition} \label{sec:PAHFIT: SL+LL spectrum}

Recovery of the galaxy PAH spectrum requires isolation of the PAH emission from the other components in the spectrum, such as the underlying stellar and dust continuum, a potential AGN continuum, the emission lines associated with molecular hydrogen and atomic species, and further account for extinction. Available codes for this task, such as \textsc{pahfit} \citep{Smith07b} and \textsc{cafe} \citep[][Diaz Santos et al. in prepr.]{Marshall2007} can have different treatment for certain components, like the modeling of the different continua, or the dust extinction parameterization, which directly impact the recovered PAH spectrum \citep[e.g.,][]{Boersma2018}. These differences are discussed and examined in Sections \ref{sec:different_codes} and \ref{sec:pahfit_vs_cafe}.

\begin{figure}
    \hspace*{-0.55cm}\includegraphics[scale=0.28]{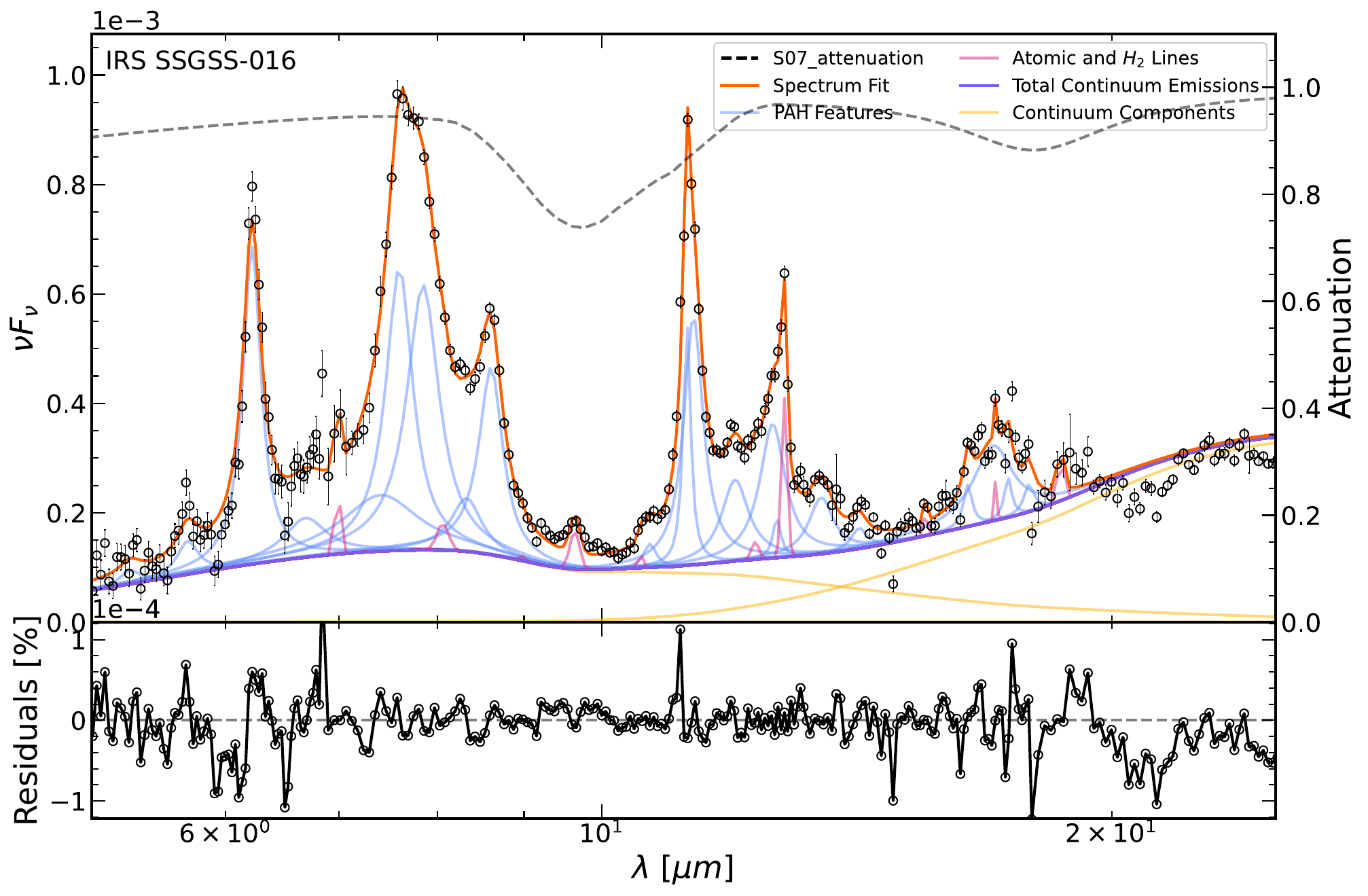}
    \caption{\textsc{pahfit} decomposition of the combined SL+LL \textit{Spitzer}-IRS spectrum of galaxy SSGSS-016. The fit (orange line) is synthesized using the following components: Dust features (light blue lines), atomic and H$_{\rm 2}$ lines (magenta lines), continuum (yellow lines; the total continuum emission is shown as a purple line), and attenuation (dashed black line).}
    \label{fig:pahfit_sl+ll}
\end{figure}

\begin{figure*}
    \centering
    \includegraphics[scale=0.5]{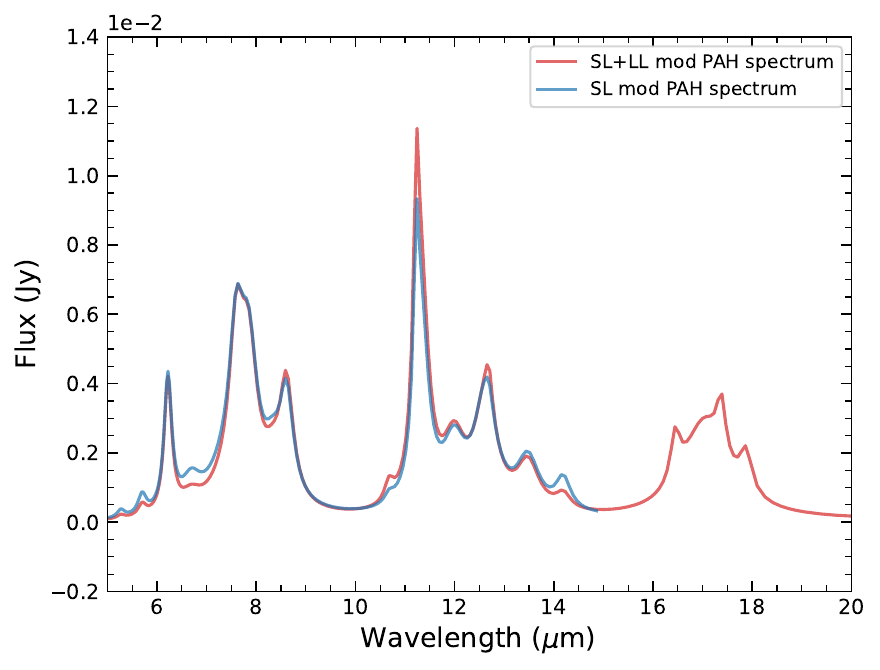}
    \includegraphics[scale=0.5]{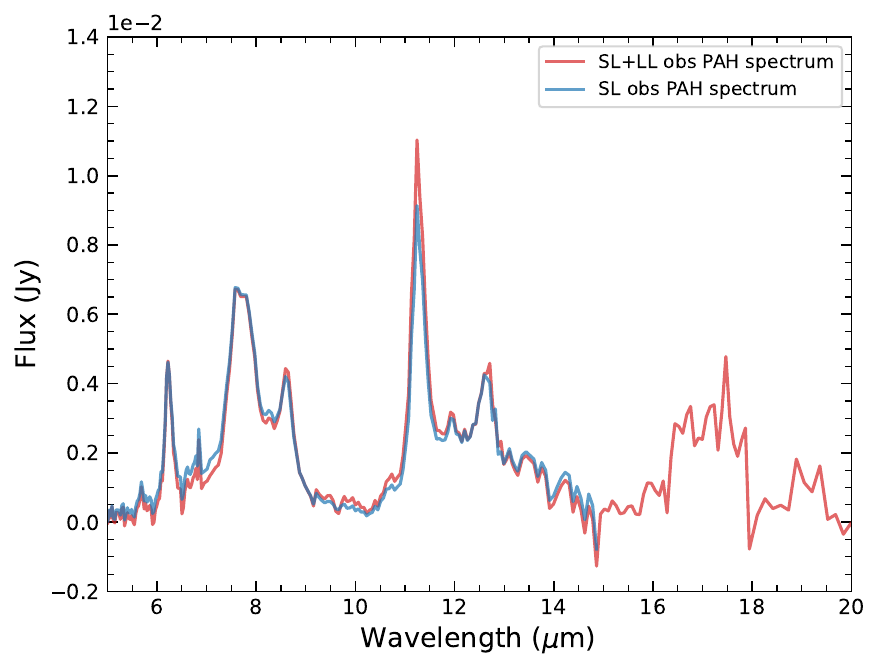}
    \caption{Comparison of the SL and SL+LL spectra (shown in blue and red lines, respectively) of the \texttt{mod} (left panel) and \texttt{obs} (right panel) PAH spectral methods (see Section \ref{sec:PAHFIT: SL+LL spectrum}). The plotted SL+LL spectra are confined in the 5--20 \micron{} range to aid comparison.}
    \label{fig:mod_obs}
\end{figure*}

In this Section, we examine the sensitivity of the recovered PAH band strengths resulting from the mid-IR spectral decomposition of \textsc{pahfit} with the inclusion of the LL (13.9--33.9 \micron) \textit{Spitzer}-IRS segments along with the SL (5--14.2 \micron) segments, the latter only used in Paper I. LL/SL offset corrections and order matching have been applied by the respective \textit{Spitzer} Legacy Programs on the Spitzer Heritage Archive delivered data. We redshift-corrected all spectra to their rest-frame wavelengths before fitting with \textsc{pahfit}. An example of decomposing a SL+LL galaxy spectrum with \textsc{pahfit} is shown in Figure \ref{fig:pahfit_sl+ll}. From the \textsc{pahfit} decomposition we obtain \textit{(i)} the modeled PAH spectrum (\texttt{mod}) by combining the different Drude components, and \textit{(ii)} the observed PAH spectrum (\texttt{obs}) isolated from other components by  subtracting out the dust continuum, stellar continuum, atomic and H$_{2}$ line, and correcting for extinction, as described in Paper I. The left and right panels of Figure \ref{fig:mod_obs} compares the SL and SL+LL PAH spectra for the \texttt{mod} and \texttt{obs} methods, respectively.

\begin{figure*}
    \begin{center}
    \includegraphics[scale=0.5]{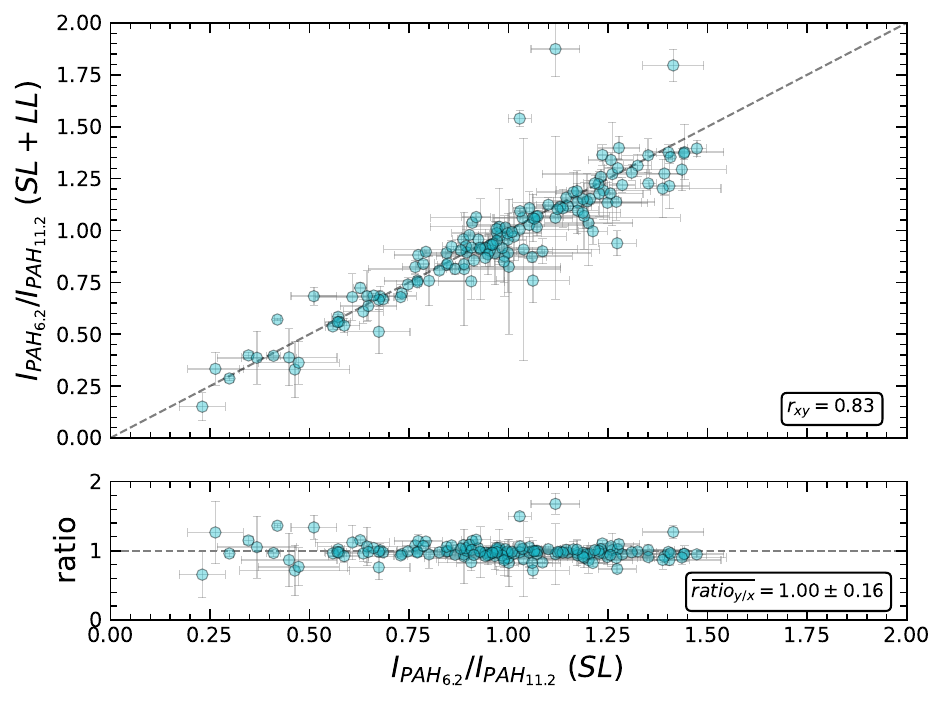}
    \includegraphics[scale=0.5]{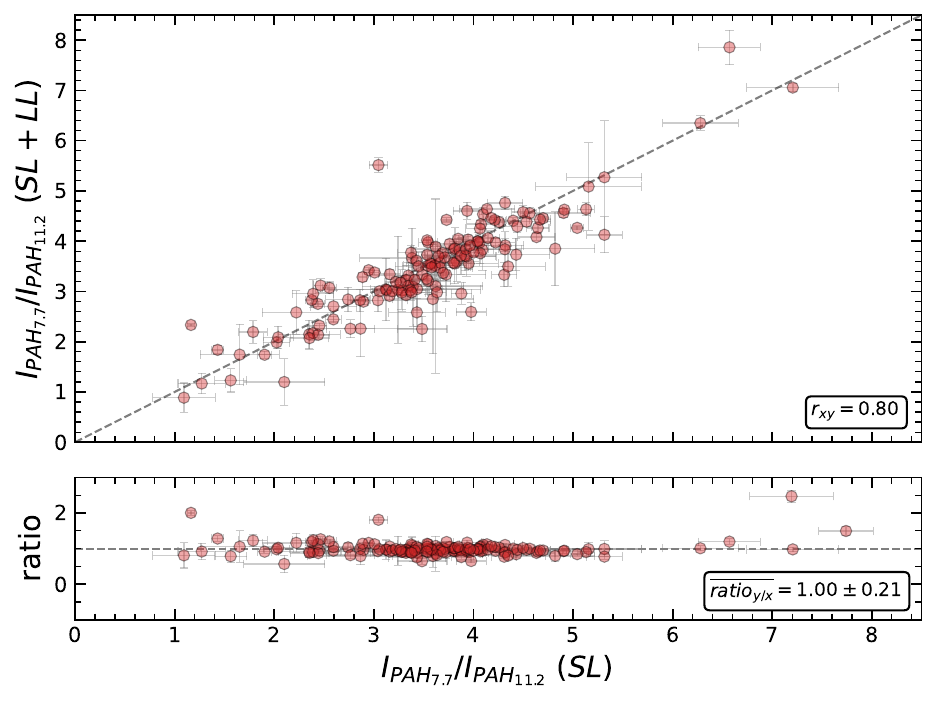}
    \caption{Comparison of \textsc{pahfit} recovered PAH band strength ratios (6.2/11.2 \micron{}, left panel; 7.7/11.2 \micron{}, right panel) when modeling the SL+LL and SL-only spectra of galaxies. Dashed lines are lines of equality. The Pearson’s correlation coefficient r$_{xy}$, along with the average ratio ($\overline{ratio}_{y/x}$) and standard deviation of the two measurements are provided.}
    \label{fig:pahfit_comparisons}
    \end{center}
\end{figure*}

A comparison between the relative PAH band intensities recovered when modeling the combined SL+LL vs the SL alone spectra, is presented in Figure \ref{fig:pahfit_comparisons} for the 6.2/11.2 and 7.7/11.2 \micron{} PAH band strength ratios, often used to probe PAH ionization. 

For the remainder of the analysis (unless noted otherwise) we make use of the SL-alone \texttt{obs} spectra to be able to make direct comparisons with results from Paper I.

\subsection{PAHdb modeling} \label{sec:PAHdb}

\begin{table}[]
    \centering
    \caption{The base run PAHdb modeling configuration adopted in \cite{Maragkoudakis2022}.}
    \begin{tabular}{lc}
    \hline
    \hline
        PAHdb & v3.20 \\
        Emission model & Cascade \\ 
        Excitation energy & 8 eV \\
        Line profile & Gaussian, FWHM=15 cm$^{-1}$ \\
        Redshift & 15 cm$^{-1}$ \\
    \hline
    \end{tabular}
    \label{tab:base_run_conf}
\end{table}

In Paper I the PAHdb modeling was performed using v3.20 of the library of computed PAH spectra, a range of excitation energies (6, 8, 10, and 12 eV), together with an emission model that takes the entire emission cascade into account, Gaussian line profiles for the PAH features with a 15 cm$^{-1}$ full width at half maximum (FWHM), and a 15 cm$^{-1}$ redshift to mimic (some) anharmonic effects and hot band emission. The impact of choosing different excitation energies were discussed in Paper I. Using an 8 eV excitation energy with the previous parameters as a reference (summarized in Table \ref{tab:base_run_conf}), referred to as base run henceforth, we perform a sensitivity analysis by changing the base run parameters, one at a time, and examine the sensitivity in the PAHdb modeling and derived PAH properties. The PAHdb properties examined are: (i) the charge breakdown, i.e., the fraction of neutral, cationic, and anionic PAHs; (ii) the size breakdown, i.e., the fraction of small (\Nc{} $<$ 50) and large (\Nc{} $\geq$ 50) PAHs; (iii) the composition breakdown, into pure PAHs and nitrogen-containing PAHs, in the case of the PAHdb v3.20 analysis; (iv) the derived \afi{} and \aNc, i.e., the fraction of cation to neutral PAH and number of C atoms in the PAHs comprising the fits; and (v) the total fitting uncertainty, $\sigma_{\rm PAHdb}$, defined as the ratio between the integral of the absolute fit residuals and that of the input spectrum (see also Paper I). 

For each galaxy in each case, we performed Monte Carlo sampling where the spectra were perturbed 1000 times within their uncertainties and ﬁtted with PAHdb. Then, for each PAHdb-derived quantity we examine their average ratio with the base run ($\overline{ratio}_{y/x}$), perform linear regression fitting to quantify the deviation, offset, and 1$\sigma$ scatter between the runs, calculate the Pearson’s correlation coefficient r$_{xy}$, and the 5\% and 10\% distance-from-the-fit lines as an additional evaluation of the scatter and the deviation from the base run.

\subsubsection{PAHdb: Emission Profiles} \label{sec:PAHdb-profiles}

The functional form used to describe the band shape along with the band width are the main parameters defining the emission profile of the individual bands of a PAH in PAHdb modeling. In Paper I, Gaussian line profiles with a FWHM of 15 cm$^{-1}$ were used. To examine the sensitivity of the modeling to the FWHM, we performed Monte Carlo fitting using Gaussian line profiles with smaller, 10 cm$^{-1}$, FWHM. Subsequently, fitting was performed taking Lorentzian line profiles with a FWHM of 15 cm$^{-1}$.

\begin{figure*}
    \includegraphics[scale=0.37]{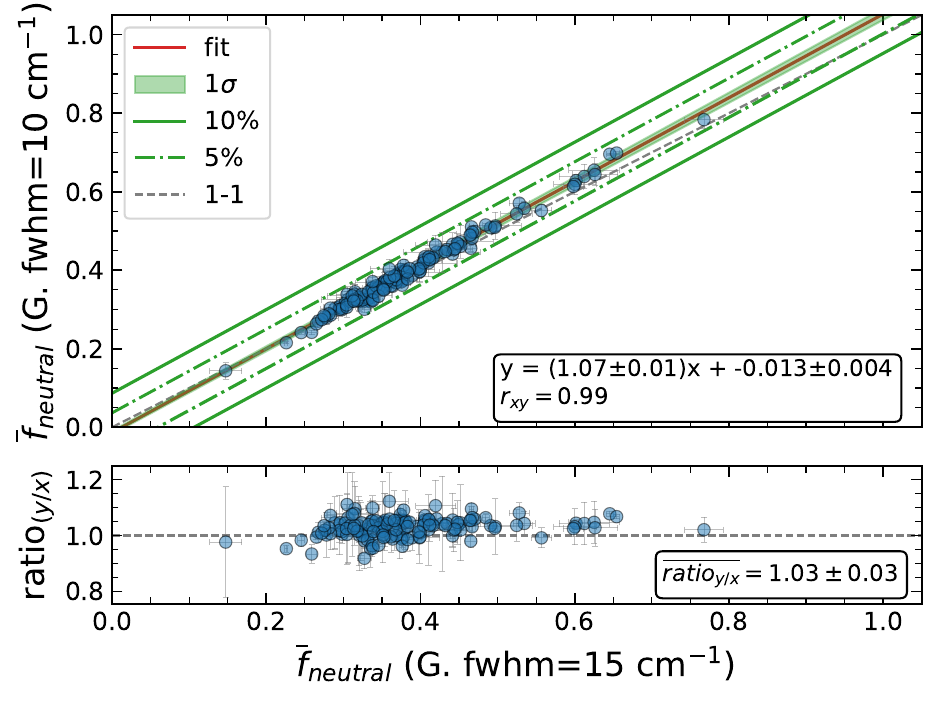}
    \includegraphics[scale=0.37]{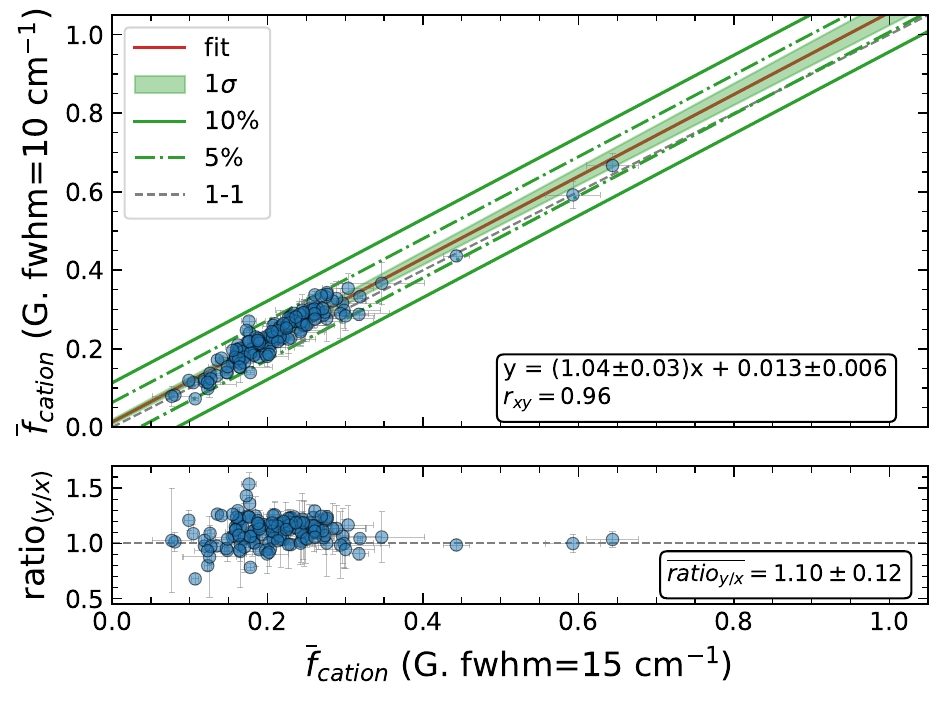}
    \includegraphics[scale=0.37]{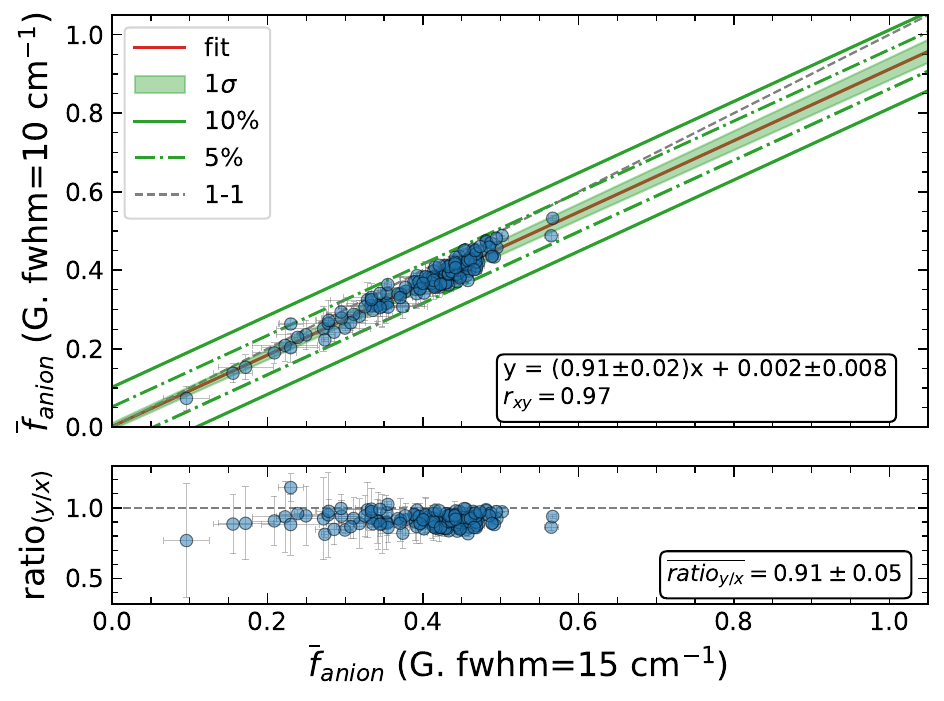} \\
    \includegraphics[scale=0.37]{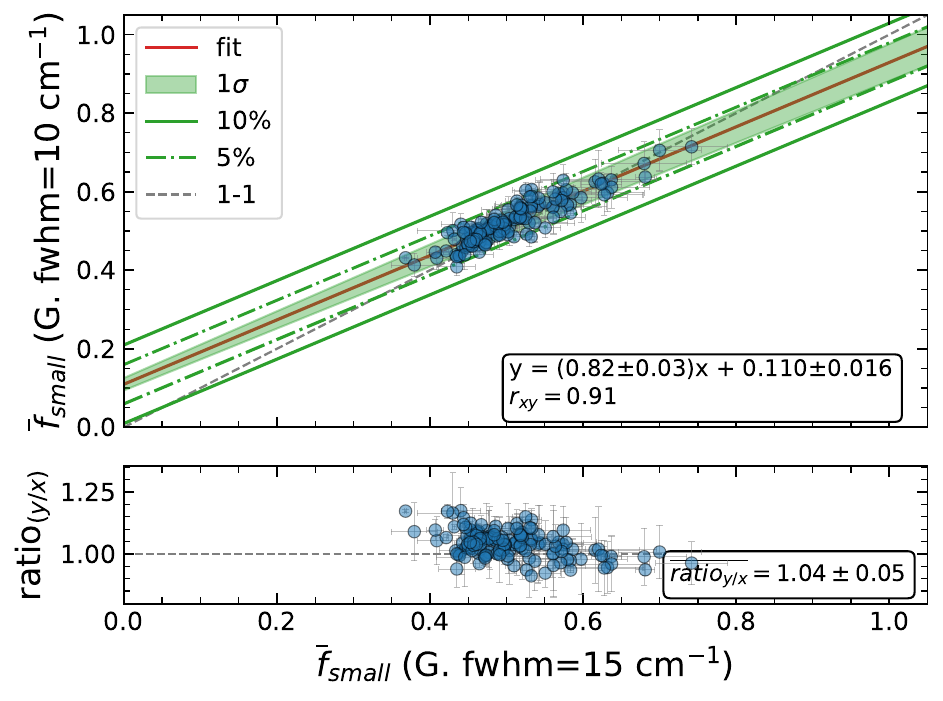}
    \includegraphics[scale=0.37]{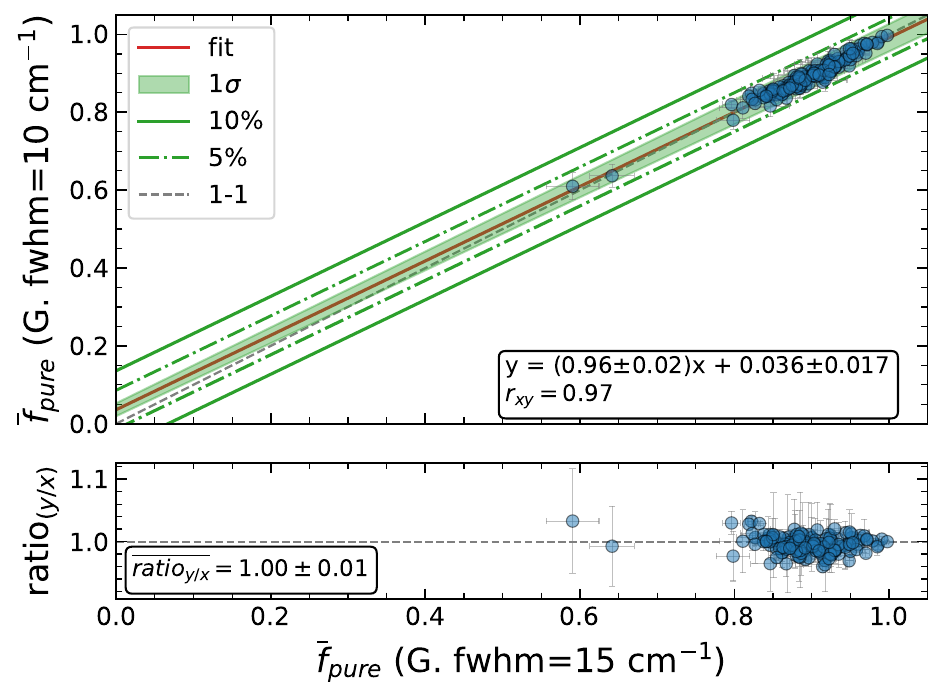}
    \includegraphics[scale=0.37]{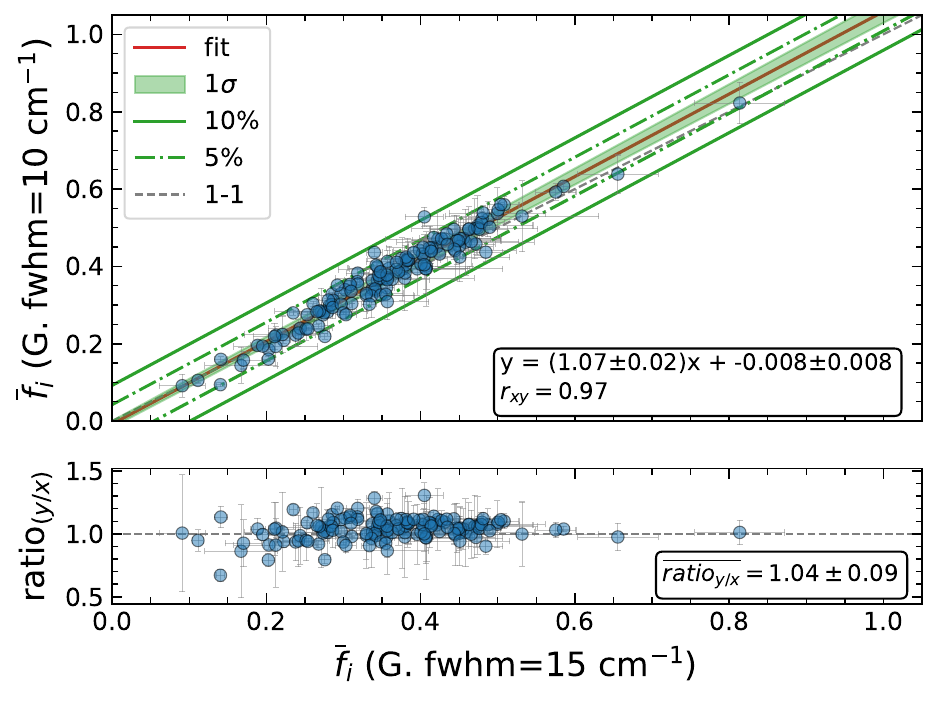} \\
    \includegraphics[scale=0.37]{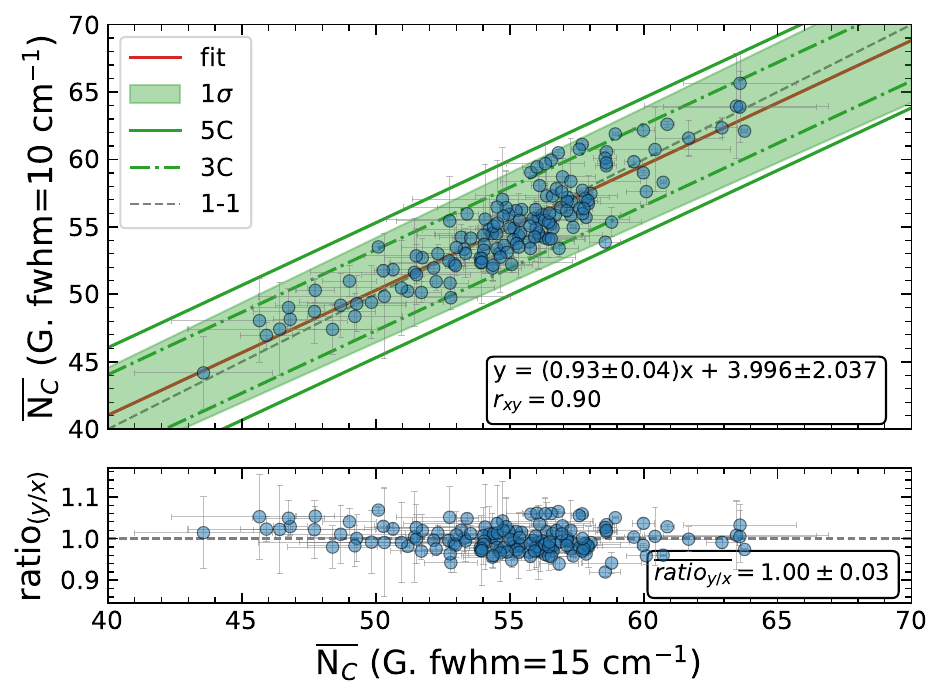}
    \includegraphics[scale=0.37]{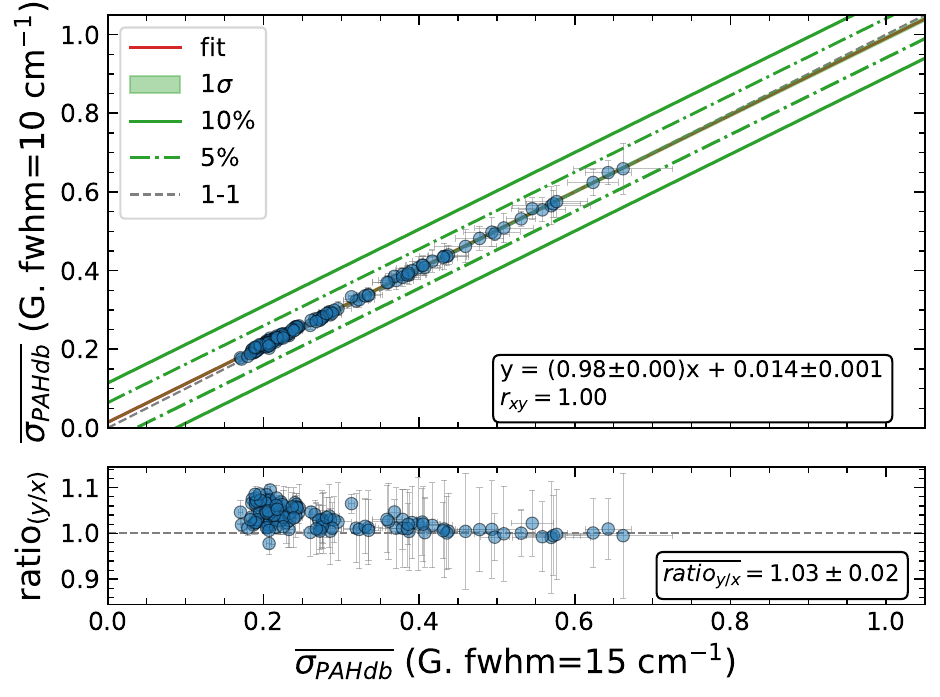}
    \caption{Comparison of the PAHdb-derived PAH properties in the base run (x-axis) and the Gaussian FWHM 10 cm$^{-1}$ emission profile run (y-axis). Top row: neutral PAH fraction (left), cation PAH fraction (middle), anion PAH fraction (right); Middle row: small PAH fraction (left), pure PAH fraction (middle), PAH ionization fraction (right); Bottom row: \aNc{} (left), $\sigma_{PAHdb}$ (middle). Linear regression fitting is shown with the red line, 1$\sigma$ dispersion with the green envelope, 5\% and 10\% distance from the fit with green dash-dotted and solid lines, respectively, and line of equality with the black dashed line. The regression parameters and the Pearson’s correlation coefficient r$_{xy}$ are provided in the inset. In each case, the ratio of the two values is plotted in the bottom sub-panels, and the average ratio ($\overline{ratio}_{y/x}$) and standard deviation are provided.}
    \label{fig:gaussian_fwhm_10_vs_15}
\end{figure*}

\begin{figure*}
    \includegraphics[scale=0.37]{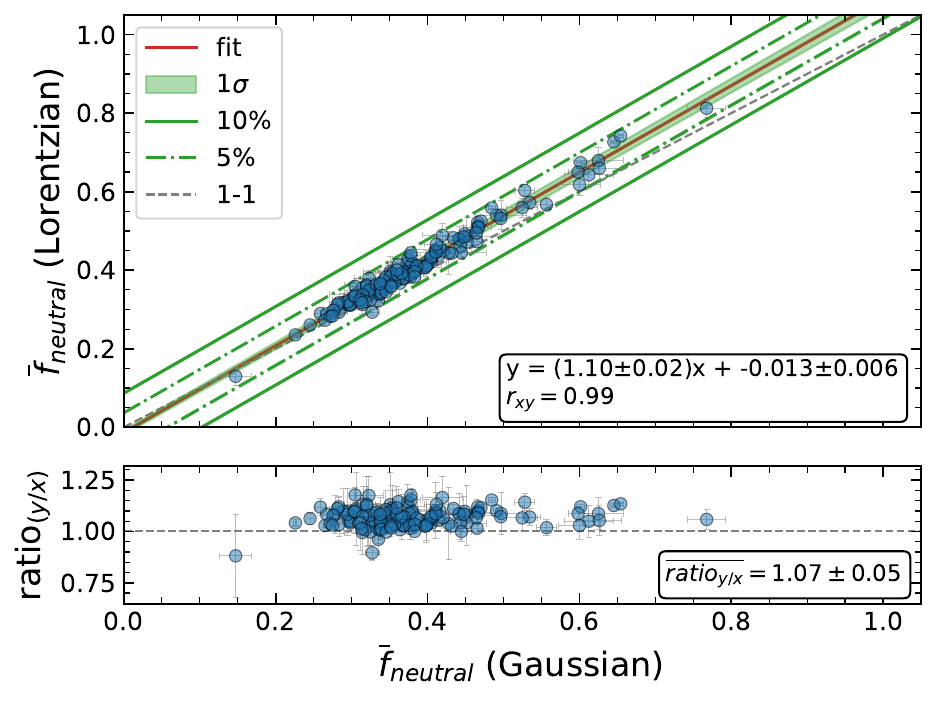}
    \includegraphics[scale=0.37]{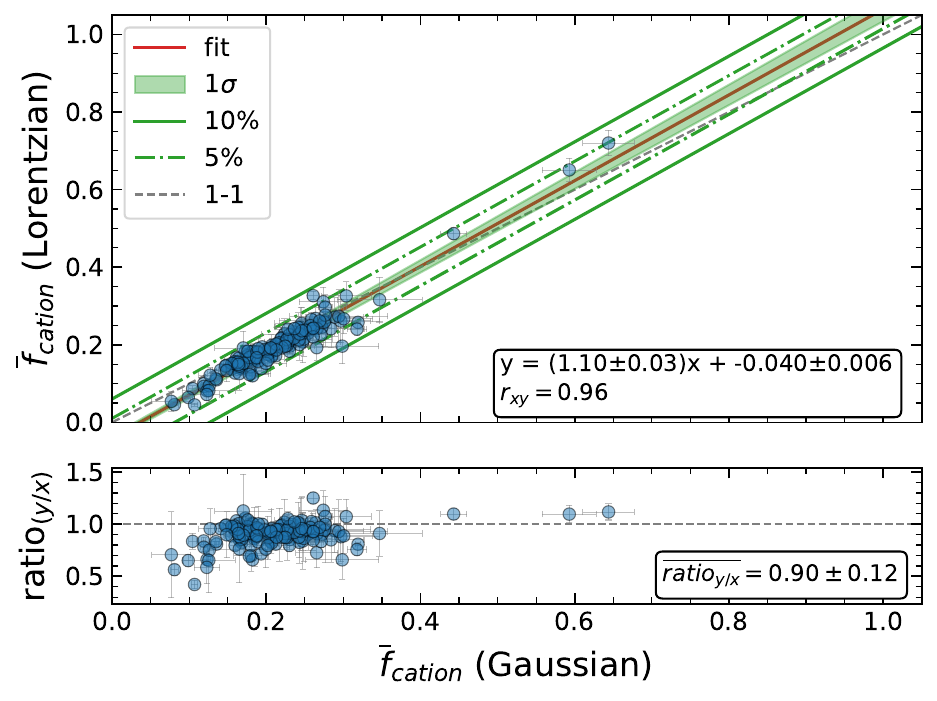}
    \includegraphics[scale=0.37]{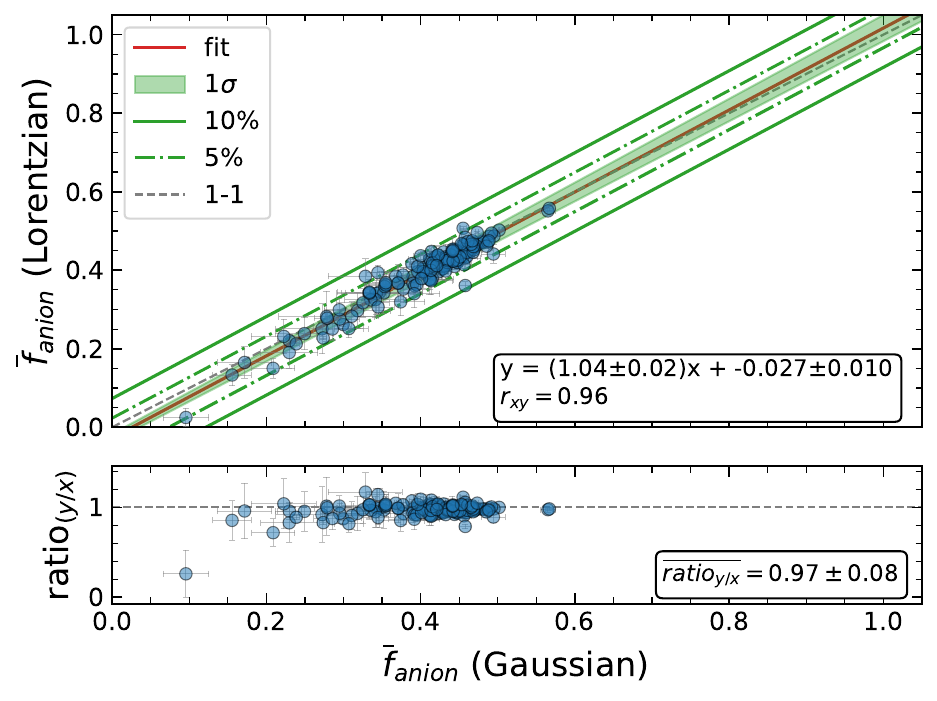} \\
    \includegraphics[scale=0.37]{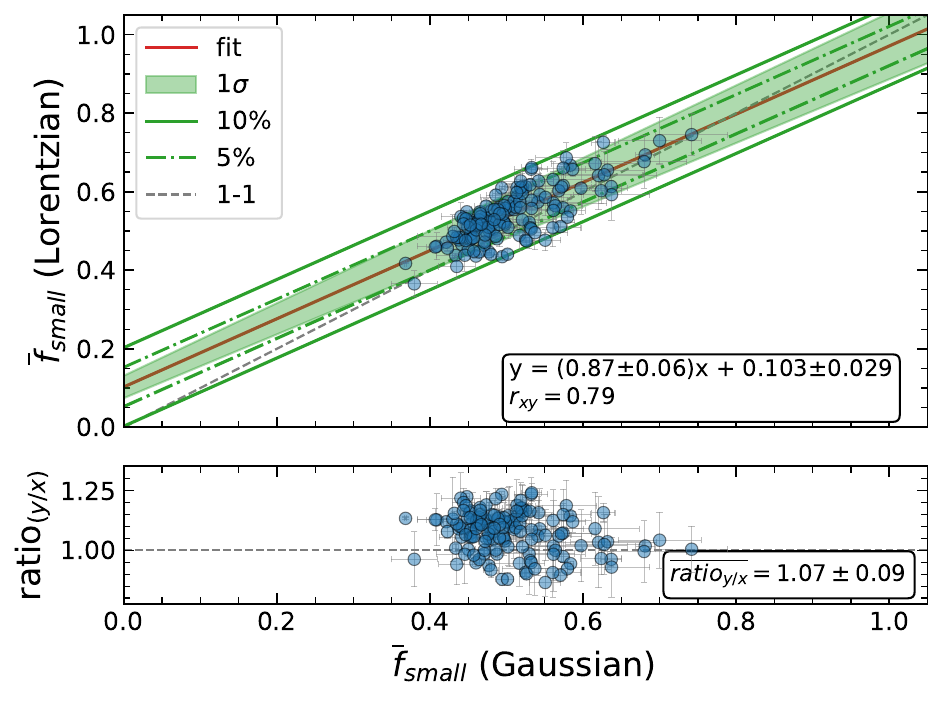}
    \includegraphics[scale=0.37]{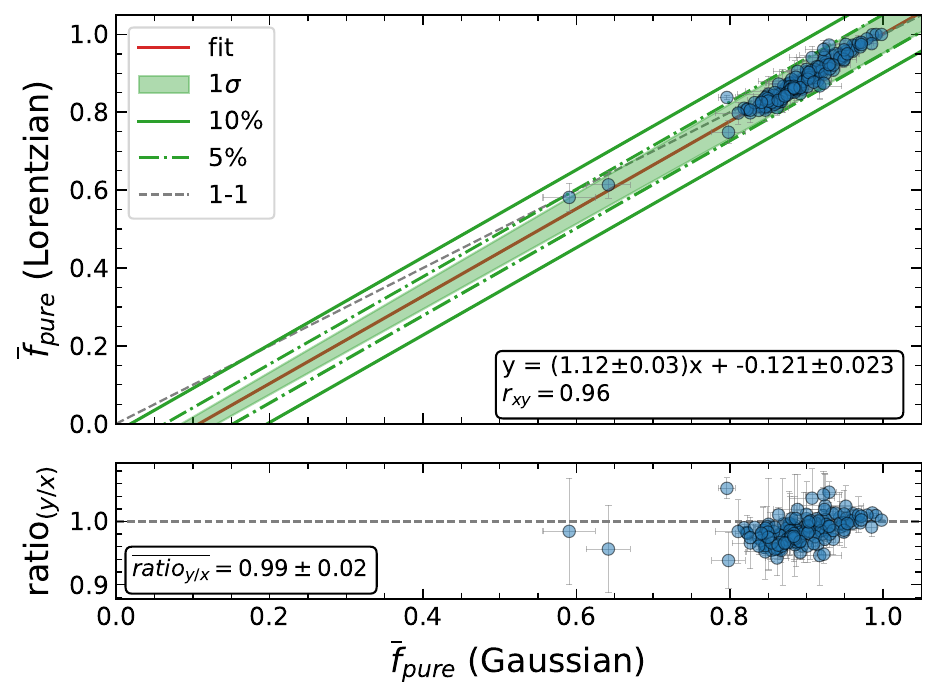}
    \includegraphics[scale=0.37]{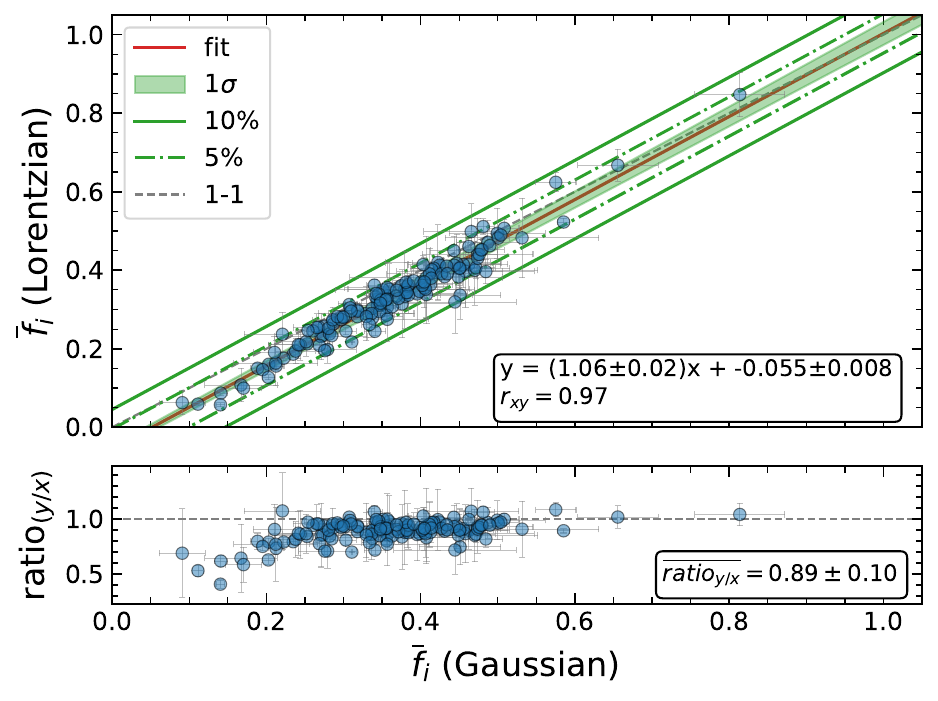} \\
    \includegraphics[scale=0.37]{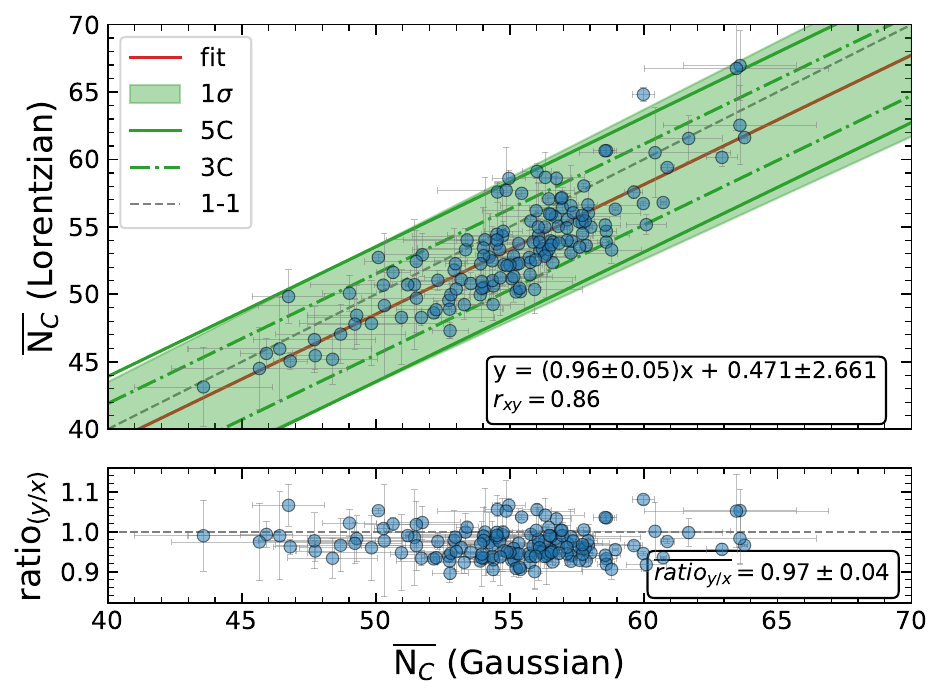}
    \includegraphics[scale=0.37]{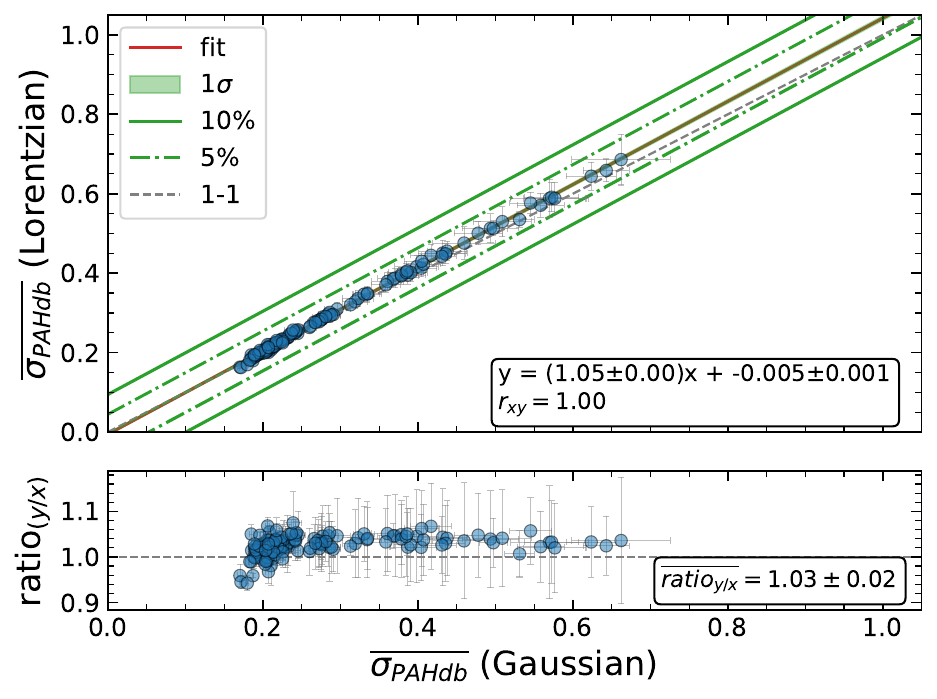}
    \caption{Comparison of the PAHdb-derived PAH properties in the base run (x-axis) and the Lorentzian emission profile run (y-axis). Top row: neutral PAH fraction (left), cation PAH fraction (middle), anion PAH fraction (right); Middle row: small PAH fraction (left), pure PAH fraction (middle), PAH ionization fraction (right); Bottom row: \aNc{} (left), $\sigma_{PAHdb}$ (middle). Linear regression fitting is shown with the red line, 1$\sigma$ dispersion with the green envelope, 5\% and 10\% distance from the best fit line with green dash-dotted and solid lines, respectively, and line of equality with the black dashed line. The regression parameters and the Pearson’s correlation coefficient r$_{xy}$ are provided in the inset. In each case, the ratio of the two values is plotted in the bottom sub-panels, and the average ratio ($\overline{ratio}_{y/x}$) and standard deviation are provided.}
    \label{fig:lorentz_vs_gauss}
\end{figure*}

\subsubsection{PAHdb: Emission Model} \label{sec:PAHdb-model}

In the cascade emission model, used for Paper I, after the absorption of a photon, each PAH will reach a different maximum temperature that depends on its heat capacity. This is followed by a step-wise radiative relaxation from that excitation level (temperature), which is taken into account when selecting this model. Here, we examine additionally results obtained from a simplified emission model, the calculated temperature model, where the maximum attained temperature for each molecule is calculated, and subsequently a blackbody distribution at that fixed temperature is used to describe the emission. 

\begin{figure*}
    \includegraphics[scale=0.37]{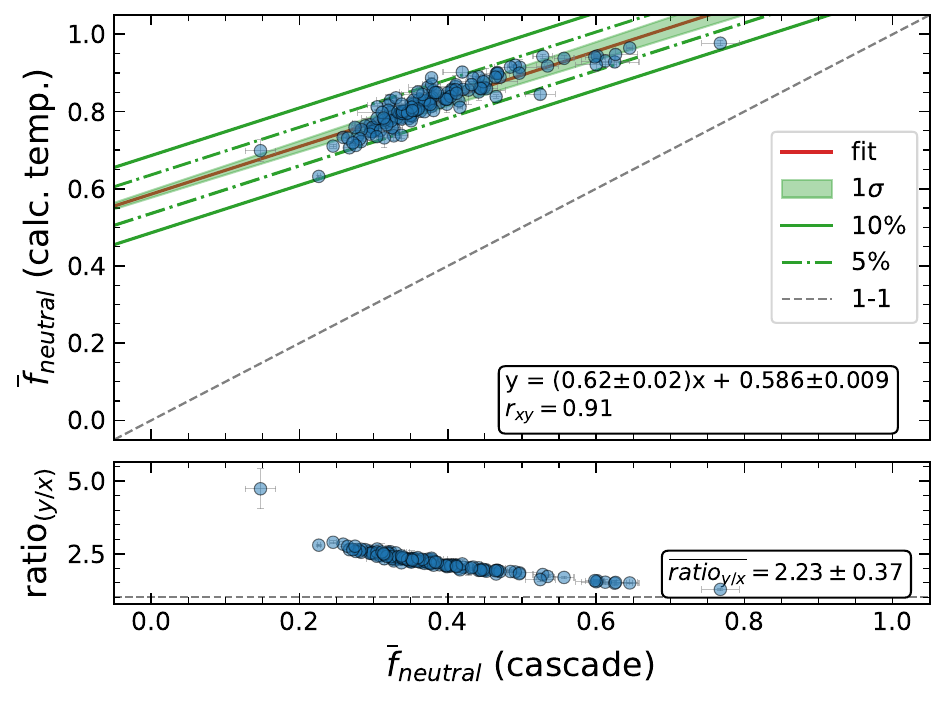}
    \includegraphics[scale=0.37]{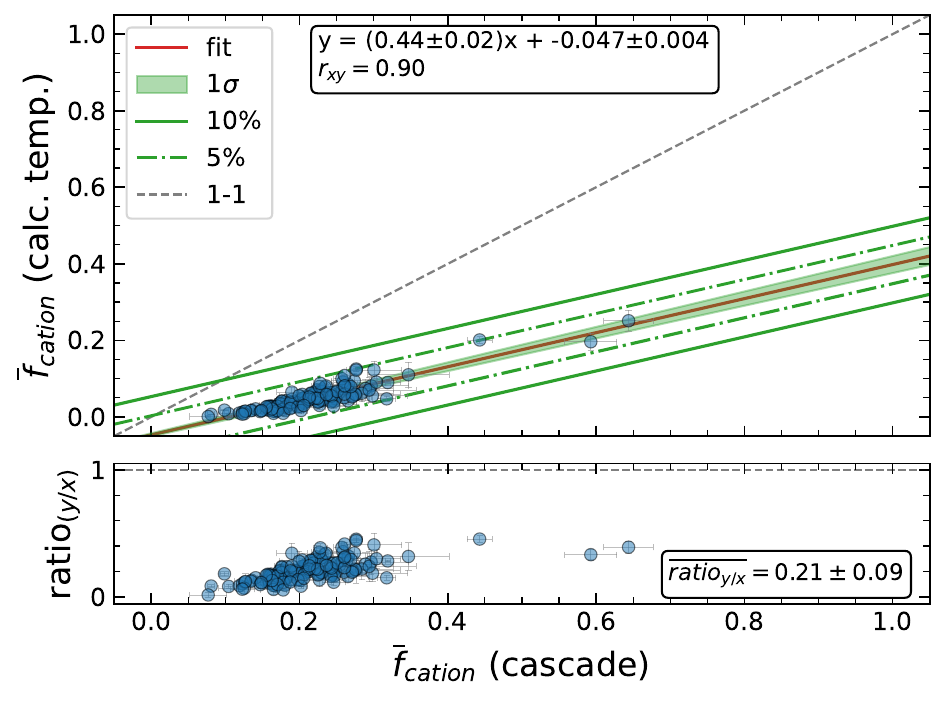}
    \includegraphics[scale=0.37]{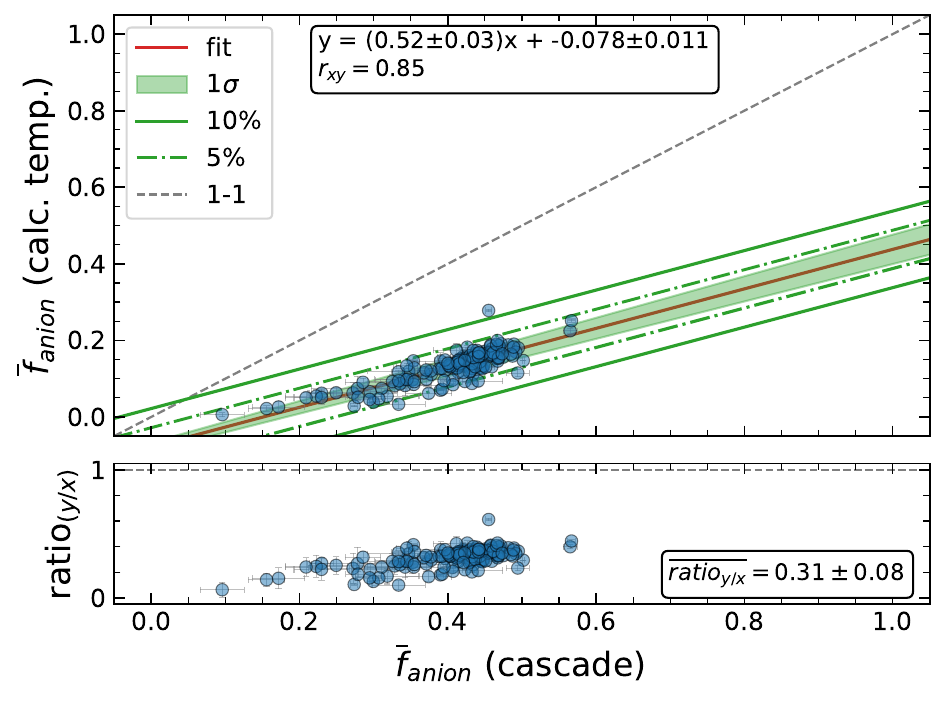} \\
    \includegraphics[scale=0.37]{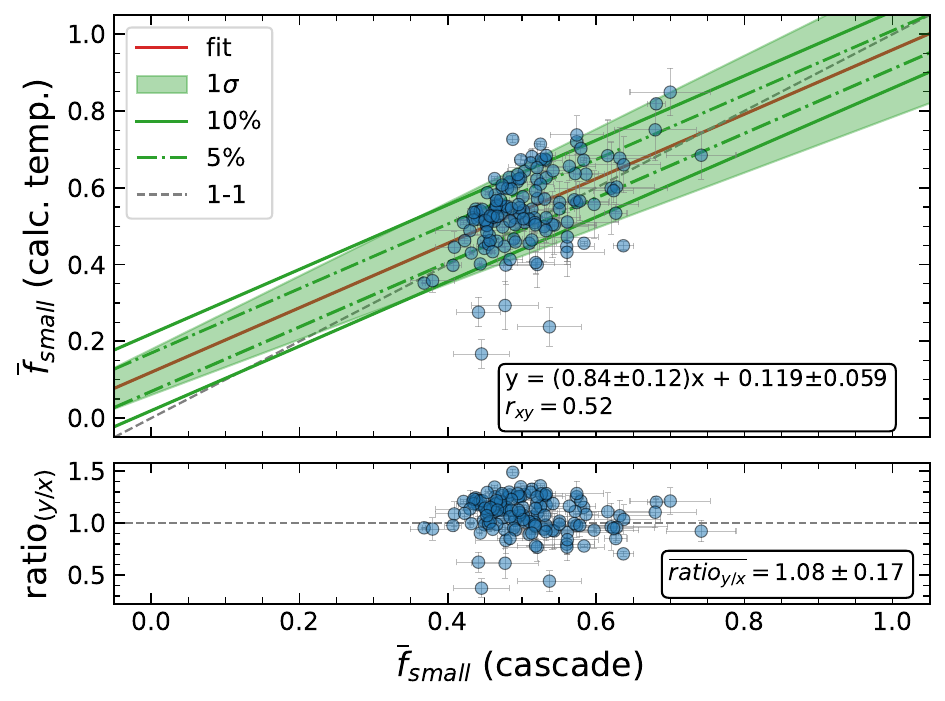}
    \includegraphics[scale=0.37]{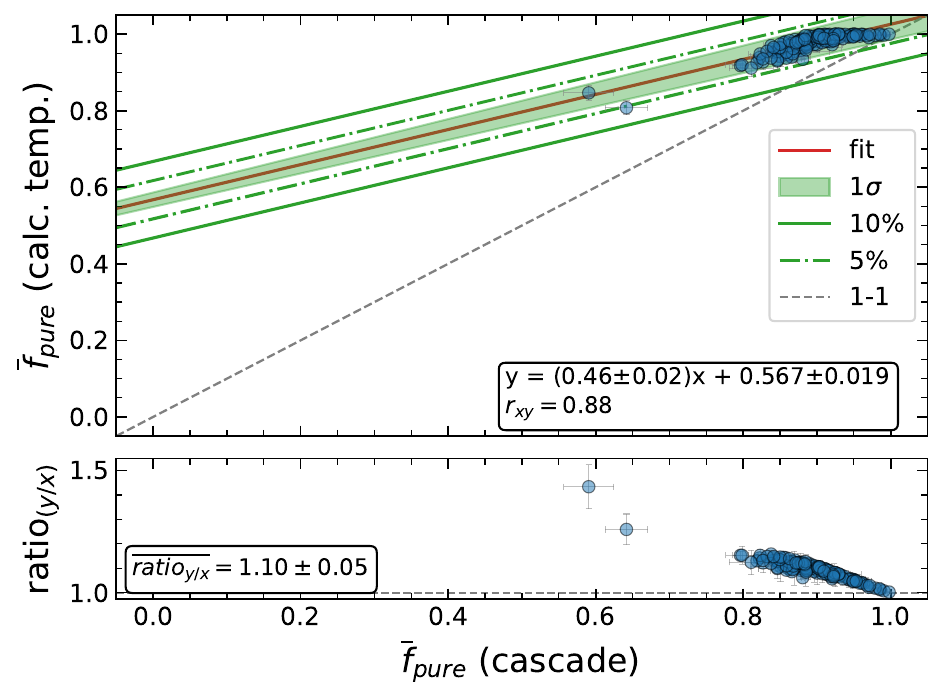}
    \includegraphics[scale=0.37]{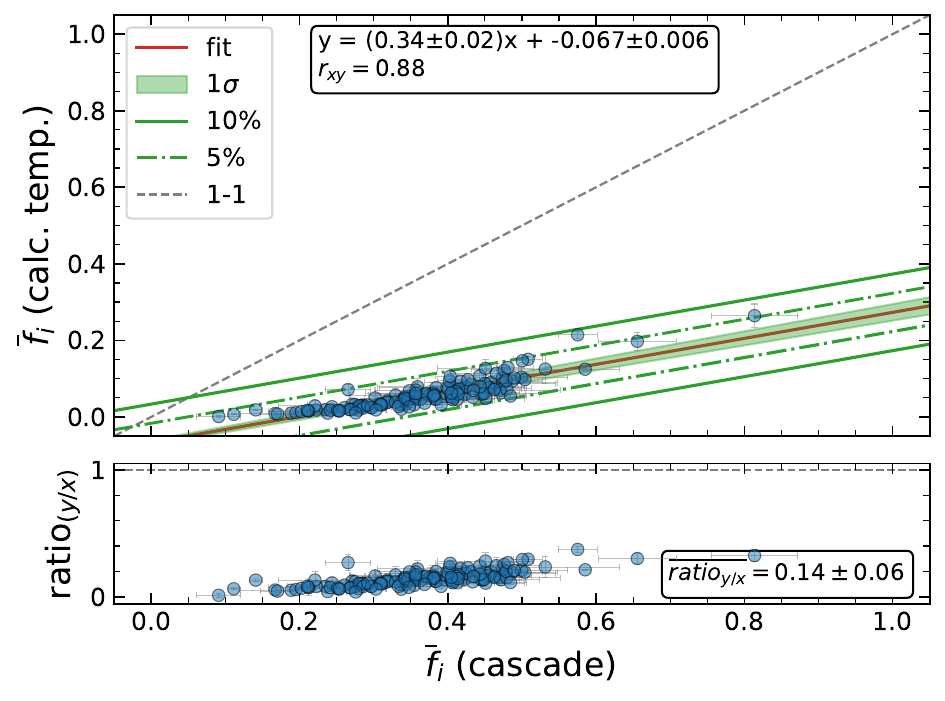} \\
    \includegraphics[scale=0.37]{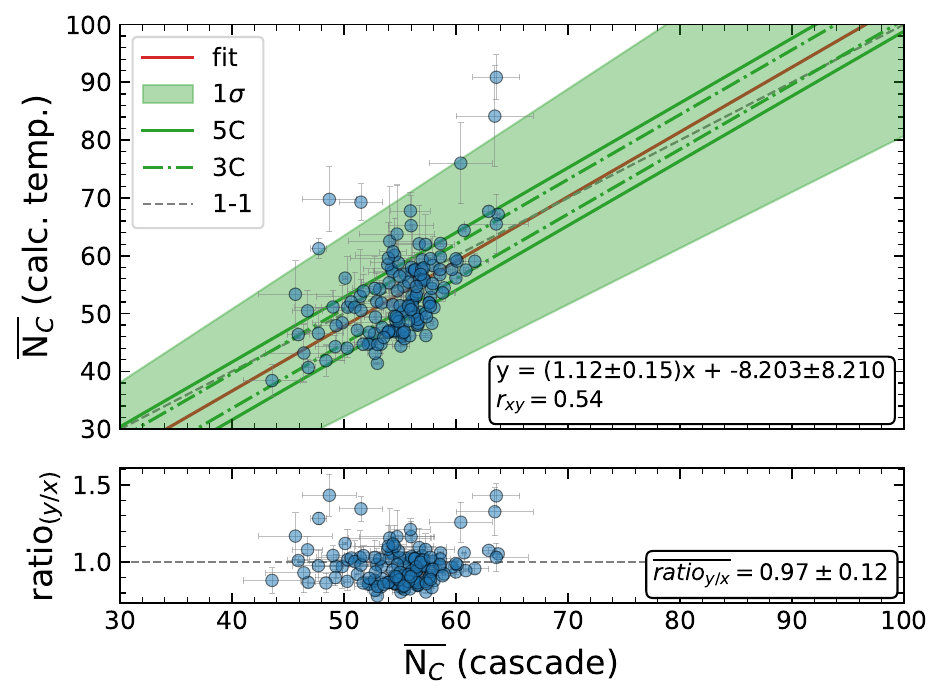}
    \includegraphics[scale=0.37]{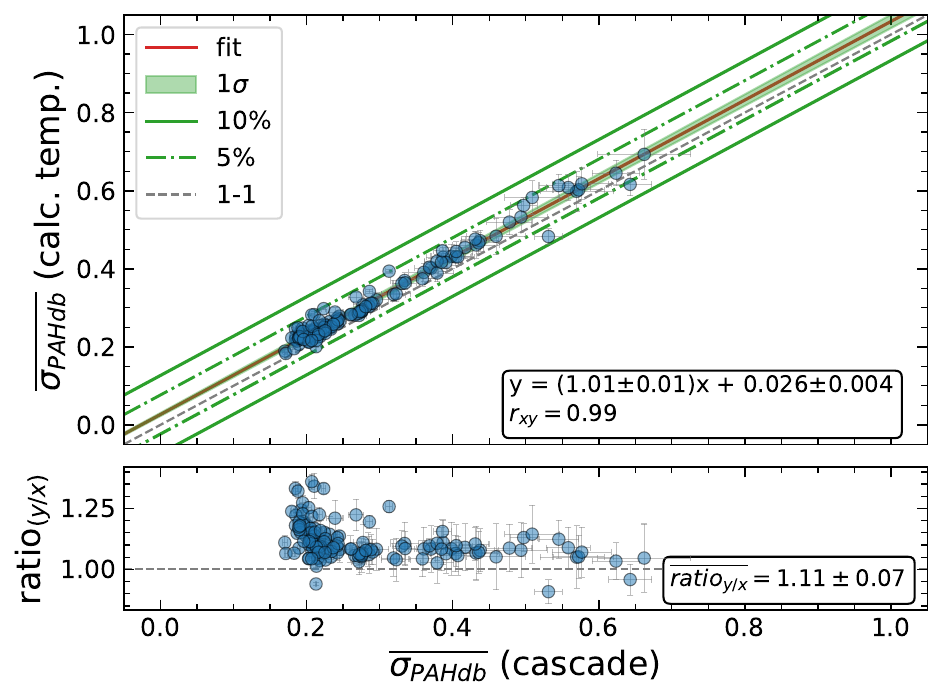}
    \caption{Comparison of the PAHdb-derived PAH properties in the base run (x-axis) and the calculated temperature emission model run (y-axis). Top row: neutral PAH fraction (left), cation PAH fraction (middle), anion PAH fraction (right); Middle row: small PAH fraction (left), pure PAH fraction (middle), PAH ionization fraction (right); Bottom row: \aNc{} (left), $\sigma_{PAHdb}$ (middle). Linear regression fitting is shown with the red line, 1$\sigma$ dispersion with the green envelope, 5\% and 10\% distance from the best fit line with green dash-dotted and solid lines, respectively, and line of equality with the black dashed line. The regression parameters and the Pearson’s correlation coefficient r$_{xy}$ are provided in the inset. In each case, the ratio of the two values is plotted in the bottom sub-panels, and the average ratio ($\overline{ratio}_{y/x}$) and standard deviation are provided.}
    \label{fig:calc_temp_vs_cascade}
\end{figure*}

\subsubsection{PAHdb: Redshift application} \label{sec:PAHdb-redshift}

\begin{figure*}
    \includegraphics[scale=0.37]{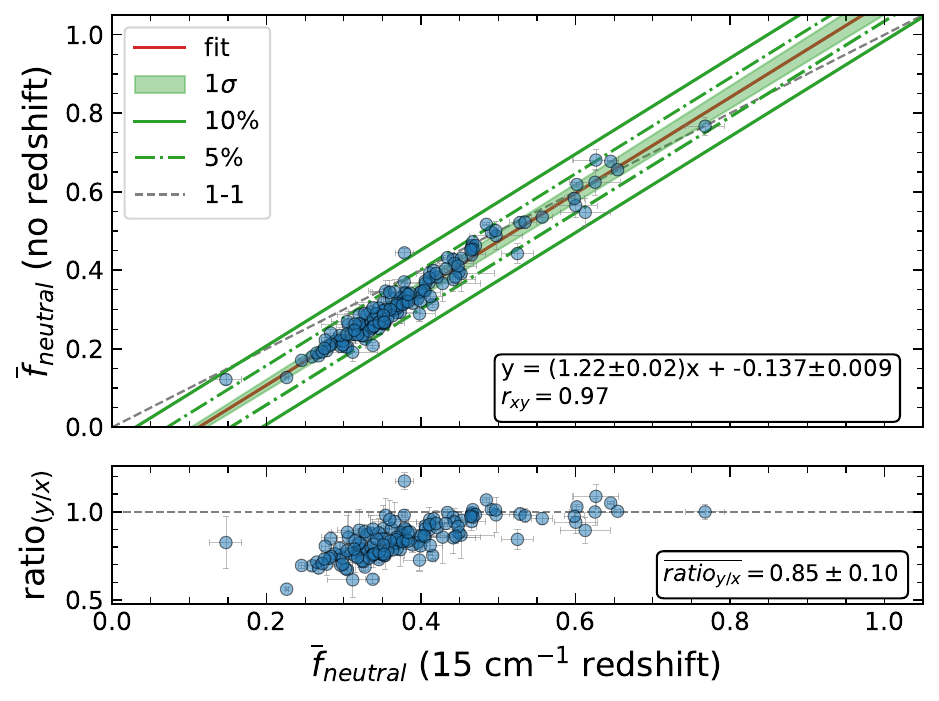}
    \includegraphics[scale=0.37]{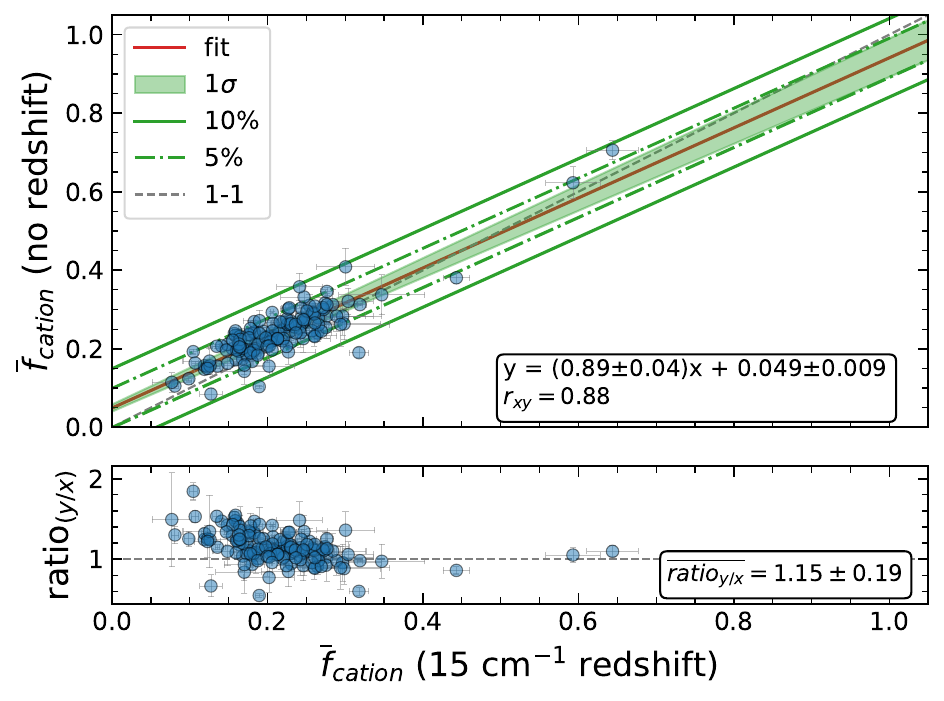}
    \includegraphics[scale=0.37]{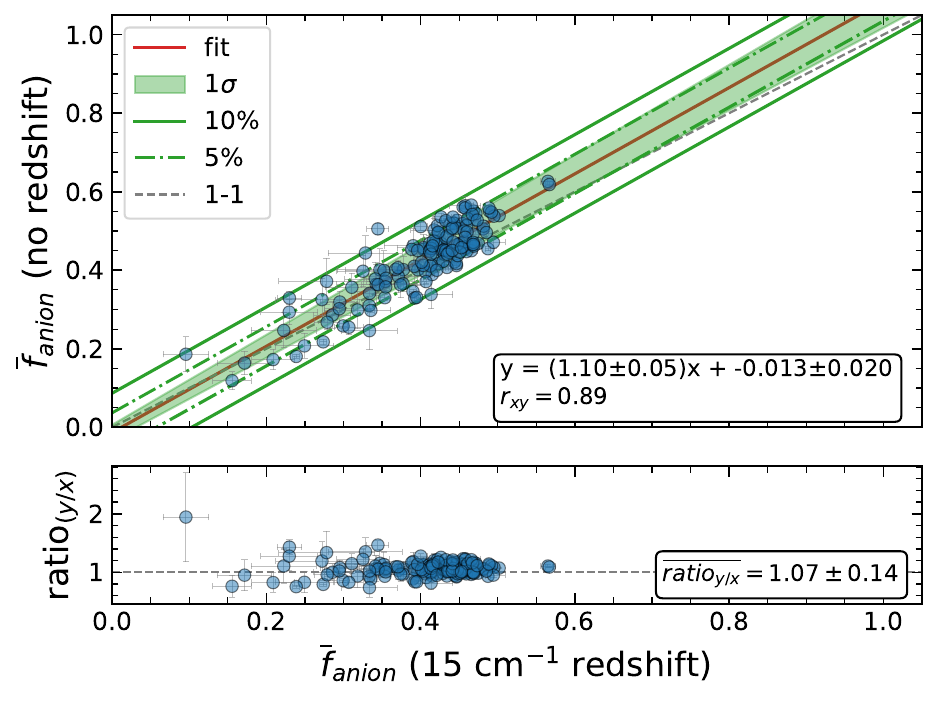} \\
    \includegraphics[scale=0.37]{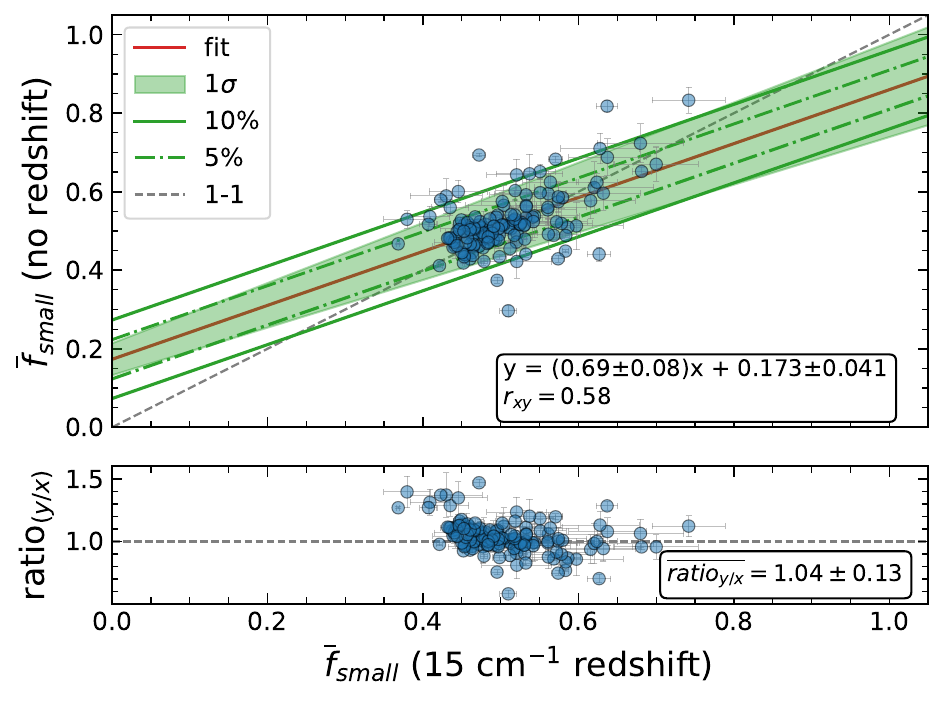}
    \includegraphics[scale=0.37]{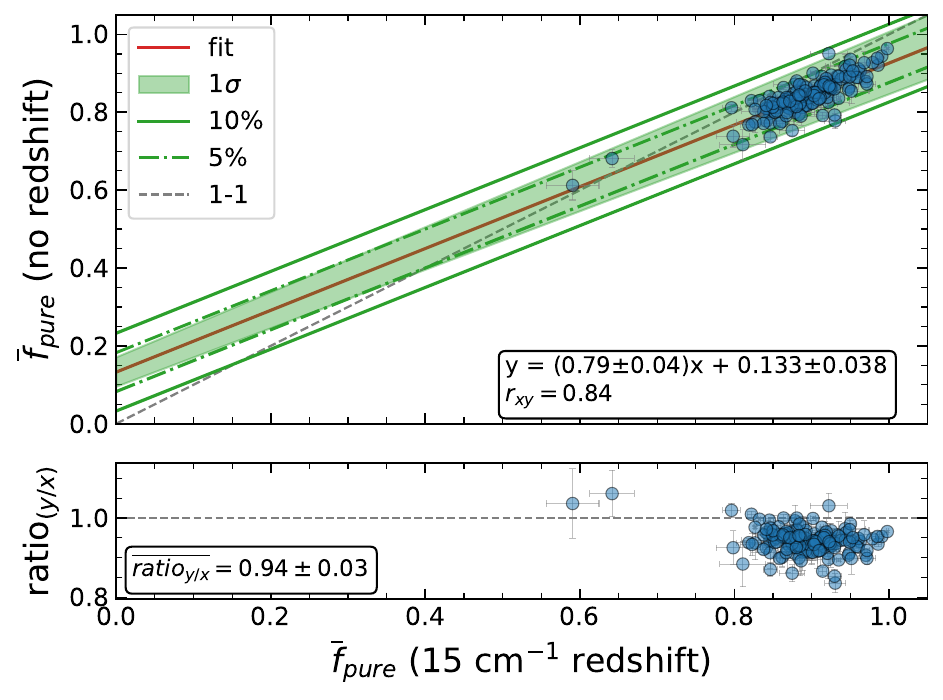}
    \includegraphics[scale=0.37]{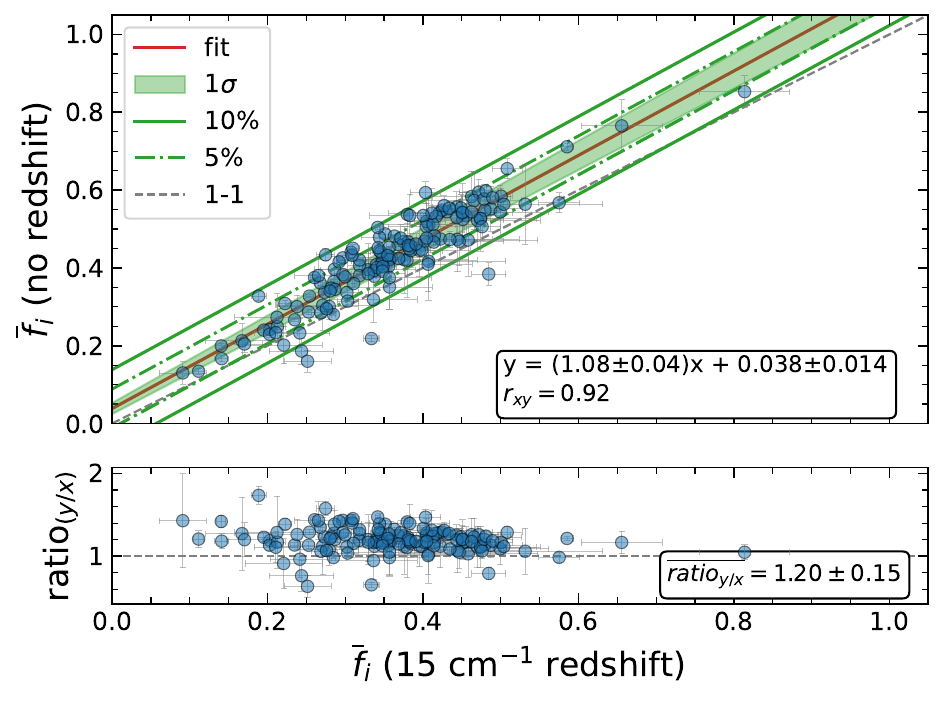} \\
    \includegraphics[scale=0.37]{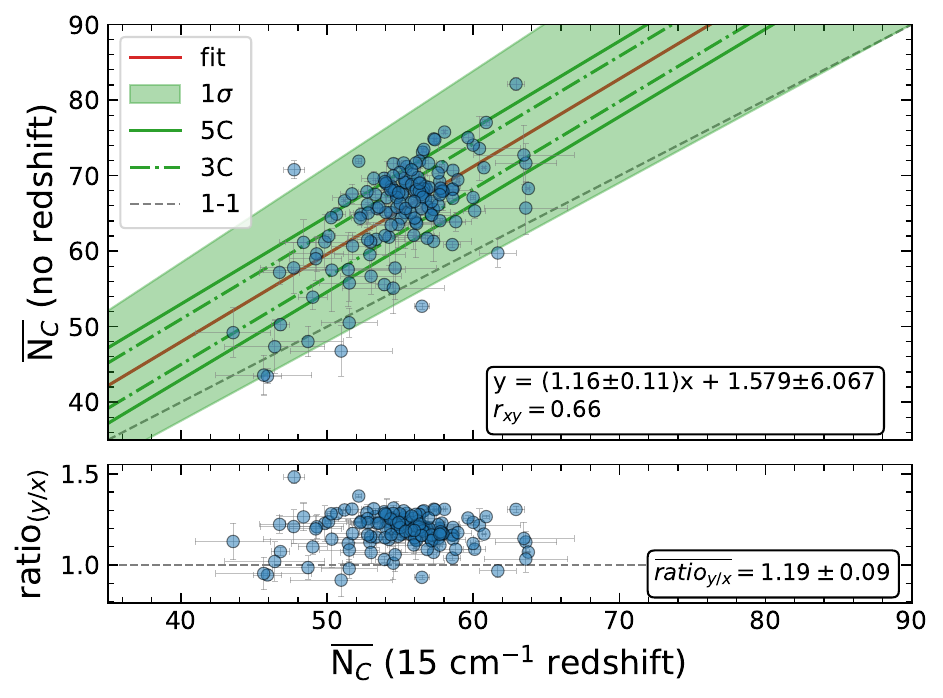}
    \includegraphics[scale=0.37]{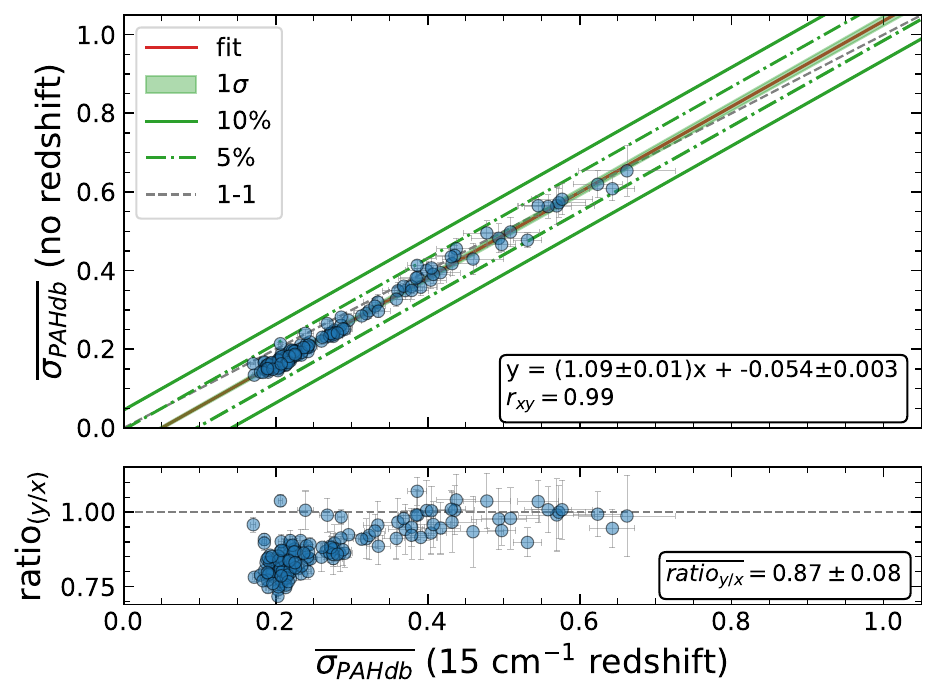}
    \caption{Comparison of the PAHdb-derived PAH properties in the base run (x-axis) and when modeling without redshift application (y-axis). Top row: neutral PAH fraction (left), cation PAH fraction (middle), anion PAH fraction (right); Middle row: small PAH fraction (left), pure PAH fraction (middle), PAH ionization fraction (right); Bottom row: \aNc{} (left), $\sigma_{PAHdb}$ (middle). Linear regression fitting is shown with the red line, 1$\sigma$ dispersion with the green envelope, 5\% and 10\% distance from the best fit line with green dash-dotted and solid lines, respectively, and line of equality with the black dashed line. The regression parameters and the Pearson’s correlation coefficient r$_{xy}$ are provided in the inset. In each case, the ratio of the two values is plotted in the bottom sub-panels, and the average ratio ($\overline{ratio}_{y/x}$) and standard deviation are provided.}
    \label{fig:no_shift_vs_shift}
\end{figure*}

The astronomical PAH spectrum is the product of highly vibrationally excited molecules, and therefore anharmonic effects are expected to introduce a small redshift to the emission band peak positions compared to the band positions in absorption. In Paper I a 15 cm$^{-1}$ redshift was applied, a typical value consistent with that measured in a number of experimental mid-IR studies \citep[e.g.,][]{Williams1995, Cook1998}. However based on a modeling study on small PAHs, \cite{Mackie2018} argue that the peak position of emission bands, in general, does not vary in cascade spectra but rather develops pronounced low frequency wings, and thus applying a post-facto redshift is unnecessary. Here, we omit a 15 cm$^{-1}$ redshift and compare the results with that from the base run.

\subsubsection{PAHdb: v4.00-\textalpha{} library \label{sec:v4.00}}

The next version of PAHdb's library of DFT-computed spectra, v4.00-\textalpha{}\footnote{The ``\textalpha'' nomenclature indicates a pre-release library version, and it can be downloaded from \href{www.astrochemistry.org/pahdb/downloads}{www.astrochemistry.org/pahdb/downloads}.}, incorporates PAHs with straight edges \citep{Ricca2018}, PAHs with armchair edges \citep[computed using the 4-31G basis set from][]{Ricca2019}, nitrogenated PAHs \citep[PANHs; computed using the 6-31G* basis set from][]{Ricca2021}, and a new set of large, irregularly edged, pure PAHs molecules \citep[computed using the 6-31G* basis set from][]{Ricca2024}. These enhancements to the library, for the first time, enables PAHdb to fit the entire astronomical 6.2 \micron{} PAH band, where earlier versions were unable to match its blue side, as well as provides a much better fit to the 10-15 \micron{} region. Library v4.00-\textalpha{} expands uppon the content of v3.20, including a larger number of PAH molecules with a better representation in terms of PAH size (although keeping the same size range with v3.20), and a comparable fraction of neutral (42\%) and cationic (47\%) PAHs. Specifically, v3.20 includes 2906 PAHs with \Nc{} $< 50$ (69\%), 939 PAHs with $50 \leq$ \Nc{} $< 75$ (22\%), and 388 PAHs with \Nc{} $\geq 75$ (9\%), while v4.00-\textalpha{} includes 5406 PAHs with \Nc{} $< 50$ (50\%), 4683 PAHs with $50 \leq$ \Nc{} $< 75$ (43\%), and 801 PAHs with \Nc{} $\geq 75$ (7\%) A comparison is made between PAHdb-derived parameters when using either v3.20 or the v4.00-\textalpha{} library versions of PAHdb, while leaving the remaining modeling parameters the same.

\begin{table}[]
    \centering
    \caption{The four configurations considered for PAHdb v4-00-\textalpha{} modeling. In all cases, the remaining modeling parameters are identical to the v3.20 base run i.e., an 8 eV excitation energy, the cascade emission model, and Gaussian line profiles with FWHM 15 cm$^{-1}$ are used.}
    \begin{tabular}{lcc}
    \hline
    \hline
    Configuration & Redshift & Composition \\
    \hline
    Case 1 & 15 cm$^{-1}$ & pure PAHs + v4.00-\textalpha{} PANHs \\
    Case 2 & - & pure PAHs + v4.00-\textalpha{} PANHs \\
    Case 3 & - & pure PAHs + v3.20 PANHs \\
    Case 4 & - & pure PAHs \\
    \hline
    \end{tabular}
    \label{tab:4cases}
\end{table}

In addition, a sensitivity analysis is performed on the v4.00-\textalpha{} library, exploring different configurations (Table \ref{tab:4cases}), based on the sensitivity analysis results on v3.20. Specifically, as discussed in Sections \ref{sec:v320-results} and \ref{sec:4.00-configurations}, the application or omission of a 15~cm$^{-1}$ redshift, as well as the inclusion or omission of PANHs, presents the highest variation in PAHdb-derived parameters. Therefore, we examine four configurations to understand the effect of each configuration on the results: \textit{Case 1}: base run parameters; \textit{Case 2}: applying no redshift; \textit{Case 3}: keeping only the v3.20 PANHs by excluding the $\sim$2000 newly added ones, and applying no redshift; and \textit{Case 4}: excluding all PANHs and applying no redshift.
In each of the four cases we examine the distributions of the derived parameters, and their PAH charge and composition breakdowns, to deduce the optimal configuration adopted for the PAHdb modeling.

\subsection{Spectral resolution} \label{sec:spitzer_vs_jwst}

Using the 4.00-\textalpha{} version of the computational library, we explore the sensitivity of the modeling and derived parameters from observations with differing spectral resolution. Specifically, for a set of galaxies observed with both \textit{Spitzer}-IRS and JWST MIRI-MRS, we perform PAHdb fitting and compare the derived PAH properties: i) directly from their decomposed spectra; ii) by degrading the resolution of the JWST observations to the \textit{Spitzer}-IRS resolution, and then compare the respective decomposed spectra. The selection of galaxies is done to ensure as much overlap between the field-of-views of the different instruments, i.e., the IRS slit and the inner MIRI-MRS IFU channels cover approximately the same area. The galaxies used are obtained from the Cylce 2 JWST GO Program 3368 ``A JWST Survey of Ultraluminous Infrared Galaxies"\footnote{\url{www.stsci.edu/jwst/science-execution/program-information?id=3368}} (PI: Lee Armus). The reduced Stage-3 1D spectra of all 4 MIRI-MRS channels were obtained from MAST using the Python \texttt{astroquery} package. The individual spectra have been scaled based on the median values of their overlapping wavelength regions, and combined to a single spectrum. Redshift correction was applied to the combined spectrum, and spectral decomposition with \textsc{pahfit} was performed prior to PAHdb modeling. For the second approach, we used the \texttt{specutils\footnote{\url{https://github.com/astropy/specutils}}} Python package to perform smoothing and resampling of the JWST spectrum to the IRS resolution and dispersion grid.  

\subsection{MIR Spectral Decomposition Methods} \label{sec:different_codes}

Different methods and modeling codes for decomposing the MIR spectrum of galaxies can produce varying results on the recovery of the PAH emission component and, subsequently, on the PAHdb-derived parameters, due to their underlying assumptions in their compound models. Here, we focus on two of the most popular and publicly available\footnote{\cite{Donnan2024} have recently presented a galaxy decomposition code, expected to become publicly available soon.} galaxy decomposition codes: i) \textsc{pahfit} and ii) \textsc{cafe: continuum and feature extraction} tool\footnote{\url{https://github.com/GOALS-survey/CAFE/}} (Diaz Santos et al. in prep.). These codes typically employ a compound model with differences in, e.g., the treatment of the dust continuum and dust extinction, properties directly affecting the spectral shape of the recovered PAH emission component. More specifically, among the differences in the compound models of the two codes is the inclusion and modeling of an AGN continuum in \textsc{cafe} (currently absent in \textsc{pahfit}), the dust continuum component treatment, where in \textsc{cafe} depends on the dust grain size and composition, as well as on the type of heating source (e.g., stellar component, starburst templates) as opposed to modified blackbodies at fixed temperatures used in \textsc{pahfit}, and the parametrization of dust extinction. \textsc{pahfit} uses a synthetic attenuation curve based on the silicate profile from \cite{Kemper2004}, with the addition of a broad 18 \micron{} feature characterized by a Lorentzian absorption profile together with a raising power law extending towards the NIR. In contrast, the attenuation curve used in \textsc{cafe} is derived from \cite{Ossenkopf1992}. It's important to note that in \textsc{pahfit}, the attenuation curve is applied uniformly to all features, including PAHs, emission lines, and various blackbody dust components. However, in \textsc{cafe}, different dust components are subject to distinct opacities, which can be adjusted by the model, with the attenuation affecting PAHs and emission lines specifically tied to the warm dust component.

The recovered PAH spectrum therefore directly depends on the underlying assumptions in the compound model in each code. As such, differences in the recovered PAHdb parameters can be expected. The decomposition under those two different codes and the subsequent PAHdb modeling is performed on the JWST spectra described in Section \ref{sec:spitzer_vs_jwst}.

\section{Results and Discussion} \label{sec:results}

We have conducted a sensitivity analysis on the spectral decomposition and PAH spectral modeling of galaxies, to assess and characterize the stability and variance introduced on the derived parameters of the average galaxy PAH population, exploring a grid of different modeling parameters, compared to a reference base run. 

\subsection{PAHFIT Decomposition}

The 5--15 \micron{} spectrum, observed with \textit{Spitzer}-IRS SL module, includes the most prominent PAH emission features, centered at 6.2, 7.7, 8.6, 11.2, and 12.7 \micron{}. In our spectral decomposition sensitivity analysis, we examine whether making use of the SL-alone spectra can adequately model and retrieve the 6--15 \micron{} PAH features. While inclusion of the LL \textit{Spitzer}-IRS spectral segment to the \textsc{pahfit} decomposition can, potentially, help better characterize the underlying continuum, the recovered PAH band ratios are in good agreement within their uncertainties with the SL-only modeling. This shows that the \textsc{pahfit} fitting of the SL segment alone can provide adequate decomposition of the PAH spectrum (Figures \ref{fig:pahfit_sl+ll}-\ref{fig:mod_obs}) and the recovered relative band intensities (Figure \ref{fig:pahfit_comparisons}). 
Overall, the SL and SL+LL segments decomposition provide consistent modeling and recovery of the 5--15 \micron{} pure PAH emission spectrum. We note that the attenuation curve modeling in galaxies with more silicate absorption can be more strongly dependent on the wavelength coverage, and for those cases, SL+LL wavelength coverage should be preferred. In Section \ref{sec:pahfit_vs_cafe} we further examine the recovery of the PAH emission spectrum on JWST observations, using the \textsc{cafe} Python galaxy spectral decomposition code, and compare with \textsc{pahfit}.

\subsection{PAHdb v3.20} \label{sec:v320-results}

PAHdb allows one to convolve the spectral bands with different line profiles including Gaussians and Lorentzians with a given FWHM. Figure \ref{fig:gaussian_fwhm_10_vs_15} shows that using narrower FWHM (10 cm$^{-1}$) Gaussian profiles compared to the base run (15 cm$^{-1}$) has a small effect on the charge and composition breakdown, but has a greater impact on the size breakdown. Band profiles of smaller width require an additional number of PAHs to compensate for and model the observed PAH bands. Indeed, $\sim$10-20 PAH molecules, on average, are used additionally in the fitting when 10 cm$^{-1}$ profiles are used, compared to the base run. PAHdb v3.20 includes a relatively larger number of small (\Nc{} $< 50$) to intermediate ($50 \leq$ \Nc{} $< 75$) sized PAHs, with respect to larger (\Nc{} $\geq 75$) PAHs (see Section \ref{sec:v4.00}). Since smaller PAHs have a much higher variety of edge structures (solo, duo, trio, quartet) than do large PAHs, there is much more variation in the 10.5 to 14 \micron{} band structure with which to choose from in fitting. Consequently, this ultimately results in a higher fraction of small (\Nc{} $<$ 50) PAHs in the fit (Figure \ref{fig:gaussian_fwhm_10_vs_15}, middle row, left panel), and a larger scatter is observed in the derived \aNc{} (Figure \ref{fig:gaussian_fwhm_10_vs_15}, bottom row, left panel). 

Regarding the charge breakdown, neutral PAHs are consistent with each other in both cases while there is a slightly higher usage of cationic PAHs vs anions in the FWHM (10 cm$^{-1}$) case (see discussion following the Lorentzian profile analysis in the previous section). As a result, the deduced \afi{} (defined as the ratio of the number density of cationic to the sum of number densities of neutral and cationic PAHs; see Paper I), is consistent between the two cases (Figure \ref{fig:gaussian_fwhm_10_vs_15}, bottom row, left panel). Fitting uncertainties are slightly higher for the FWHM (10 cm$^{-1}$) case, for values between 0.15 and 0.3, but overall comparable to the base run. Note that fitting uncertainties $>$ 0.3 are predominantly due to the under-fitting of the blue side of the 6.2 \micron{} feature and is not representative of the overall 5-15 \micron{} PAH spectrum modeling (see discussion in Paper I, and Section \ref{sec:v4.00}).

Lorentzian emission profiles with 15 cm$^{-1}$ FWHM provide a consistent charge breakdown with the base run (Figure \ref{fig:lorentz_vs_gauss}). The main trade-off is observed between the fraction of neutral and cationic PAHs ($\sim$10\%), yet consistent within their uncertainties. Similarly, the deduced \afi{} is consistent with the base run within their uncertainties, varying $\sim$10\%. A similar trend is observed for the fraction of pure vs nitrogen-containing PAHs. Although the average ratio of $\overline{f}_{small}$ between the two runs is $\sim$10\%, the Pearson's coefficient is r$_{xy} = 0.8$ indicative of moderate scatter (Figure \ref{fig:lorentz_vs_gauss}, middle row, left panel), however it is justified given the comparable uncertainties of the quantities in both runs. This also projects, though to a lesser extent, to \aNc{} (r$_{xy} = 0.86$). PAHdb modeling uncertainties are slightly higher in the Lorentzian case but overall consistent with the base run.

When modeling the 5-15 \micron{} PAH spectrum, the contribution of neutral PAHs to the fit is mainly driven by the 11.2 \micron{} PAH band, as the bands between 6--9 \micron{} are primarily due to charged PAHs, with a smaller contribution from neutral PAHs. Therefore, changes in the FWHM or the emission line profile should not introduce significant variation to the fraction of neutral PAHs because the 11.2 \micron{} PAH feature is reproduced predominantly by neutral PAHs. Similarly, the charged PAH fraction (cation and anion) in PAHdb decomposition is mainly controlled by the 6--9 \micron{} PAH bands, produced primarily by PAH cations, consisting of multiple sub-components and blended features \citep[e.g.,][]{Peeters2017}. Due to this increased complexity in the spectrum from multiple sub-features and feature blending, the charged PAH fraction is more sensitive to profile shape and width variations. 

\begin{figure*}
    \centering
    \includegraphics[scale=0.55]{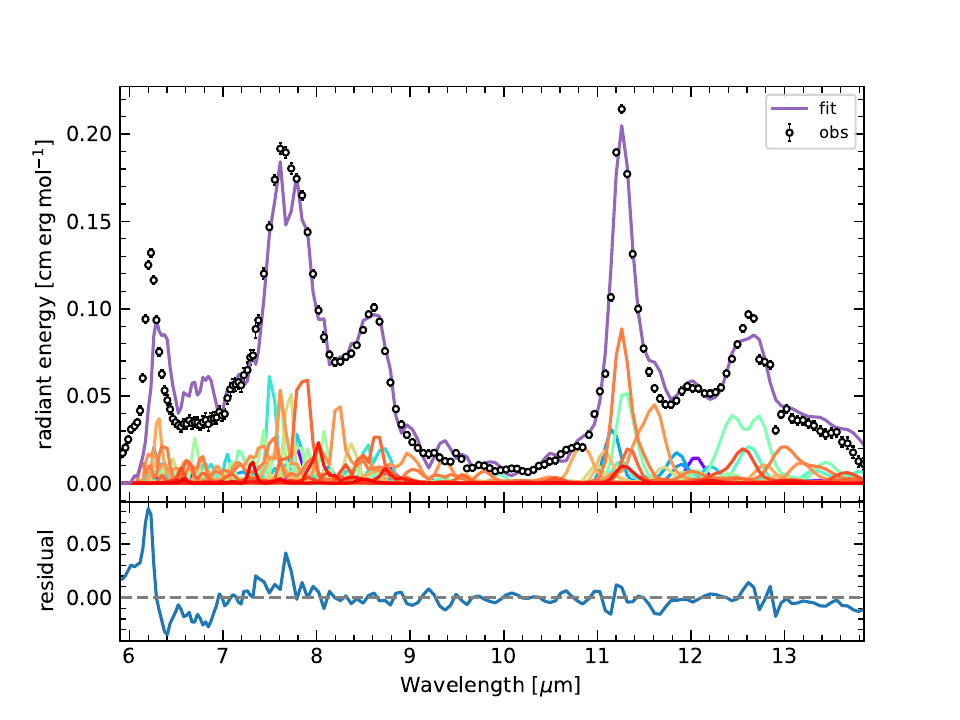}
    \includegraphics[scale=0.55]{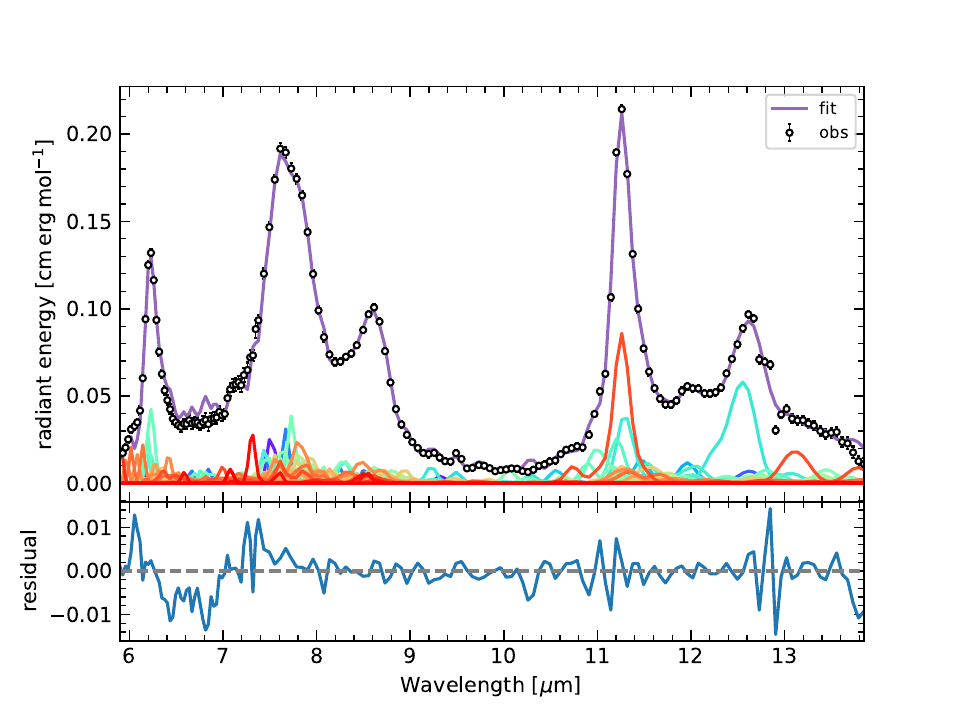}
    \caption{Comparison between the PAHdb v3.20 (left panel) and PAHdb v4.00-\textalpha{} library version (right panel) modeling of the 6--15 \micron{} PAH spectrum of galaxy IRAS 05129+5128. A considerable improvement of the modeling of the 6.2 \micron{} feature is now achieved with the v4.00-\textalpha{} library.}
    \label{fig:example_320_vs_400a}
\end{figure*}

In the simplified description of the calculated temperature model, the stepwise radiative relaxation from the initial excitation level, fully accounted for in the cascade emission model, is not considered. The fitting uncertainties are $\sim$10\% higher (Figure \ref{fig:calc_temp_vs_cascade}, bottom row, middle panel). The most notable deviation from the base run is the significantly higher fraction of neutral PAHs used in the fit, at the cost of a lower fraction of charged PAHs, which further translates to an $\sim$80\% drop in the deduced \afi. Examination of the respective neutral and charged PAH contribution to the fits reveals a significant emission component from neutral PAHs used to model the 6.2 \micron{} feature, a band predominantly attributed to charged PAHs \citep[e.g.,][]{Allamandola1999}. Furthermore, features at 11.9, 12.8, 13.6, 14.2, and 14.8 \micron{}, typically ascribed to a combination of neutral and charged PAHs \citep[e.g.,][]{Allamandola1999}, are modeled using almost exclusively neutral PAHs in the calculated temperature model. The elevated usage of neutral PAHs is also linked to the $\sim$10\% decrease of the nitrogen-containing PAHs in the fit, due to a database bias in the content of neutral ($\sim$40\%) vs cationic ($\sim$60\%) nitrogen-containing PAHs. There is a $\sim$10\% increase in the small PAHs fraction, consistent within the uncertainties, however with significantly more scatter (r$_{xy}=0.52$) also present in the deduced \aNc. This is because in the PAHdb breakdown small PAHs are considered those with $20 <$ \Nc{} $ < 50$, a somewhat broad range, and therefore variance in the usage of small PAHs will introduce scatter in the derived \aNc.  

\begin{figure*}
    \includegraphics[scale=0.37]{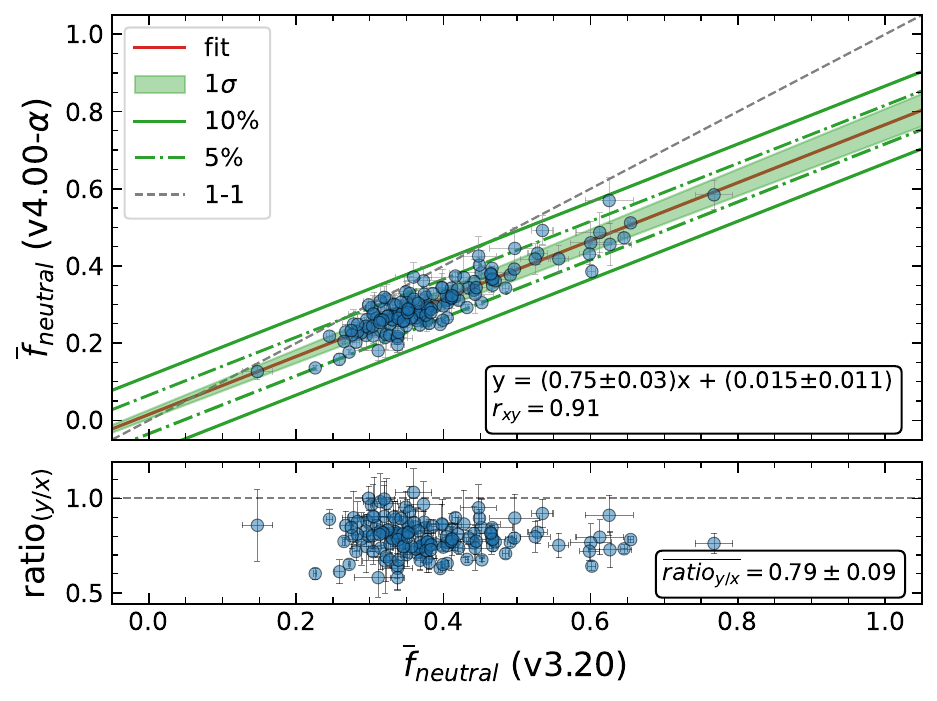}
    \includegraphics[scale=0.37]{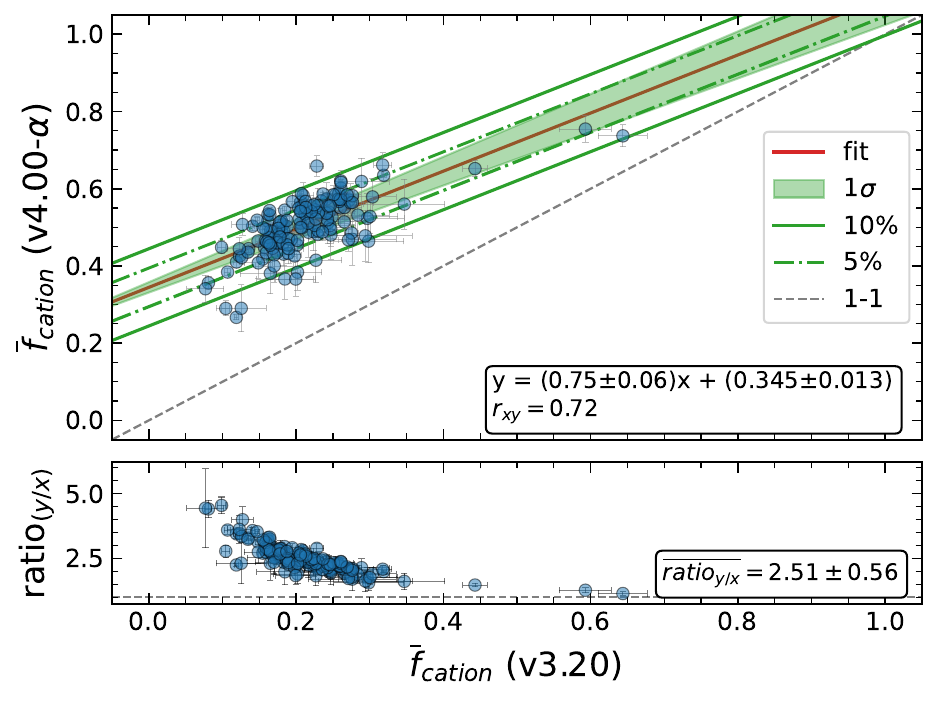}
    \includegraphics[scale=0.37]{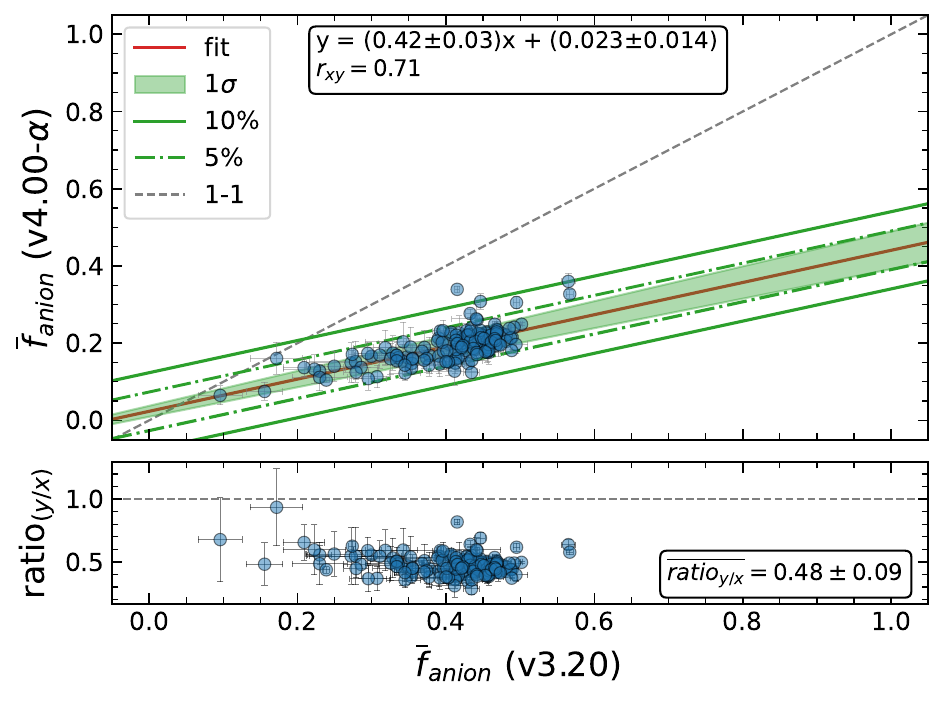} \\
    \includegraphics[scale=0.37]{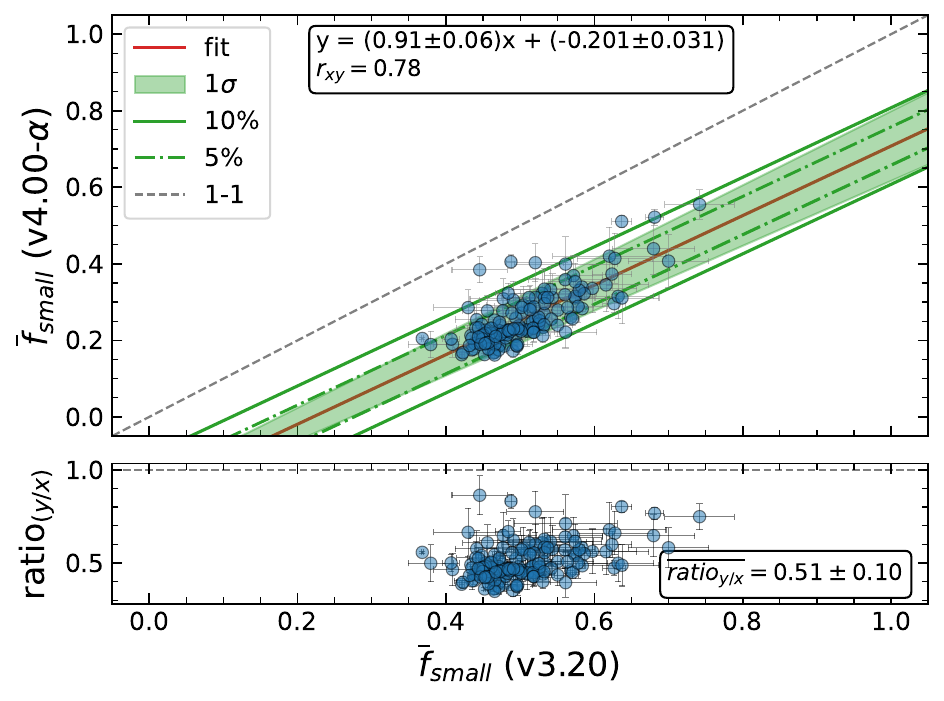}
    \includegraphics[scale=0.37]{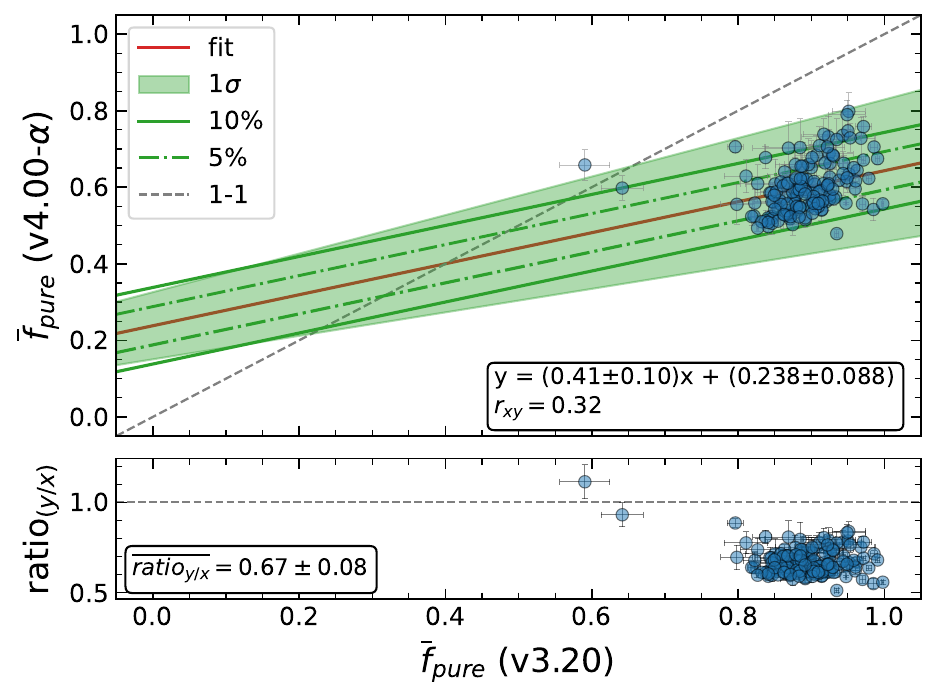}
    \includegraphics[scale=0.37]{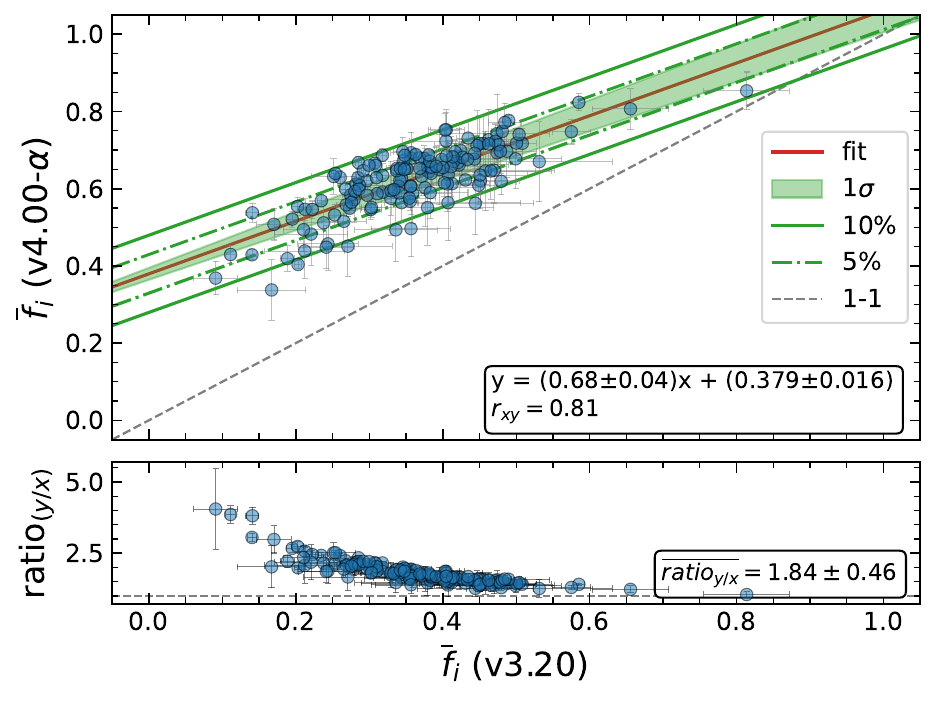} \\
    \includegraphics[scale=0.37]{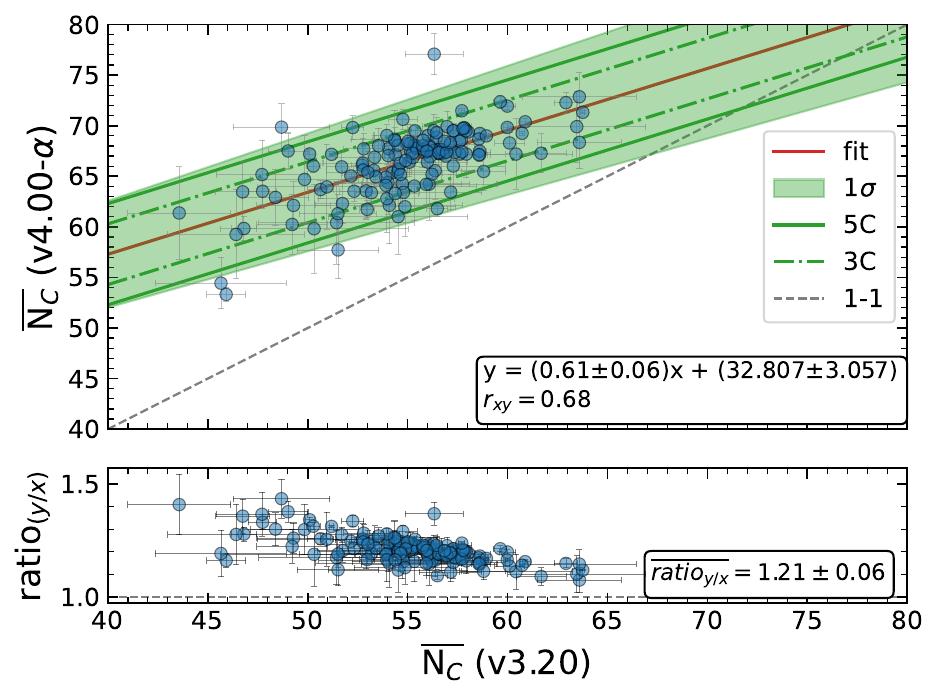}
    \includegraphics[scale=0.37]{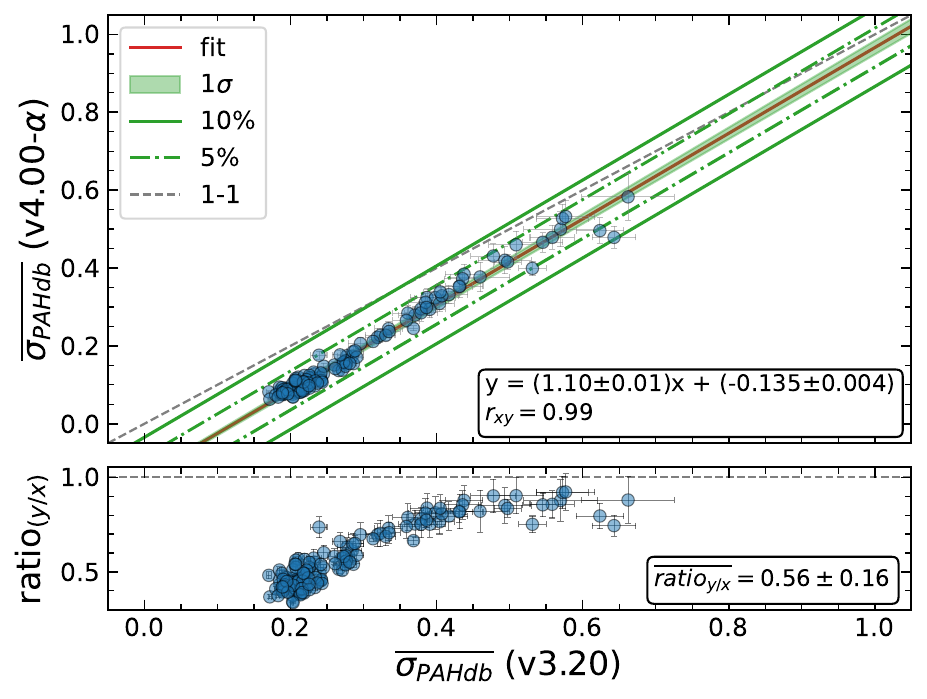}
    \caption{Comparison of the PAHdb-derived PAH properties in the base run (x-axis) and modeling with PAHdb v4.00-\textalpha{} library version (y-axis) under the same configuration as in the base run. Top row: neutral PAH fraction (left), cation PAH fraction (middle), anion PAH fraction (right); Middle row: small PAH fraction (left), pure PAH fraction (middle), PAH ionization fraction (right); Bottom row: \aNc{} (left), $\sigma_{PAHdb}$ (middle). Linear regression fitting is shown with the red line, 1$\sigma$ dispersion with the green envelope, 5\% and 10\% distance from the best fit line with green dash-dotted and solid lines, respectively, and line of equality with the black dashed line. The regression parameters and the Pearson’s correlation coefficient r$_{xy}$ are provided in the inset. In each case, the ratio of the two values is plotted in the bottom sub-panels, and the average ratio ($\overline{ratio}_{y/x}$) and standard deviation are provided.}
    \label{fig:v401_vs_v320}
\end{figure*}

Not applying a 15 cm$^{-1}$ redshift to the bands positions (Figure \ref{fig:no_shift_vs_shift}) results in a $\sim$15\% fluctuation in the decomposition results, and a $20 \%$ increase on the derived \afi{} and \aNc, although with moderate scatter for \aNc{} (r$_{xy}=0.66$), in comparison to the base run. The fluctuation in the charge breakdown is driven by the modeling of the 6.2 \micron{} PAH band. In PAHdb v3.20 (and previous library versions) there has been an ongoing difficulty in matching the blue side of the astronomical 6.2 μm PAH band. Although the introduction of nitrogen into a PAH's carbon-carbon structure does shift its 6.2 \micron{} spectra to the blue \citep{Hudgins2005}, an unrealistically large amount of nitrogen would need to be locked up in PANHs to account for the spectral features in the observational spectra \citep{Ricca2021}. Therefore, application of a redshift to the bands further hinders the matching of the 6.2 \micron{} band, while absence of redshift provides better matching, with the usage of additional cationic PAHs. The average fitting uncertainty in absence of redshift is also reduced by $\sim$13\%. Thus, redshift has a perceptible impact on the derived parameters, with no redshift resulting in improved fits. Although a redshift application attempts to mimic the anharmonic shifts and red tails of the PAH bands, absent under the harmonic approximation in the DFT computations, such effects are expected to be more pronounced in smaller molecules at shorter wavelengths, and specifically for the 3.3 \micron{} PAH band \citep{maltseva_high-resolution_2015,maltseva_high-resolution_2018,mackieAnharmonicQuarticForce2015,Mackie2016,Mackie2018,esposito_anharmonic_2024, esposito_anharmonicity_2024, esposito_infrared_2024}, and less salient for the 6--15 \micron{} PAH bands.

Examining the trends of the sensitivity analysis results we see that although quantitatively the PAHdb-derived parameters change under different modeling configurations, their variation follows, on average, positive linear scaling. For example, a galaxy with a higher PAH quantity compared to another galaxy in one configuration will show, on average, a higher PAHdb quantity value in another modeling configuration. Table \ref{tab:comparison} summarizes the average variation (ratio) of the PAHdb-derived parameters from each modeling configuration of v3.20 compared to the v3.20 base run.  This provides an indication of the difference between previous literature results with PAHdb v3.20 under the base run configuration, if a different modeling configuration was adopted.  

Finally, we note that in all the examined cases of the sensitivity analysis, the PAH size, either measured through the size fraction or by the \aNc, presented the largest amount of scatter compared to the base run, as opposed to the PAH charge state. This indicates that PAH charge is principally driving the spectral decomposition and fitting, while the PAH size has a secondary effect. However, this could be attributed to the lack of intermediate and large PAHs with irregular edge structures in the v3.20 library, without of which a lot of the 10 to 14 \micron{} PAH spectral structure cannot be modeled. In such case small PAHs that have rich structure in this regions are used in the fit. With a large number of small PAHs compared to intermediate or larger PAHs in PAHdb v3.20, there is more versatility in the fit to employ PAHs between 20--50 \Nc, resulting in larger scatter in the derived PAH size.    

\subsection{PAHdb v4.00-\textalpha} \label{sec:pahdbv4.00-alpha}

The v4.00-\textalpha{} library of DFT-computed spectra offers a significant improvement over v3.20 in matching the entire observed 6--15 \micron{} PAH spectrum \citep[Figure \ref{fig:example_320_vs_400a}; see also][]{Ricca2024}. Table \ref{tab:comparison} shows the average variation of the PAHdb-derived parameters between v4.00-\textalpha{} library version compared to the v3.20 base run. This improvement is reflected in the $\sim$44\% decrease of the total average uncertainty, $\overline{\sigma_{\rm PAHdb}}$, which has dropped  from 0.28 to 0.16, as well as in band-specific errors. Perhaps most notably is the 6.2 \micron{} PAH band, where the error $\overline{\sigma_{\rm 6.2}}$ drops from 0.75 to 0.08. With such a dramatic improvement in the quality of the fits, some differences are expected for the PAHdb-derived parameters when compared to those derived using v3.20 of the library. With PAHdb v4.00-\textalpha , the recovered PAH cation fraction has systematically increased, as a result of the 6.2 \micron{} PAH band being produced mostly by PAH cations. Consequently, the neutral PAH fraction is decreased and the \afi{} is increased. Furthermore, the usage of PAH anions has considerably decreased ($\sim$50\%), which is in line with the expectancy of low anion fractions in star-forming regions and PDRs versus higher anion abundances in the diffuse ISM \citep{BakesTielens1994}. With the majority of the sample being SFGs, low anion fractions are expected there. With more intermediate-to-large PAHs added in v4.00-\textalpha{} library version, the usage of large PAHs in the fit has increased to $\sim$50\%, leading to an increase in the \aNc{} by $\sim$20\%. Under the base run configuration though, an excessive amount of PANHs ($\sim$33\%) are employed in the fit, as the large number of newly added PANHs ($\sim$2000) are used to model the blue side of the 6.2 \micron{} PAH feature \citep{Ricca2021}.

\subsubsection{PAHdb v4.00-\textalpha{} Configurations} \label{sec:4.00-configurations}

\begin{table*}
    \centering
    \caption{Average scale factors from the regression analysis (Figures \ref{fig:gaussian_fwhm_10_vs_15}--\ref{fig:no_shift_vs_shift}, and Figure \ref{fig:v401_vs_v320}) showing the variance between the PAHdb parameters derived under different modeling configurations and database version, compared to the base run configuration. v4.00-\textalpha{} version employs the base run configuration, while v4.00-\textalpha$\dagger$ is the optimal configuration for modeling (see Section \ref{sec:4.00-configurations}).}
    \begin{tabular}{lcccccccc}
        \hline
        \hline
        Case &$\overline{f}_{neut}$ & $\overline{f}_{cat}$ & $\overline{f}_{an}$ & $\overline{f}_{small}$ & $\overline{f}_{pure}$ & $\overline{f}_{i}$ & $\overline{N}_{C}$ & $\overline{\sigma_{PAHdb}}$  \\
        \hline
        FWHM=10 cm$^{-1}$ & $1.03 \pm 0.03$ & $1.10 \pm 0.12$ & $0.91 \pm 0.05$ & $1.04 \pm 0.05$ & $1.00 \pm 0.01$ & $1.04 \pm 0.09$ & $1.00 \pm 0.03$ & $1.03 \pm 0.02$  \\
        Lorentzian & $1.07 \pm 0.05$ & $0.90 \pm 0.12$ & $0.97 \pm 0.08$ & $1.07 \pm 0.09$ & $0.99 \pm 0.02$ & $0.89 \pm 0.10$ & $0.97 \pm 0.04$ & $1.03 \pm 0.02$  \\
        Calc. Temp. & $2.23 \pm 0.37$ & $0.21 \pm 0.09$ & $0.31 \pm 0.08$ & $1.08 \pm 0.17$ & $1.10 \pm 0.05$ & $1.14 \pm 0.06$ & $0.97 \pm 0.12$ & $1.11 \pm 0.07$  \\
        No redshift & $0.85 \pm 0.10$ & $1.15 \pm 0.19$ & $1.07 \pm 0.14$ & $1.04 \pm 0.13$ & $0.94 \pm 0.03$ & $1.20 \pm 0.15$ & $1.19 \pm 0.09$ & $0.87 \pm 0.08$  \\
        v4.00-\textalpha & $0.79 \pm 0.09$ & $2.51 \pm 0.56$ & $0.48 \pm 0.09$ & $0.51 \pm 0.10$ & $0.67 \pm 0.08$ & $1.84 \pm 0.46$ & $1.21 \pm 0.06$ & $0.56 \pm 0.16$  \\
        v4.00-\textalpha$\dagger$ & $1.12 \pm 0.08$ & $2.42 \pm 0.51$ & $0.22 \pm 0.11$ & $0.74 \pm 0.09$ & $1.00 \pm 0.00$ & $1.55 \pm 0.26$ & $1.24 \pm 0.07$ & $0.51 \pm 0.19$  \\
        \hline
        
    \end{tabular}
    \label{tab:comparison}
\end{table*}

\begin{figure*}
    \centering
    \includegraphics[scale=0.45]{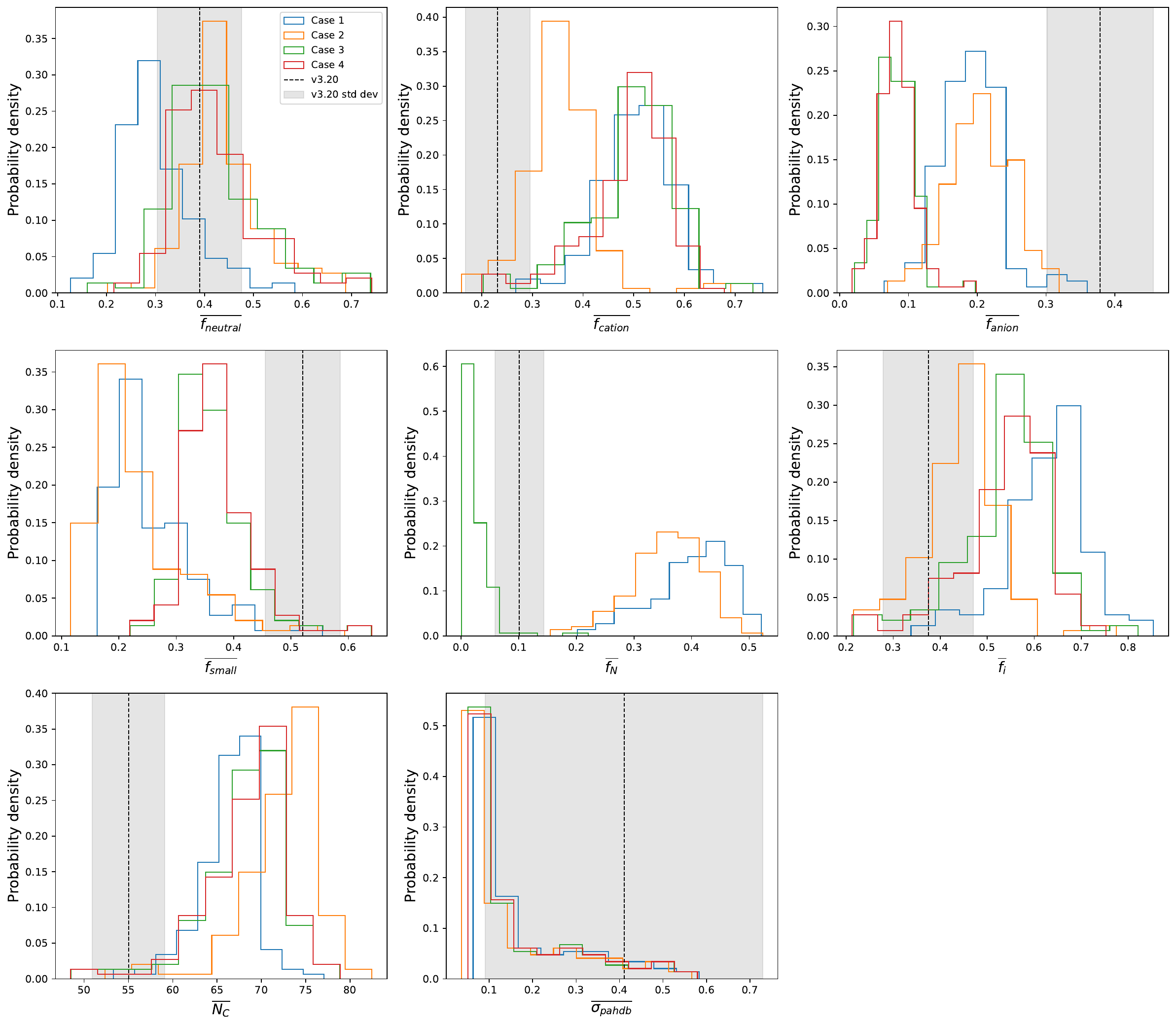}
   \caption{The distributions of the characteristics of the PAH population derived using the PAHdb v4.00-\textalpha{} library for different sets of parameters (Cases 1-4; Section \ref{sec:v4.00}). The PAHdb v3.20 average values (vertical dashed line) and standard deviations (gray area) are shown for reference.}
    \label{fig:histograms}
\end{figure*}

For the v4.00-\textalpha{} version of the library of DFT-computed spectra, we focus on determining the optimal modeling configuration for galaxy decomposition, based upon the results of the PAHdb v3.20 sensitivity analysis. Specifically, not applying a redshift to the bands presented considerable deviations from the base-run configuration (up to $\sim$20\%), while improving the fits by $\sim$13\%. Furthermore, modeling with the v4.00-\textalpha{} with the base run parameters results in an increased PANH fraction. Yet, nitrogen abundances in galaxies are found significantly lower compared to hydrogen or oxygen \citep[e.g.,][]{Chiappini2003, Perez-Montero2013, Oliveira2024}, thus nitrogen-substituted PAHs, if present, are expected to contribute at much lower rates. Therefore, we examine modeling configurations that concentrate on the application or absence of redshift, and the PANH content in the database. 

Figure \ref{fig:histograms} presents the distributions of the PAHdb-derived parameters under the four different PAHdb v4.00-\textalpha{} library version configurations (Table \ref{tab:4cases}) described in Section \ref{sec:v4.00}, as well as the average values and standard deviations from the v3.20 base run modeling. One of the most notable results is that the vast majority of galaxies are very well fitted under all four configurations with $\sigma_{\rm PAHdb}<0.1$ for most and a considerable improvement compared to using v3.20 with $\overline{\sigma_{PAHdb}}\sim0.4$ for most. However, the distributions of the derived parameters in the different configurations are distinct with different average values, exposing fitting degeneracies.

\begin{figure*}
    \centering
    \includegraphics[scale=0.7]{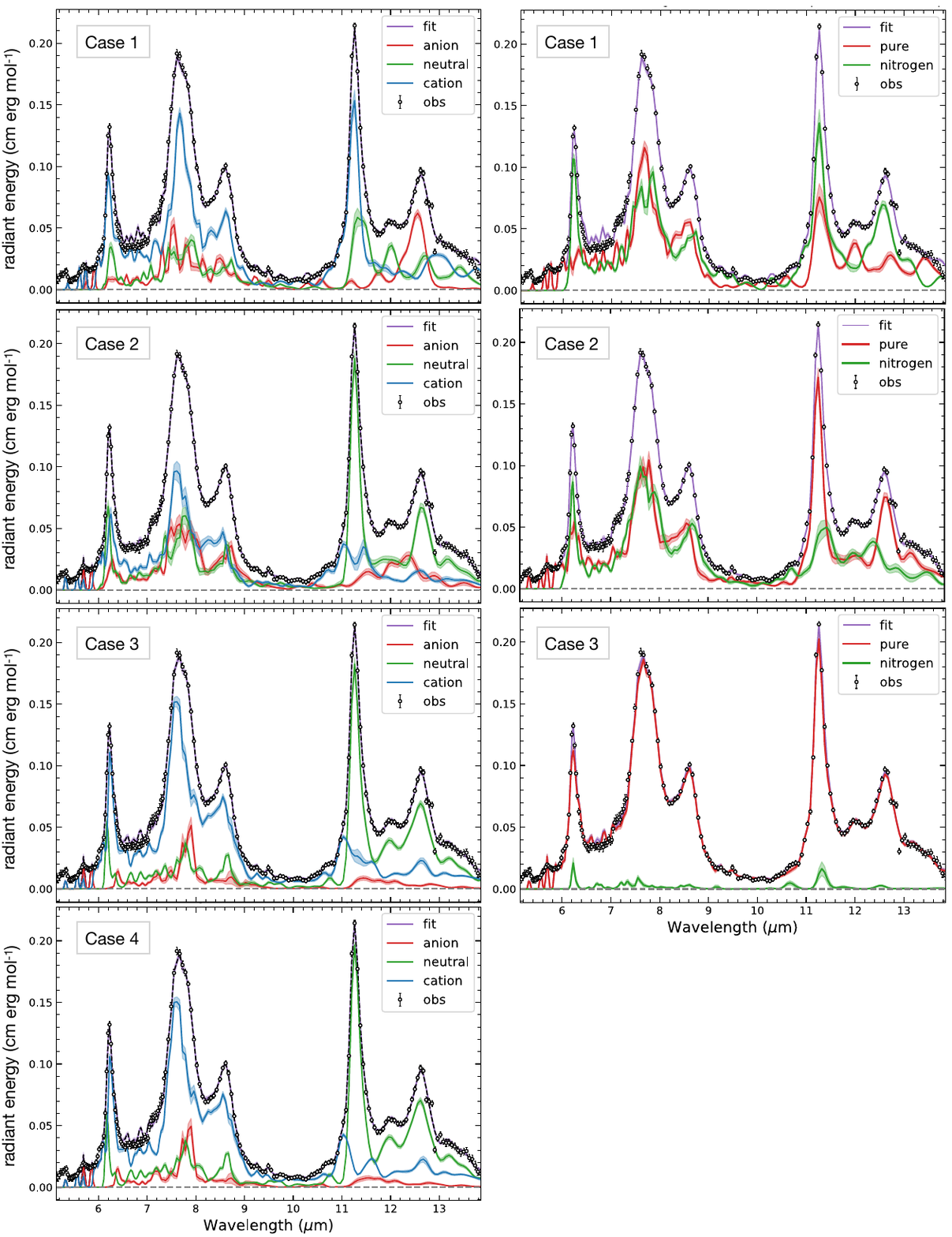}  
    \caption{The PAH charge (left column) and composition (right column) breakdown for galaxy IRAS 05129+5128 under different PAHdb v4.00-\textalpha{} library modeling configurations. Cases 1--4 (Section \ref{sec:v4.00}) are shown in succession starting from the top row.}
    \label{fig:breakdown_comparison}
\end{figure*}

To understand modeling degeneracies, we take a closer look and examine the PAH charge breakdown across the different PAH bands, as well as the fractions of pure vs nitrogen-containing PAHs used. We use the PAH charge and composition breakdown for the galaxy IRAS 05129+5128 (Figure \ref{fig:breakdown_comparison}) as an example to discuss the average trends under the four cases considered for the PAHdb v4.00-\textalpha{} library across the entire sample. Case 1 results in a high fraction of PANHs ($\overline{f_{N}} \sim$0.45) and an increase in $\overline{f}_{anion}$ ($\sim$0.2) when compared to the other three cases, but, most strikingly, the neutral 11.2 \micron{} PAH band is matched almost exclusively using PAH cations. For Case 2, which uses the complete v4.00-\textalpha{} library without applying a redshift, still results in a high PANH fraction ($\overline{f_{N}} \sim$0.35) and $\overline{f}_{anion}$. This is similar to Case 1, but now the 11.2 \micron{} PAH band is mainly reproduced by neutral PAHs. However, the PAH bands between  6--9 \micron{} (at 6.2 \micron{}, 7.7 \micron{} complex, and 8.6 \micron{}) that are principally attributed to PAH cations, are reproduced with comparable contributions of neutral and cation PAH mixtures. Case 3, which excludes the newly added PANHs and applies no redshift, shows little contribution from the v3.20 PANHs and has a need for fewer PAH anions ($\overline{f}_{anion} < 0.1$). Moreover, the charge breakdown per band now better represents that established in earlier work, i.e., the 11.2 \micron{} band is reproduced by neutral PAHs and the 6-9 \micron{} bands is reproduced principally by cationic PAHs. Lastly, Case 4, which excludes all PANHs and applies no redshift, provides results comparable to those from Case 3. In addition, it provides a clearer separation between the PAH cations carrying the 11.0 \micron{} band and the neutral PAHs responsible for the 11.2 \micron{} band. Therefore, the Case 4 configuration i.e., modeling with pure PAHs without applying a redshift to the bands, is considered as the optimal configuration for the modeling of the astronomical PAH spectrum with the PAHdb v4.00-\textalpha{} library.

The evaluation of the model parameter space with the v4.00-\textalpha{} library indicates that pure PAHs alone can fit the 6--15 \micron{} PAH emission spectrum of galaxies well. However, this does not mean that PANHs are not present in space. To the contrary, nitrogen should be readily incorporated into the PAHs when they form in circumstellar envelopes where nitrogen is present in high abundances \citep[e.g.,][]{Hudgins2005}. Moreover, the unequivocal detection of cyano-PAHs \citep[][]{McGuire2018, McGuire2021, Agundez2023} in the radio point to ongoing nitrogen chemistry. Yet, the relative abundances of cyano-PAHs and PANHs compared to pure PAHs on galaxy-wide scales is yet undetermined. Unfortunately, at present the inclusion of the newly added PANH spectra when fitting results in a nitrogenated PAH fraction is at odds with our current understanding.

\subsection{The average PAH population properties} \label{sec:average_properties}

\begin{table}
    \caption{The average PAH population parameters for the galaxies in the sample, as derived with PAHdb v3.20 with the base run configuration (Paper I), and the optimal modeling configuration (i.e., pure PAHs without applying a redshift, indicated with the $\dagger$ symbol) deduced in this work using the PAHdb v4.00-\textalpha{} library.}
    \begin{tabular}{lcc}
    \hline
    \hline
        Parameter &3.20 (Paper I) & v4.00-\textalpha$\dagger$ \\ 
    \hline
        $\overline{f}_{neut}$ & 0.38 $\pm$ 0.09 & 0.43 $\pm$ 0.09 \\
        $\overline{f}_{cat}$ &0.21 $\pm$ 0.07 &0.49 $\pm$ 0.08 \\
        $\overline{f}_{an}$ &0.40 $\pm$ 0.08 &0.08 $\pm$ 0.03 \\
        $\overline{f}_{small}$ &0.51 $\pm$ 0.06 &0.37 $\pm$ 0.06 \\
        $\overline{f}_{pure}$ &0.90 $\pm$ 0.05 &1.00 $\pm$ 0.00 \\
        $\overline{f}_{i}$ &0.36 $\pm$ 0.10 &0.54 $\pm$ 0.10 \\
        $\overline{N}_{C}$ &55 $\pm$ 3.7 &68 $\pm$ 4.6 \\
        $\overline{\sigma_{PAHdb}}$ &0.28 $\pm$ 0.11 &0.16 $\pm$ 0.13\\ 
    \hline
    \end{tabular}
    \label{tab:pah_properties_v320_vs_400alpha}
\end{table}

Under PAHdb v3.20 modeling, we found (Paper I) that the average PAH population within galaxies consists of mid-sized PAHs with an average number of carbon atoms of \aNc{} = 55 and average PAH ionization fractions of $\overline{f}_i=0.36$. In this work, utilizing the latest library of PAH computed spectra (v4.00-\textalpha) along with the optimal PAHdb modeling configuration (Section \ref{sec:4.00-configurations}) we present the most accurate and robust determination of the average PAH population properties in galaxies. Much of this improvement is due to addition of a significant number of irregularly shaped, large PAHs to the spectral libraries. Without these uneven edges, large PAHs do not have bands between ~11.5 to 13.5 \micron{} \citep[e.g.,][]{Hudgins1999, Hony2001}. Since small PAHs have rich structure in this region, the earlier fits are biased to small PAHs. Including larger PAHs also significantly impacts results involving the 6.2 \micron{}, pure C-C stretching band. It directly impacts PAH size since the number of C atoms goes up roughly as the square of the area, and charge balance as the 6.2 \micron{} band is dominated by PAH cations. 

Table \ref{tab:pah_properties_v320_vs_400alpha} summarizes the average PAHdb-derived properties for the galaxies under the two modeling configurations. The successful modeling of the 6.2 \micron{} PAH band with PAHdb v4.00-\textalpha{} elevates the usage of PAH cations in the fit and consequently the deduced PAH ionization fraction to $\overline{f}_i=0.54$, which is more in line with the ionizing radiation field in SFGs \citep[e.g.,][]{Bolatto2024}. Furthermore, the addition of more intermediate-to-large PAHs in v4.00-\textalpha{} library achieved a more precise reproduction of the bands in the 11-15 \micron{} part of the spectrum (see Figure \ref{fig:example_320_vs_400a}), which are typically attributed to PAHs with \Nc{} $> 50$, thus increasing the \aNc{} to 68. This increase in the average PAH size is consistent with the increase in the PAH ionization fraction which is controlled by the more intense radiation fields within star-forming regions in galaxies, where smaller PAHs are typically subjected to photodestruction \citep[e.g.,][]{Allain1996, LePage2003}. 

The advancement in PAHdb's library content and spectral computation's fidelity, which include accounting for the effect of polarization in the computations and the inclusion of a new set of large, irregularly edged PAHs molecules \citep[see][]{Ricca2024}, has allowed the improvement and better characterization of the average PAH population properties in galaxies. Earlier results were dominated by systematic uncertainties because of the inability to satisfactory reproduce the 6.2 \micron{} PAH band as well as a systematic issue with reproducing the 10-15 \micron{} region. Although numerically the PAHdb-derived parameters have changed when compared to modeling results with PAHdb v3.20, our current analysis revealed that their variation remains linear. As such, previously reported trends on the PAH parameters in the literature remain qualitatively valid, especially under the relative comparison of PAH properties among similar type of sources, or within different regions of extended sources.


\subsection{Modeling of JWST Spectra} \label{sec:pahfit_jwst_modeling}

JWST, with its 1-2 orders of magnitude gain in spectral and spatial resolution over \textit{Spitzer}, is returning detailed spectroscopic observations, allowing examination of astronomical PAH spectra in high fidelity. The increased resolving power of MIRI-MRS observations allows clear separation of atomic and molecular gas emission lines from nearby PAH features, revealing the presence or even loss of sub-components previously attributed to the PAH family \citep[e.g.,][]{Chown2024}. As such, we attempt an initial assessment and examine whether such higher resolution observations result in differing PAHdb-derived parameters when compared to lower resolution data. Ideally, spectra extracted from the exact same physical regions are required, a study we will conduct in detail in a future work. Here, we perform an initial examination of the modeling of the PAH emission spectrum for two galaxies observed with both \textit{Spitzer}-IRS and JWST MIRI-MRS, selected to ensure maximum overlap between the field-of-views (FOVs) of the different instruments. The selected galaxies are IRAS~F09111-1007 and IRAS~09022-361, from the JWST GO Program 3368. In addition, to eliminate effects from physical region mismatches and achieve a more straightforward examination of the effects of spectral resolution on the PAHdb-derived parameters, we have performed spectral smoothing and resampling of the JWST observations to the \textit{Spitzer}-IRS resolution ($R\simeq2500\rightarrow100$) and dispersion ($n_{\rm wave}\simeq10000\rightarrow200$).

\begin{figure*}
    \centering
    \includegraphics[trim={1 1 1 3cm}, clip, scale=0.24]{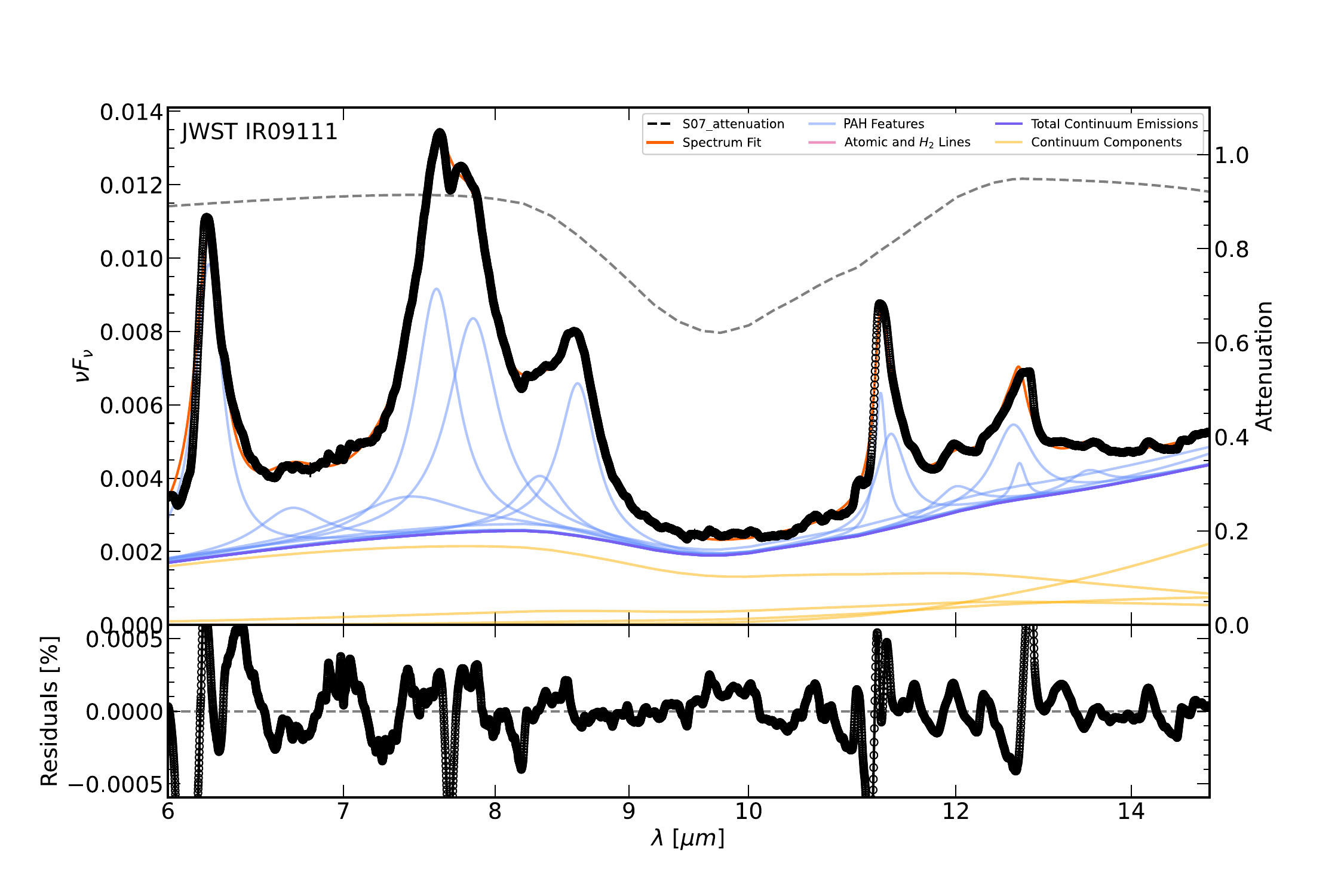}
    \includegraphics[trim={1 1 1 1cm}, clip, scale=0.5]{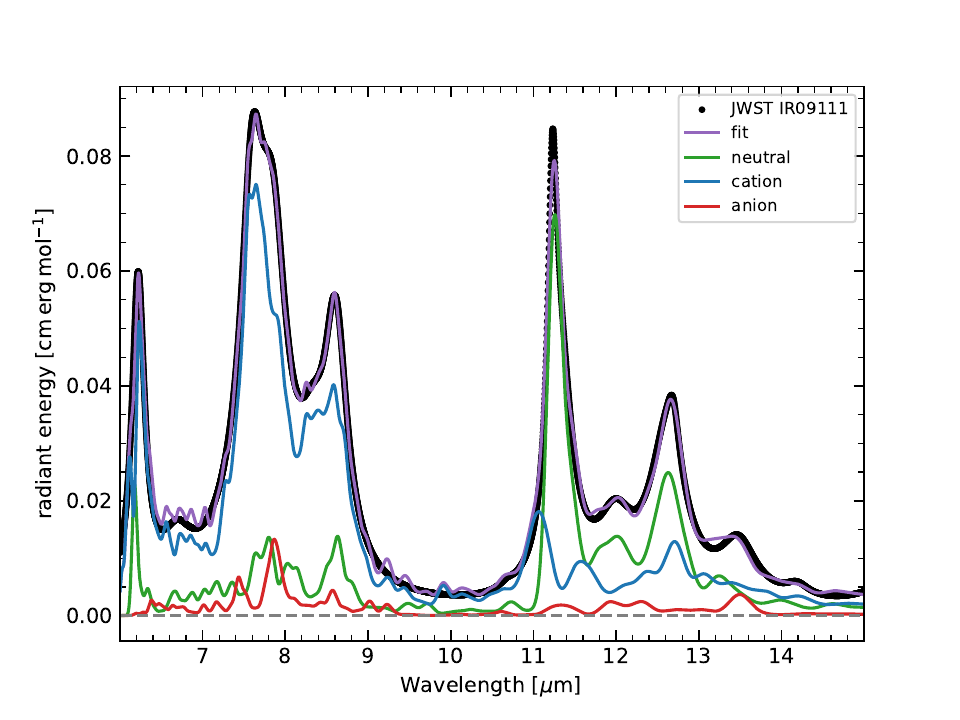} \\
    \includegraphics[trim={1 1 1 3cm}, clip, scale=0.24]{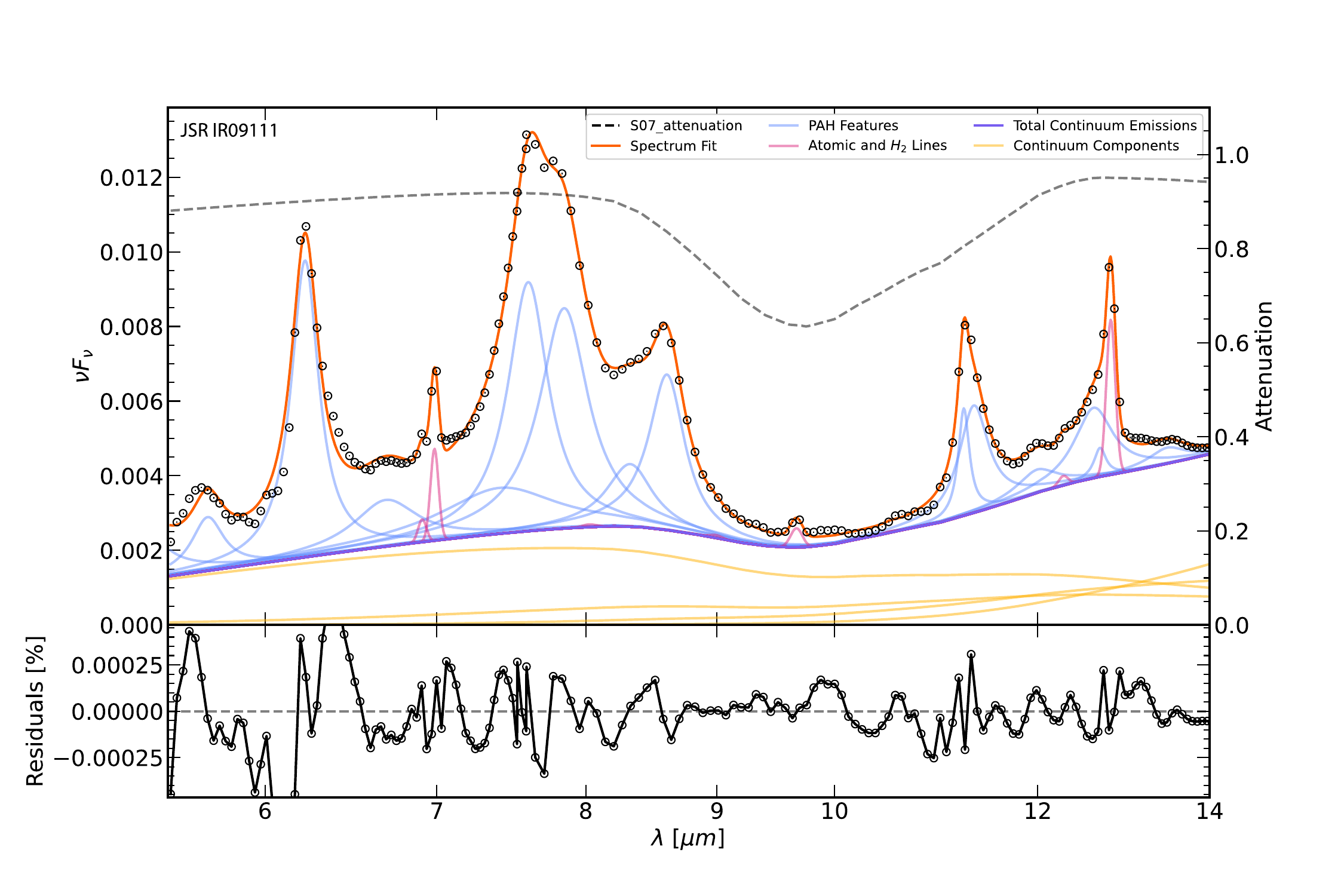} 
    \includegraphics[trim={1 1 1 1cm}, clip, scale=0.5]{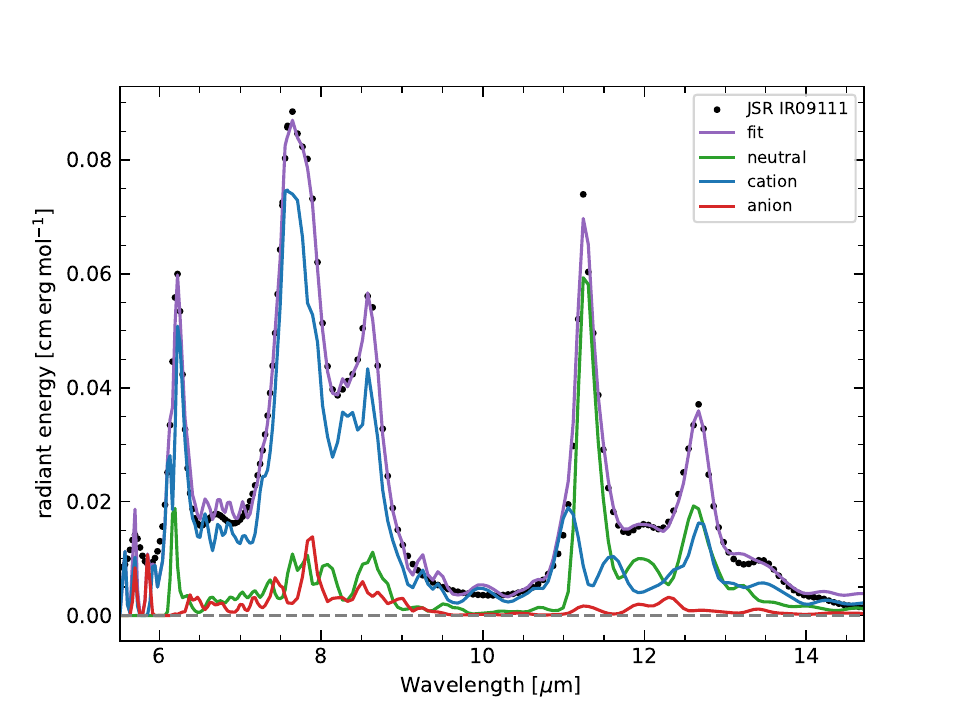} \\
    \includegraphics[trim={1 1 1 3cm}, clip,scale=0.24]{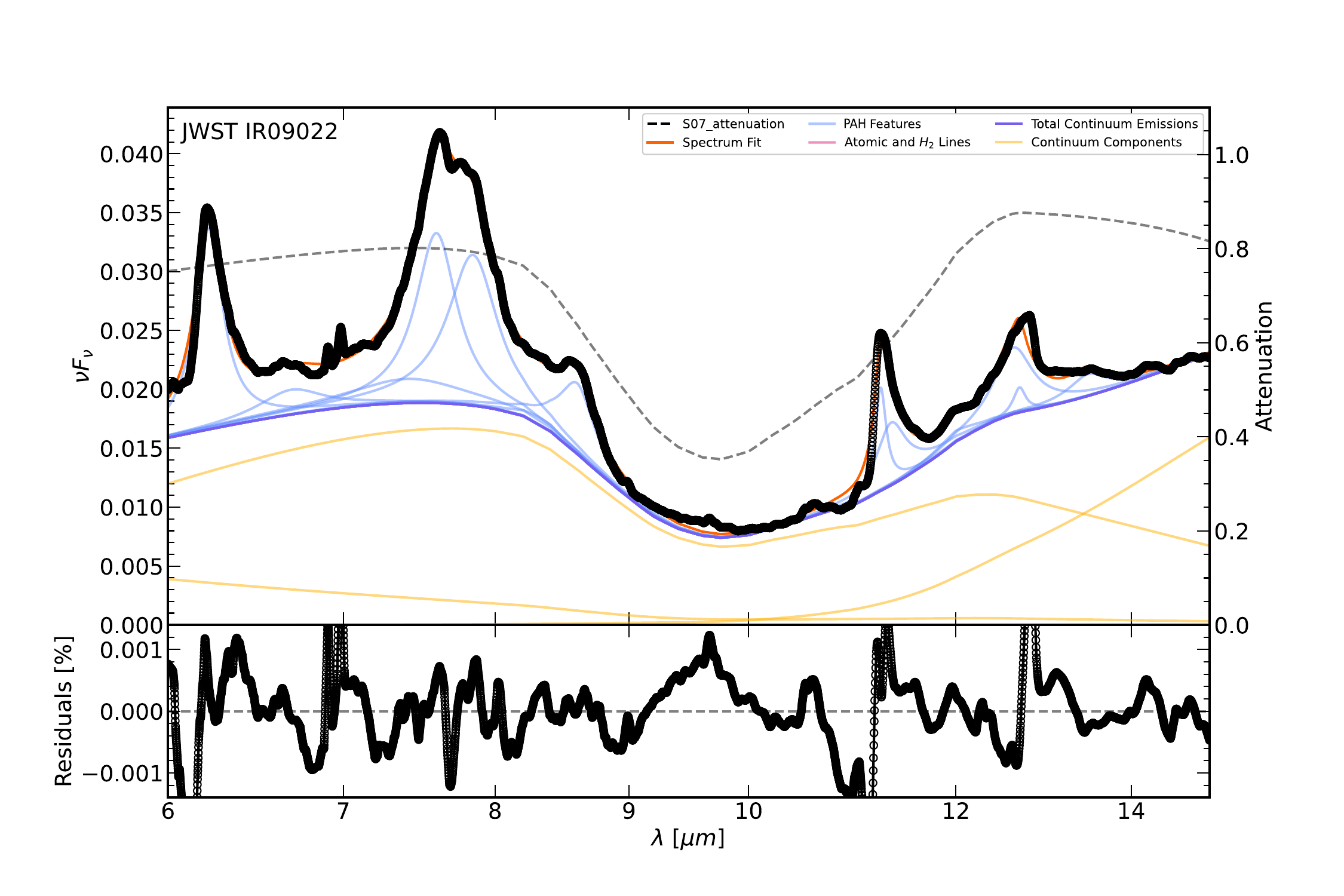}
    \includegraphics[trim={1 1 1 1cm}, clip,scale=0.5]{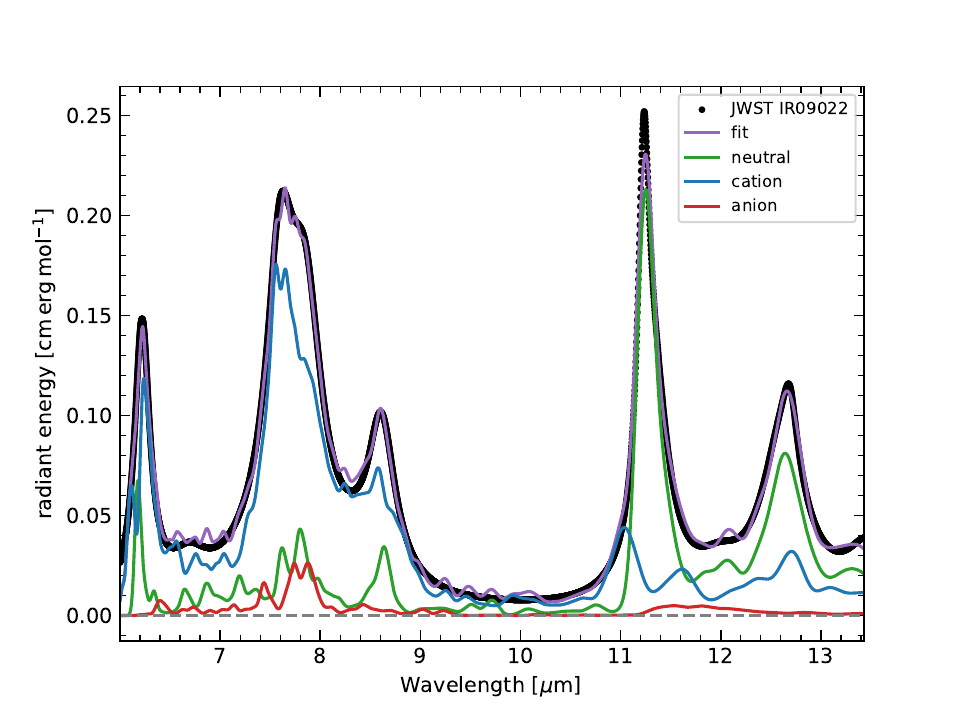} \\
    \includegraphics[trim={1 1 1 3cm}, clip,scale=0.24]{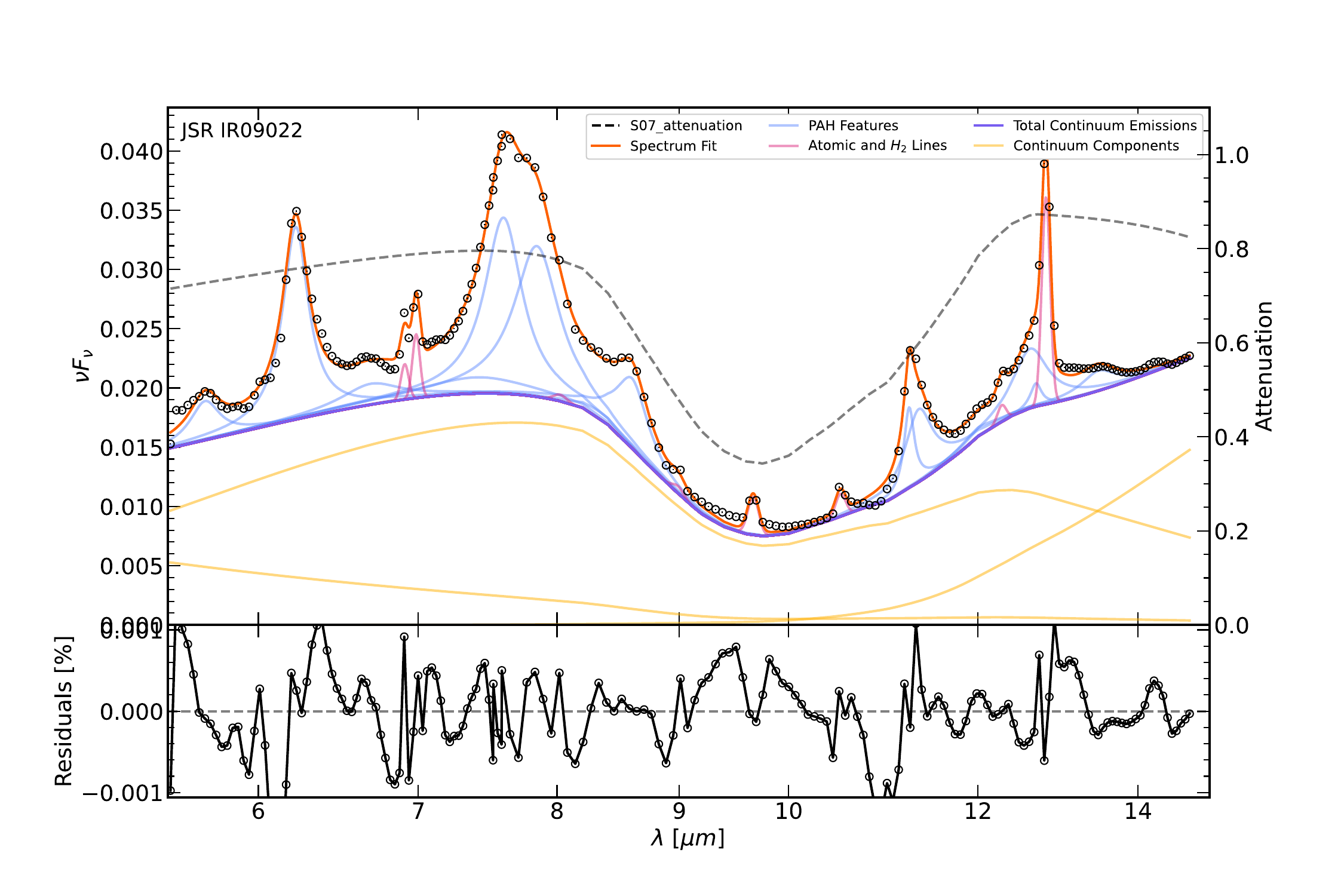}
    \includegraphics[trim={1 1 1 1cm}, clip,scale=0.5]{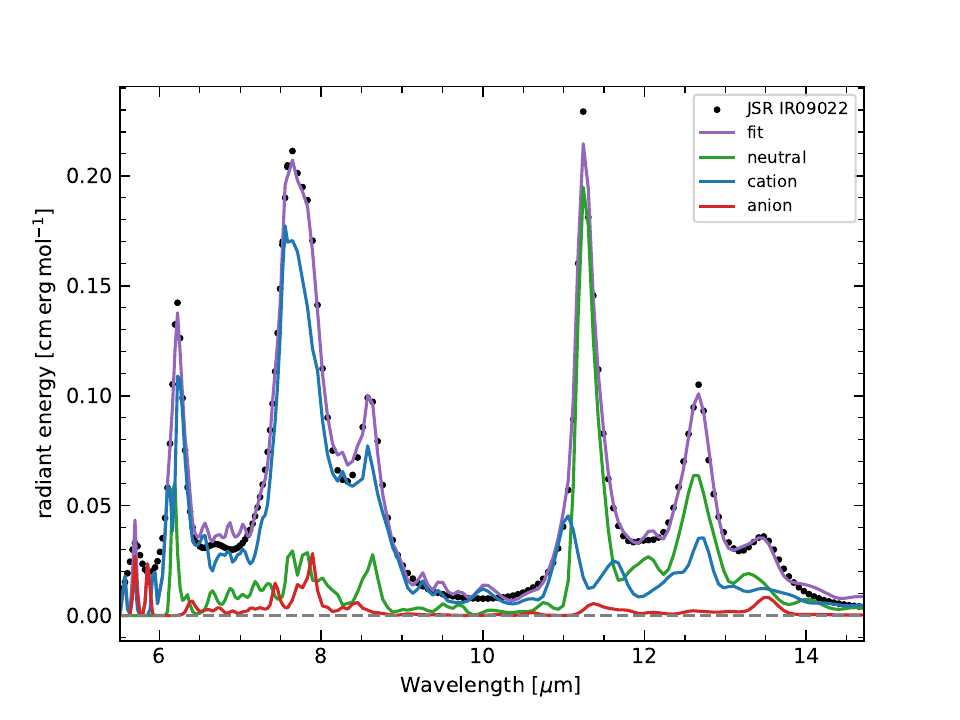} 
    \caption{MIR decomposition with \textsc{pahfit} and PAHdb v4.00-\textalpha{} for the JWST MIRI-MRS and the JWST smoothed and resampled (JSR) observations of galaxies IRAS F09111-1007 (top) and IRAS 09022-3615 (bottom). In the JWST spectra, emission lines were removed prior to \textsc{pahfit} fitting. Fitting of the two spectra shows consistent results, with a $\sim5$\% difference between the neutral and anion fractions.}
    \label{fig:jwst_pahfit_pahdb}
\end{figure*}

\begin{table*}
    \caption{Comparison of the PAHdb v4.00-\textalpha{} derived PAH charge fractions and \Nc{} between JWST, JWST smoothed and resampled (JSR), and \textit{Spitzer}-IRS observations, for galaxies IRAS~F09111-1007 and IRAS~09022-361.}
    \begin{tabular}{cccccc}
        \hline
        \hline
        Galaxy & Obs & $f_{neut}$ & $f_{cat}$ & $f_{an}$ & \aNc \\
        \hline
        \multirow{3}{*}{IRASF~09111-1007} & JWST & 0.2982 $\pm$ 0.0001 & 0.6588 $\pm$ 0.0001 & 0.0430 $\pm$ 0.0002& 74.38 $\pm$ 0.01 \\ 
        & JSR & 0.2470 $\pm$ 0.0002 & 0.6590 $\pm$ 0.0002 & 0.0940 $\pm$ 0.0003 & 76.24 $\pm$ 0.03 \\ 
        & IRS & 0.39 $\pm$ 0.02 & 0.55 $\pm$ 0.02 & 0.06 $\pm$ 0.02 & 73.25 $\pm$ 1.86 \\ 
        \hline
        \multirow{3}{*}{IRAS~09022-361} & JWST & 0.3333 $\pm$ 0.0001 & 0.6113 $\pm$ 0.0001 & 0.0554 $\pm$ 0.0001 & 73.31 $\pm$ 0.01 \\
        & JSR & 0.3085 $\pm$ 0.0001 & 0.6236 $\pm$ 0.0002 & 0.0679 $\pm$ 0.0002 & 73.53 $\pm$ 0.01 \\
        & IRS & 0.34 $\pm$ 0.01 & 0.57 $\pm$ 0.01 & 0.09 $\pm$ 0.01 & 72.81 $\pm$ 1.02 \\ 
        \hline
    \end{tabular}
    \label{tab:irs_jwst}
\end{table*}

Figure \ref{fig:jwst_pahfit_pahdb} shows the \textsc{pahfit} decomposition and PAHdb v4.00-\textalpha{} modeling of the JWST MIRI-MRS, and the JWST smoothed and resampled (JSR) spectra for the two galaxies, where emission lines were removed to aid the \textsc{pahfit} modeling. In Appendix \ref{sec:IRS-appendix} we present the decomposition of the corresponding \textit{Spitzer}-IRS spectra. Table \ref{tab:irs_jwst} summarizes the PAHdb-derived parameters from the modeling of the JWST, JSR, and \textit{Spitzer}-IRS PAH spectra. A $\sim$5\% variance between the neutral and anion fractions is found between the JWST and JSR spectra, showing otherwise consistent decomposition on the resolved PAH features even at the \textit{Spitzer}-IRS resolution.

\subsection{Comparison of Galaxy MIR Decomposition Codes} \label{sec:pahfit_vs_cafe}

\begin{figure*}
    \centering
    \includegraphics[scale=0.45]{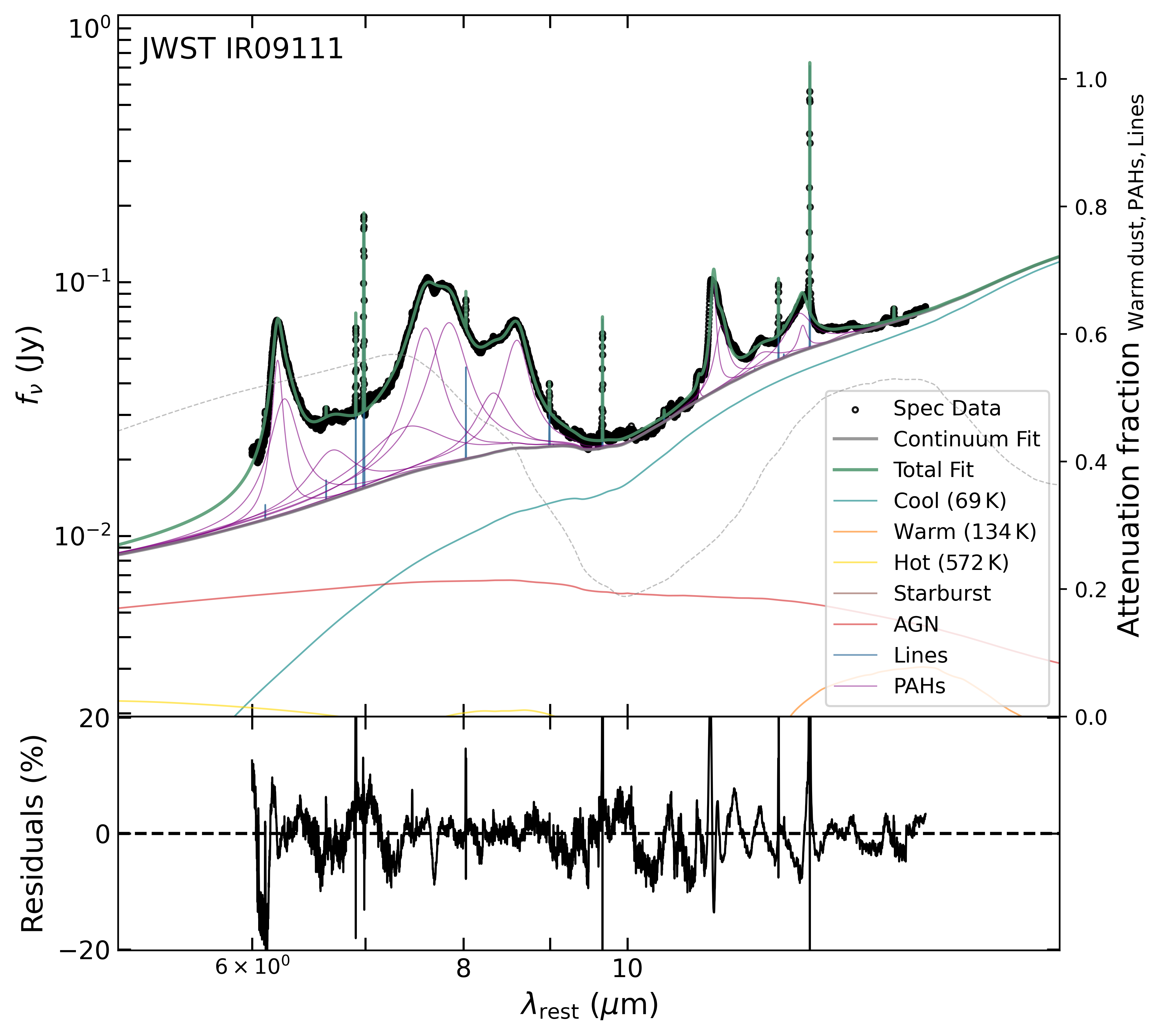}
    \includegraphics[trim={1 1 1 1cm}, clip, scale=0.55]{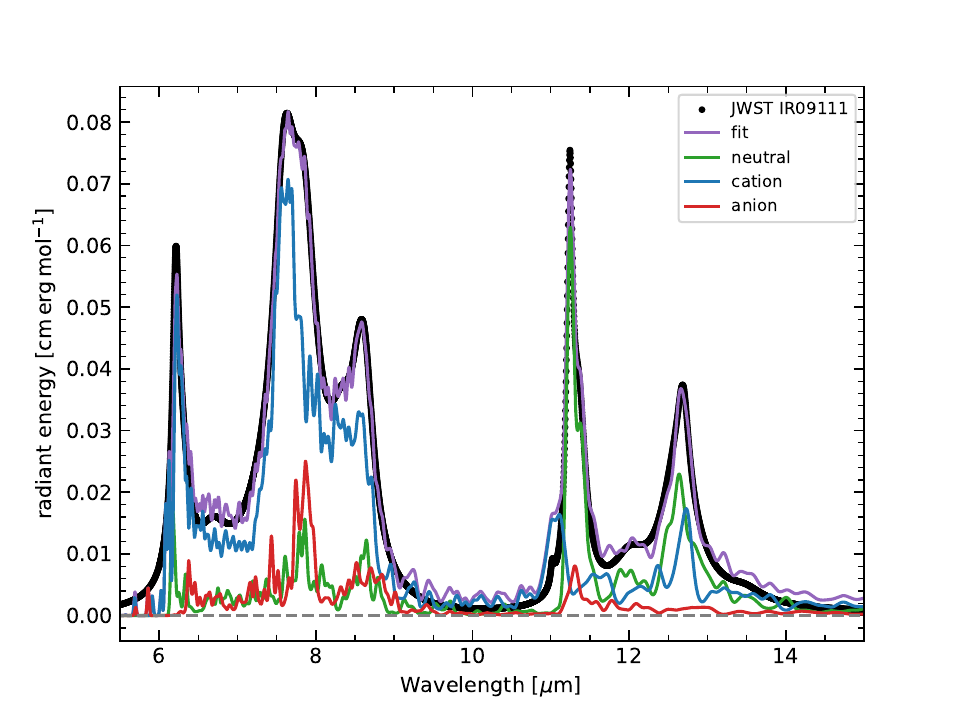} \\
    \includegraphics[scale=0.45]{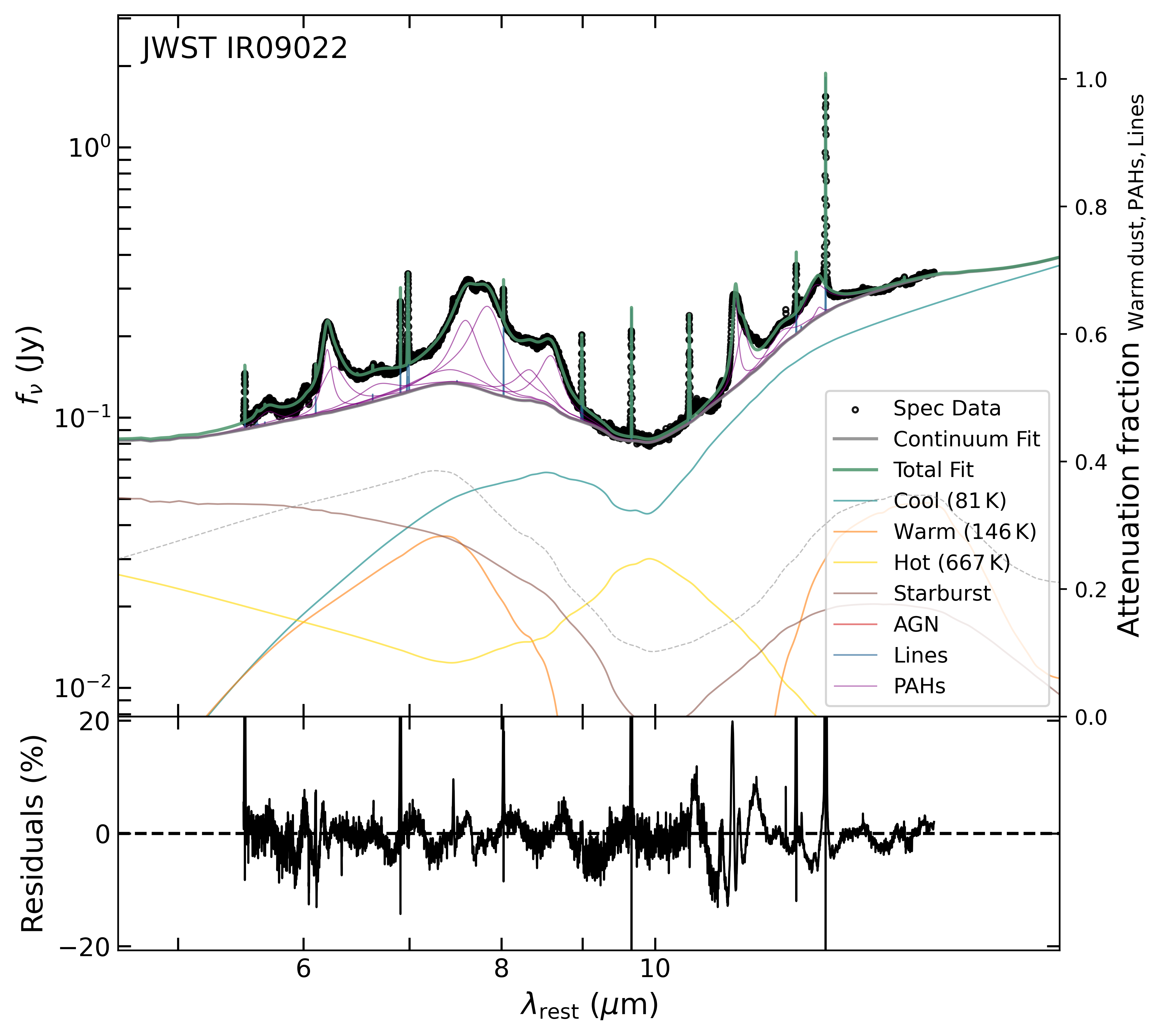}
    \includegraphics[trim={1 1 1 1cm}, clip,scale=0.55]{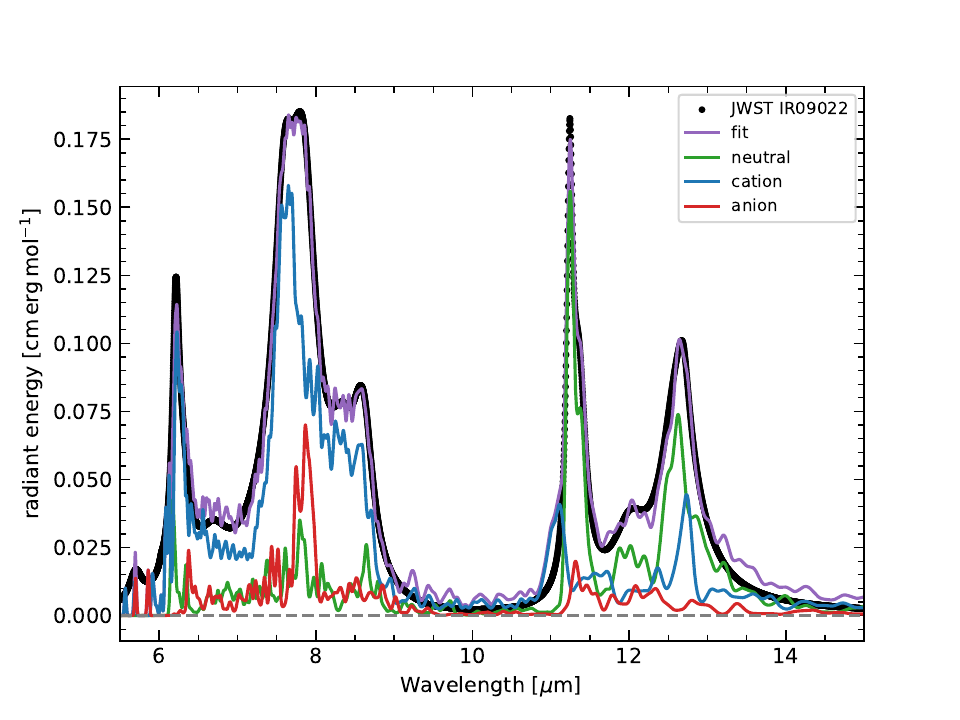}
    \caption{MIR decomposition with \textsc{cafe} and PAHdb v4.00-\textalpha{} for the JWST MIRI-MRS spectra of galaxies IRAS F09111-1007 (top row) and IRAS 09022-3615 (bottom row).}
    \label{fig:jwst_cafe_pahdb}
\end{figure*}

The PAH emission spectrum recovered from galaxy spectral decomposition will be dependent on the adopted compound model (Section \ref{sec:different_codes}), and as such different decomposition codes can produce different PAH emission spectra. Here, we explore the differences in the recovered PAHdb parameters when modeling the PAH spectrum from the MIRI-MRS observations of IRAS~F09111-1007 and IRAS~09022-361, resulting from the decompositions with \textsc{pahfit} (Section \ref{sec:pahfit_jwst_modeling}) and the recent release of Python \textsc{cafe}.

\begin{figure*}
    \centering
    \includegraphics[scale=0.55]{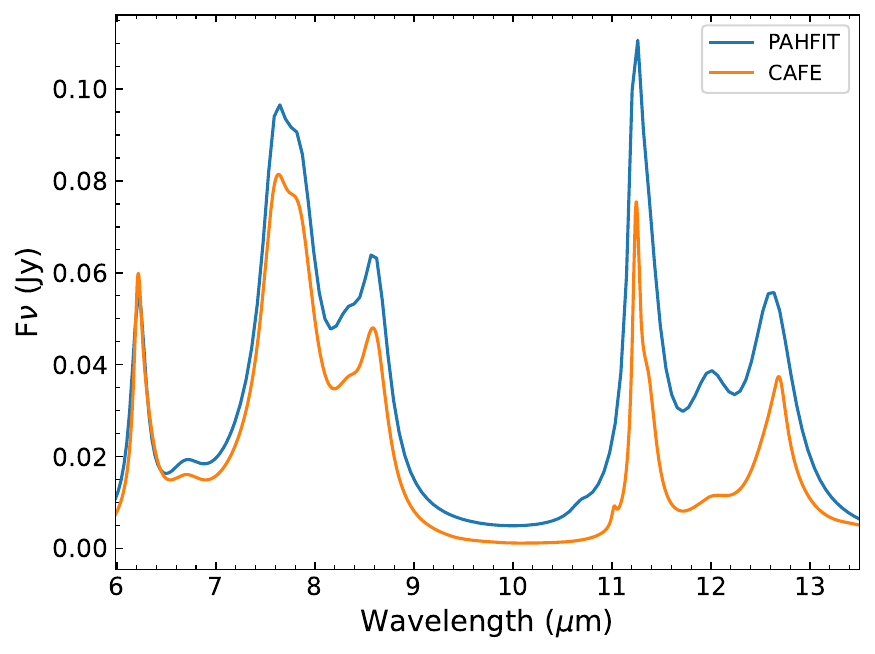}
    \includegraphics[scale=0.55]{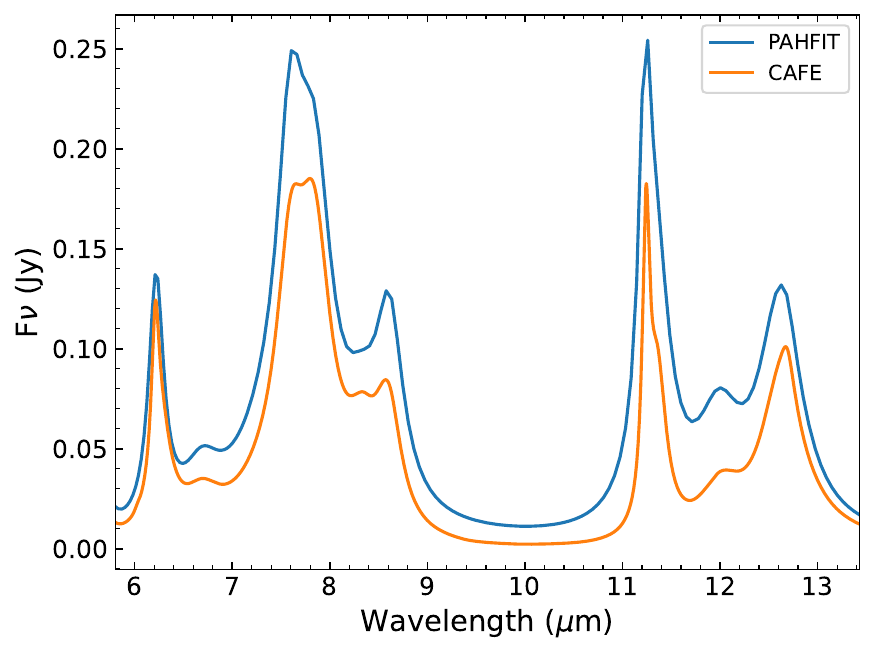}
    \caption{Comparison of the recovered PAH emission spectrum for galaxies IRAS~F09111-1007 (left) and IRAS~09022-361 (right), from the decomposition of their JWST MIRI-MRS observations with \textsc{pahfit} and \textsc{cafe} codes.}
    \label{fig:pahfit_vs_cafe}
\end{figure*}

Figure \ref{fig:jwst_cafe_pahdb} shows the \textsc{cafe} decomposition and PAHdb modeling of the two galaxies, and a comparison of the recovered PAH spectra from the \textsc{cafe} and \textsc{pahfit} codes is presented in Figure \ref{fig:pahfit_vs_cafe}. From the two examples, the PAH spectrum recovered from \textsc{cafe} has a lower intensity with more profound spectral differences in the features longward of 11 \micron{} compared to \textsc{pahfit}. This is a combined result of the modeling differences in the continuum, such as the inclusion and modeling of an AGN continuum (\textsc{cafe}), the PAH components used (e.g., at 11.0 \micron{}), but most importantly different treatment of the dust extinction model as discussed in Section \ref{sec:different_codes}.

Table \ref{tab:cafe_vs_pahfit} summarizes the PAHdb-derived charge breakdown and \Nc{} from the modeling of the respective PAH emission spectra. The fraction of PAH cations is identical in both type of spectra, as it is driven by the features in the 6--9 \micron{} region which show the least spectral variance, whereas the $\sim$7\% difference in neutral PAHs is driven by the larger variation in the 11--14 \micron{} region bands, and especially that of the 11.2 \micron{} feature. A more comprehensive comparison of the two codes, including quantification of the variance of the PAHdb-derived results when employing either of the extracted PAH emission spectra, is planned in a future work using a larger sample of galaxies.

\begin{table*}
    \caption{Comparison of the PAHdb v4.00-\textalpha{} derived PAH charge fractions and \Nc{}, between \textsc{pahfit} and \textsc{cafe} spectral decomposition of the MIRI-MRS JWST observations for galaxies IRAS~F09111-1007 and IRAS~09022-361.}
    \begin{tabular}{lccccc}
        \hline
        \hline
        Galaxy & Code & $f_{neut}$ & $f_{cat}$ & $f_{an}$ & \aNc \\
        \hline
        \multirow{2}{*}{IRASF~09111-1007} & \textsc{pahfit} & 0.2982 $\pm$ 0.0001 & 0.6588 $\pm$ 0.0001 & 0.0430 $\pm$ 0.0002& 74.38 $\pm$ 0.01 \\ 
        & \textsc{cafe} & 0.2153 $\pm$ 0.0001 & 0.6257 $\pm$ 0.0001 & 0.1590 $\pm$ 0.0001 & 71.03 $\pm$ 0.01 \\ 
        \hline
        \multirow{2}{*}{IRAS~09022-361} & \textsc{pahfit} & 0.3333 $\pm$ 0.0001 & 0.6113 $\pm$ 0.0001 & 0.0554 $\pm$ 0.0001 & 73.31 $\pm$ 0.01 \\
        & \textsc{cafe} & 0.2570 $\pm$ 0.0001 & 0.5905 $\pm$ 0.0001 & 0.1525 $\pm$ 0.0001 & 71.48 $\pm$ 0.01 \\
        \hline
    \end{tabular}
    \label{tab:cafe_vs_pahfit}
\end{table*}

\section{A new library of galaxy PAH emission templates} \label{sec:Templates}

\begin{figure*}
    \centering
    \includegraphics[width=0.5\linewidth]{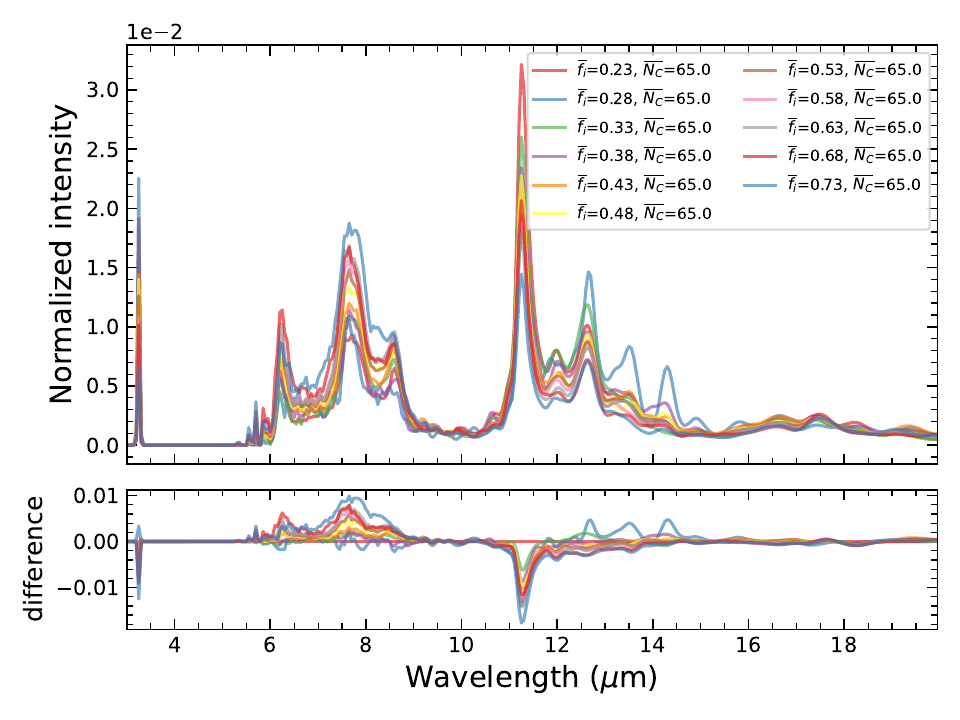}\hfill
    \includegraphics[width=0.5\linewidth]{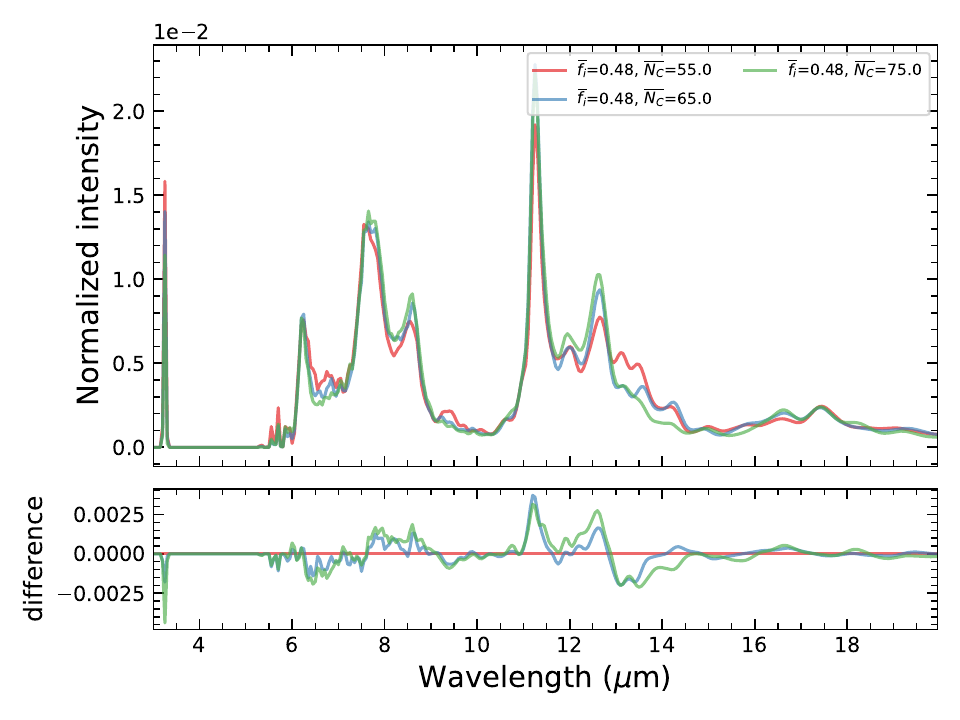}
    \caption{Example of the new PAHdb 4.00-\textalpha{} library PAH emission templates. Left: Templates at ﬁxed \aNc{} and varying $\overline{f_{i}}$. Right: Templates at ﬁxed $\overline{f_{i}}$ and varying \aNc{}. The spectra have been normalized (in $F_{\nu}$ units) on the total integrated ﬂux. The bottom panels show the spectral differences between templates by subtracting each spectrum from the ﬁrst spectrum in the set. The complete library is available at \url{www.astrochemistry.org/pahdb/templates_gal/v4.00-alpha/}.}
    \label{fig:templates}
\end{figure*}

In Paper I we introduced a library of PAH emission templates, generated from the modeling of galaxies with PAHdb v3.20. Here, we have created an updated library of PAH emission templates, using PAHdb 4.00-\textalpha{} library version and adopting the optimal configuration for galaxy modeling (Section \ref{sec:4.00-configurations}). These templates can be utilized e.g., in galaxy SED modeling to model the PAH emission component in dust emission models. The templates are parameterized on \aNc{} and $\overline{f_{i}}$, which are the average of each extrapolated galaxy spectrum in each 2D bin. Figure \ref{fig:templates} compares a number of template spectra at ﬁxed \aNc{} and varying $\overline{f_{i}}$ (left), and vice versa (right). At ﬁxed \aNc{}, the PAH features in the 6–9 \micron{} wavelength range increase with increasing $\overline{f_{i}}$, whereas the 3.3 \micron{} PAH band and those between 11–13 \micron{} decrease. At ﬁxed $\overline{f_{i}}$, the 3.3 \micron{} PAH band and those between 11–13 \micron{} increase with increasing \aNc{}, with no substantial variation between 6-9 \micron. The templates are offered for different radiation ﬁelds (6, 8, 10, and 12 eV) and are available online\footnote{\url{www.astrochemistry.org/pahdb/templates_gal/v4.00-alpha/}}.

\section{Summary and Conclusions} \label{sec:summary}

We have conducted a sensitivity analysis of the modeling of the PAH emission in galaxies, in order to assess the variance on the PAHdb derived parameters under different galaxy decomposition approaches, codes, and modeling configurations. The main conclusions from our work are summarize below.

(i) Decomposition of the SL and SL+LL \textit{Spitzer}-IRS spectral segments with \textsc{pahfit} provide consistent modeling and recovery of the 5--15 \micron{} isolated PAH emission spectrum, with the PAH band strength ratios being consistent within their uncertainties.  

(ii) PAHdb modeling with either Gaussian or Lorentzian line profiles, or small variations (5 cm$^{-1}$) in FWHM, produces consistent charge, composition, and size breakdown ($\sim$10\% variance), with moderate scatter in the size fractions and \aNc{}. Uncertainties when fitting with Lorentzian line profiles or using a FWHM = 10 cm$^{-1}$ are slightly higher, but overall consistent with the base run.   

(iii) Using a simplified emission model like the ``calculated temperature" model, results in PAHdb-fits that use predominantly neutral PAHs, even for those bands usually attributed to PAH cations (e.g., the 6.2 \micron{} band). The fits under the ``calculated temperature" model have larger uncertainty when compared to the base run, which makes use of the``cascade" emission model.   

(iv) Not applying a 15 cm$^{-1}$ redshift has considerable impact on the PAHdb-derived parameters, with an average of $\sim$15\% fluctuation in the charge fractions and a 20\% increase on \afi{} and \aNc{} in comparison to the base run. Without a redshift, the average fitting uncertainty is also reduced by $\sim$13\%, resulting in improved fits.

(v) The 4.00-\textalpha{} library version of PAHdb achieves the complete modeling of the 6–15 \micron{} PAH spectrum. The full matching of the the 6.2 \micron{} PAH band, previously under-fitted with PAHdb v3.20, results in an elevated average PAH cation and ionization fraction in galaxies, compared to v3.20 modeling. Similarly, with more intermediate-to-large PAHs added in v4.00-\textalpha{} library version, previously misrepresented in v3.20, the usage of large PAHs in the fit has increased leading to an increase in the deduced \aNc{} by $\sim$20\%. 

(vi) Different modeling configurations with PAHdb 4.00-\textalpha{} library version produces overall good fits but with varying derived parameters for the PAH population, exposing fitting degeneracies. 

(vii) Using the PAHdb 4.00-\textalpha{} library and the results from the v3.20 sensitivity analysis, we explore the optimal PAHdb configuration for modeling and fitting the PAH emission spectrum of galaxies. The unconstrained inclusion of PANHs in the fit, can lead to inaccurate parameter determination. Usage of pure PAHs, without nitrogen substitutions and without applying a redshift to the bands, is most consistent with our current understanding on the PAH emission spectrum, and is considered the optimal configuration for galaxy modeling, producing accurate PAH charge state breakdowns in all PAH bands. 

(viii) Modeling with PAHdb v4.00-\textalpha{} and the optimal configuration shows that average PAH ionization fraction in galaxies is $\overline{f}_i = 0.54$, and the \aNc{} $= 68$. 

(ix) While quantitatively the PAHdb-derived parameters may change under different modeling configurations or database versions, their variation follows on average a positive linear scaling, and previously reported trends on the PAH parameters remain qualitatively valid.

(x) We perform modeling on two JWST galaxy spectra, covering similar FOVs with previous \textit{Spitzer}-IRS observations, in order to examine the variance on the PAHdb derived parameters when higher resolution observations are modeled, that may hold additional spectroscopic information. To mitigate any differences in physical regions, we also smooth and resample the JWST observations to the \textit{Spitzer}-IRS resolution and dispersion. The PAH emission spectrum from the JWST observations show a $\sim$5\% variance in the $f_{neut}$ and $f_{anion}$ with the JWST smoothed spectra, and a $\sim$5-10\% variance on the recovered charge fractions, with the $f_{cat}$ increasing and the $f_{neut}$ decreasing in the MIRI-MRS observations. A larger set of JWST and \textit{Spitzer}-IRS observations on similar regions is required to fully quantify PAHdb modeling variations from different spectral resolution observations.

(xi) Decomposition of JWST spectra with the two currently available galaxy MIR spectral decomposition codes, \textsc{pahfit} and \textsc{cafe}, produce PAH emission spectra of different intensity and noticeable spectral variation in the 11--14 \micron{} region, a product of the difference in the underlying assumptions in the compound model of each code. This 11--14 \micron{} spectral variation is driving the recovered $\sim$7\% difference in the neutral PAHs fraction in the modeling of the respective PAH spectra with PAHdb. A statistically significant sample is required to fully quantify variations between the two spectral decomposition codes. 

(xii) A new library of galaxy PAH emission templates is delivered from the modeling of galaxies with PAHdb v4.00-\textalpha{} library version, parameterized on the average excitation energy, \aNc, and $\overline{f_{i}}$.

\begin{acknowledgments}

We would like to thank the referee for providing constructive comments and suggestions that have improved the clarity of this paper. The authors thank Thomas S.-Y. Lai for helpful discussion about \textsc{cafe}. A.M., C.B., P.T., J.D.B., L.J.A., V.J.E., and A.R. acknowledge support from the Internal Scientist Funding Model (ISFM) Laboratory Astrophysics Directed Work Package at NASA Ames (22-A22ISFM-0009). A.M., J.D.B., and L.J.A. are thankful for an appointment at NASA Ames Research Center through the Bay Area Environmental Research Institute (80NSSC19M0193). C.B. is grateful for an appointment at NASA Ames Research Center through the San Jos\'e State University Research Foundation (80NSSC22M0107). V.J.E.'s research has been supported by an appointment to the NASA Postdoctoral Program at NASA Ames Research Center, administered by the Oak Ridge Associated Universities through a contract with NASA. A.R. is grateful for an appointment at NASA Ames Research Center through SETI (80NSSC23M0046). E.P. acknowledges support from the Natural Sciences and Engineering Research Council of Canada. Usage of the Metropolis HPC Facility at the Crete Center for Quantum Complexity and Nanotechnology of the University of Crete, supported by the European Union Seventh Framework Programme (FP7-REGPOT-2012-2013-1) under grant agreement no. 316165, is also acknowledged.

\end{acknowledgments}

\appendix

\section{Decomposition of the Spitzer-IRS spectra with available JWST observations} \label{sec:IRS-appendix}

\begin{figure*}
    \centering
    \includegraphics[trim={1 1 1 3cm}, clip, scale=0.24]{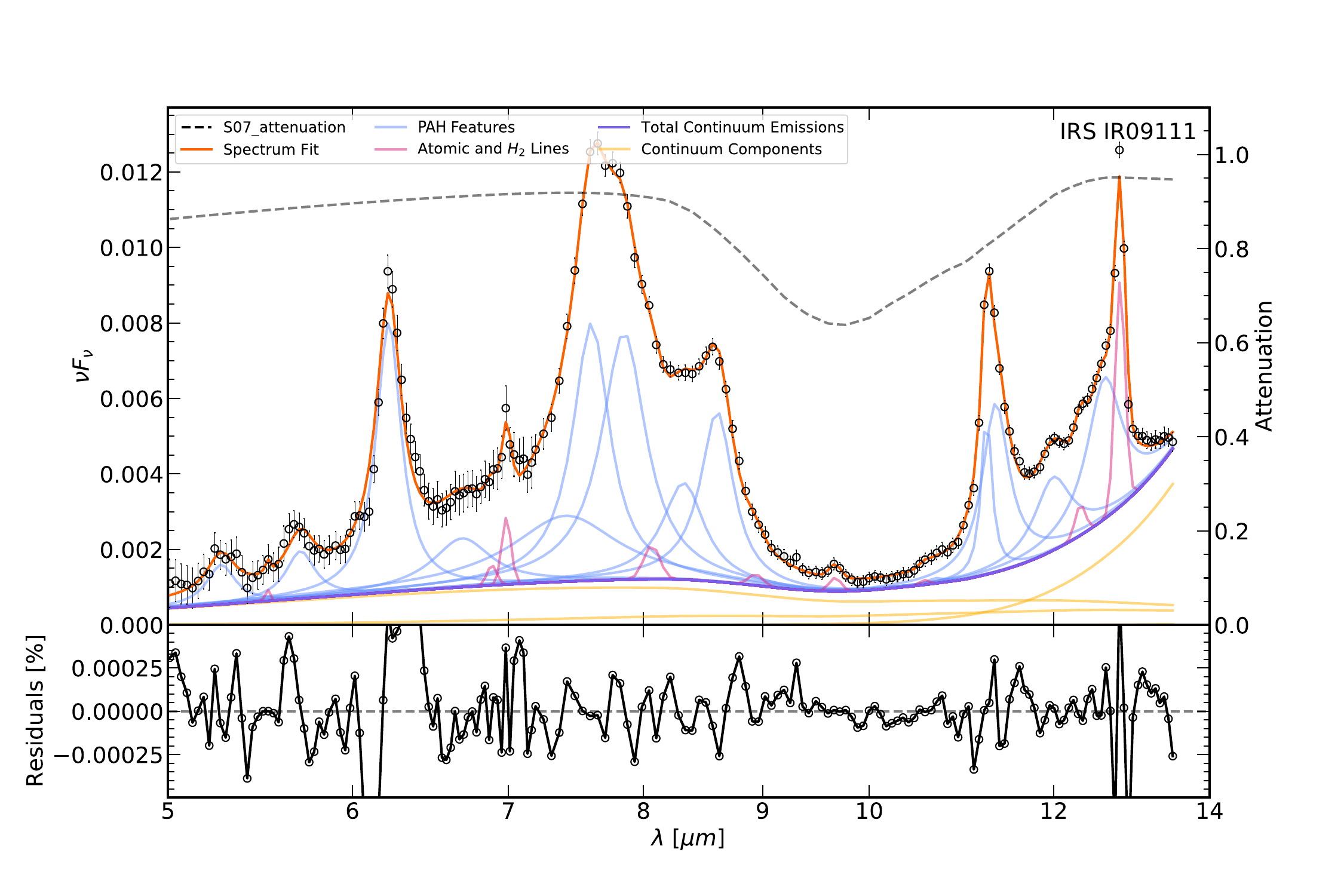}
    \includegraphics[trim={1 1 1 1cm}, clip, scale=0.5]{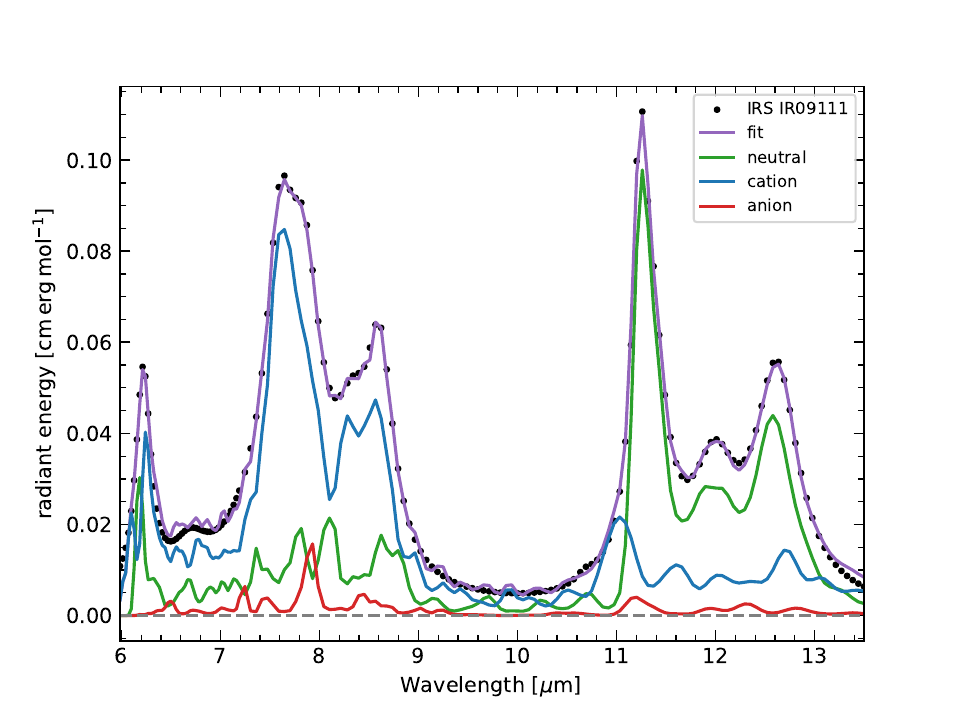} \\
    \includegraphics[trim={1 1 1 3cm}, clip, scale=0.24]{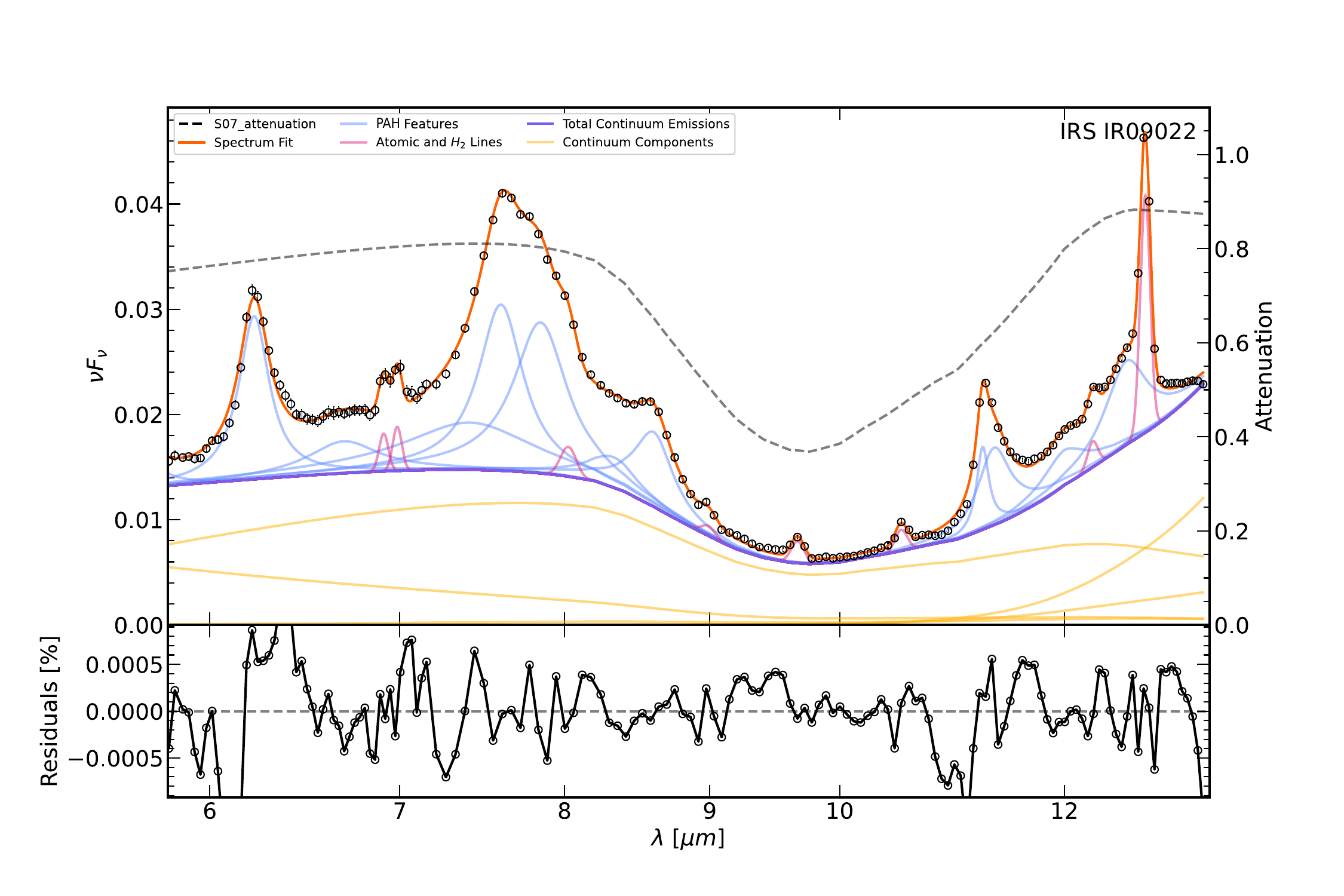} 
    \includegraphics[trim={1 1 1 1cm}, clip, scale=0.5]{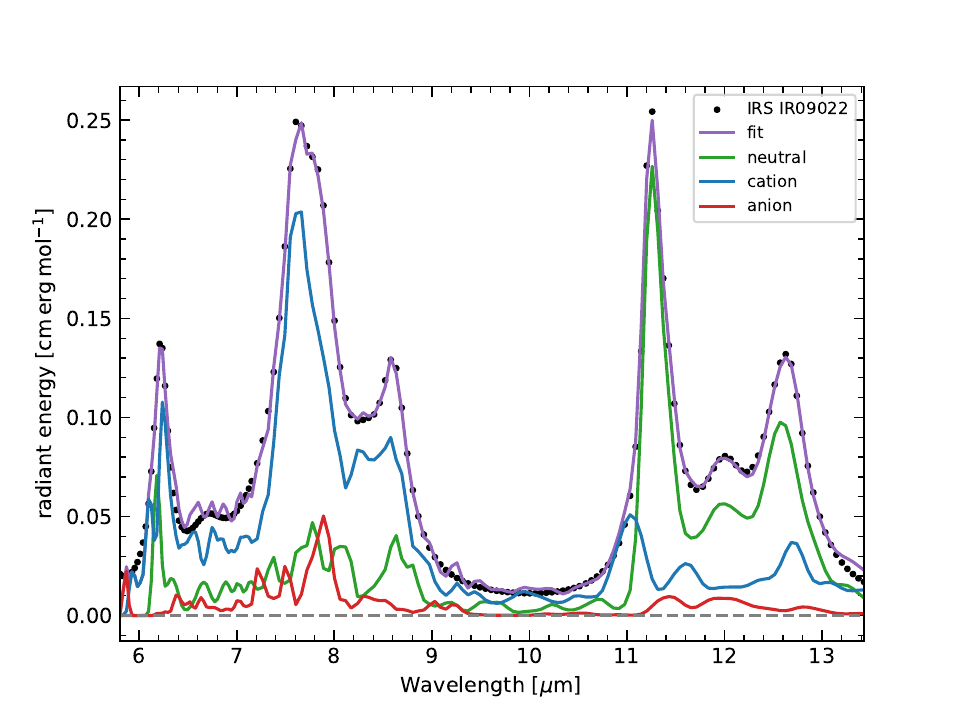}
    \caption{MIR decomposition with \textsc{pahfit} and PAHdb v4.00-\textalpha{} for the \textit{Spitzer}-IRS observations of galaxies IRAS F09111-1007 (top) and IRAS 09022-3615 (bottom).}
    \label{fig:irs-appendix}
\end{figure*}

The \textit{Spitzer}-IRS \textsc{pahfit} decomposition and PAHdb modeling of galaxies IRAS~F09111-1007 and IRAS~09022-361 (Figure \ref{fig:irs-appendix}), providing an additional comparison with their respective JWST observations, as well as the JWST smoothed and resampled spectra presented in Section \ref{sec:pahfit_jwst_modeling}. Table \ref{tab:irs_jwst} summarizes the PAHdb-derived parameters from the modeling of the JWST, JWST smoothed and resampled (JSR), and \textit{Spitzer}-IRS PAH spectra. Although the selection of galaxies was done to ensure as much overlap between the field-of-views of the different instruments, this comparison serves as an initial assessment of the expected difference in the PAHdb-derived properties between JWST and \textit{Spitzer}-IRS observations. A $\sim$5--10\% variance between the neutral and cation fractions is detected in the PAH emission spectra between the JWST and the actual \textit{Spitzer}-IRS observations, with the $f_{cat}$ increasing and the $f_{neut}$ decreasing in the MIRI-MRS observations.

\bibliography{bibliography}

\begin{thebibliography}{}
\expandafter\ifx\csname natexlab\endcsname\relax\def\natexlab#1{#1}\fi
\providecommand{\url}[1]{\href{#1}{#1}}
\providecommand{\dodoi}[1]{doi:~\href{http://doi.org/#1}{\nolinkurl{#1}}}
\providecommand{\doeprint}[1]{\href{http://ascl.net/#1}{\nolinkurl{http://ascl.net/#1}}}
\providecommand{\doarXiv}[1]{\href{https://arxiv.org/abs/#1}{\nolinkurl{https://arxiv.org/abs/#1}}}

\bibitem[{{Ag{\'u}ndez} {et~al.}(2023){Ag{\'u}ndez}, {Marcelino}, {Tercero}, \& {Cernicharo}}]{Agundez2023}
{Ag{\'u}ndez}, M., {Marcelino}, N., {Tercero}, B., \& {Cernicharo}, J. 2023, \aap, 677, L13, \dodoi{10.1051/0004-6361/202347524}

\bibitem[{{Allain} {et~al.}(1996){Allain}, {Leach}, \& {Sedlmayr}}]{Allain1996}
{Allain}, T., {Leach}, S., \& {Sedlmayr}, E. 1996, \aap, 305, 602

\bibitem[{{Allamandola} {et~al.}(1999){Allamandola}, {Hudgins}, \& {Sandford}}]{Allamandola1999}
{Allamandola}, L.~J., {Hudgins}, D.~M., \& {Sandford}, S.~A. 1999, \apj, 511, L115, \dodoi{10.1086/311843}

\bibitem[{{Allamandola} {et~al.}(1985){Allamandola}, {Tielens}, \& {Barker}}]{Allamandola1985}
{Allamandola}, L.~J., {Tielens}, A.~G.~G.~M., \& {Barker}, J.~R. 1985, \apjl, 290, L25, \dodoi{10.1086/184435}

\bibitem[{{Bakes} \& {Tielens}(1994)}]{BakesTielens1994}
{Bakes}, E.~L.~O., \& {Tielens}, A.~G.~G.~M. 1994, \apj, 427, 822, \dodoi{10.1086/174188}

\bibitem[{{Bakes} {et~al.}(2001){Bakes}, {Tielens}, \& {Bauschlicher}}]{Bakes2001}
{Bakes}, E.~L.~O., {Tielens}, A.~G.~G.~M., \& {Bauschlicher}, Charles~W., J. 2001, \apj, 556, 501, \dodoi{10.1086/321501}

\bibitem[{{Bauschlicher} {et~al.}(2010){Bauschlicher}, {Boersma}, {Ricca}, {Mattioda}, {Cami}, {Peeters}, {S{\'a}nchez de Armas}, {Puerta Saborido}, {Hudgins}, \& {Allamandola}}]{Bauschlicher2010}
{Bauschlicher}, Jr., C.~W., {Boersma}, C., {Ricca}, A., {et~al.} 2010, \apjs, 189, 341, \dodoi{10.1088/0067-0049/189/2/341}

\bibitem[{{Bauschlicher} {et~al.}(2018){Bauschlicher}, {Ricca}, {Boersma}, \& {Allamandola}}]{Bauschlicher2018}
{Bauschlicher}, Charles~W., J., {Ricca}, A., {Boersma}, C., \& {Allamandola}, L.~J. 2018, \apjs, 234, 32, \dodoi{10.3847/1538-4365/aaa019}

\bibitem[{{Boersma} {et~al.}(2014{\natexlab{a}}){Boersma}, {Bregman}, \& {Allamandola}}]{Boersma2014b}
{Boersma}, C., {Bregman}, J., \& {Allamandola}, L.~J. 2014{\natexlab{a}}, \apj, 795, 110, \dodoi{10.1088/0004-637X/795/2/110}

\bibitem[{{Boersma} {et~al.}(2015){Boersma}, {Bregman}, \& {Allamandola}}]{Boersma2015}
---. 2015, \apj, 806, 121, \dodoi{10.1088/0004-637X/806/1/121}

\bibitem[{{Boersma} {et~al.}(2018){Boersma}, {Bregman}, \& {Allamandola}}]{Boersma2018}
---. 2018, \apj, 858, 67, \dodoi{10.3847/1538-4357/aabcbe}

\bibitem[{{Boersma} {et~al.}(2013){Boersma}, {Bregman}, \& {Allamandola}}]{Boersma2013}
{Boersma}, C., {Bregman}, J.~D., \& {Allamandola}, L.~J. 2013, \apj, 769, 117, \dodoi{10.1088/0004-637X/769/2/117}

\bibitem[{{Boersma} {et~al.}(2024){Boersma}, {Bregman}, {Allamandola}, {Temi}, \& {Maragkoudakis}}]{Boersma2024}
{Boersma}, C., {Bregman}, J.~D., {Allamandola}, L.~J., {Temi}, P., \& {Maragkoudakis}, A. 2024, \apj, 975, 177, \dodoi{10.3847/1538-4357/ad7d08}

\bibitem[{{Boersma} {et~al.}(2014{\natexlab{b}}){Boersma}, {Bauschlicher}, {Ricca}, {Mattioda}, {Cami}, {Peeters}, {S{\'a}nchez de Armas}, {Puerta Saborido}, {Hudgins}, \& {Allamandola}}]{Boersma2014a}
{Boersma}, C., {Bauschlicher}, Jr., C.~W., {Ricca}, A., {et~al.} 2014{\natexlab{b}}, \apjs, 211, 8, \dodoi{10.1088/0067-0049/211/1/8}

\bibitem[{{Bolatto} {et~al.}(2024){Bolatto}, {Levy}, {Tarantino}, {Boyer}, {Fisher}, {Cronin}, {Leroy}, {Klessen}, {Smith}, {Berg}, {B{\"o}ker}, {Boogaard}, {Ostriker}, {Thompson}, {Ott}, {Lenki{\'c}}, {Lopez}, {Dale}, {Veilleux}, {van der Werf}, {Glover}, {Sandstrom}, {Skillman}, {Chisholm}, {Villanueva}, {Lai}, {Lopez}, {Mills}, {Emig}, {Armus}, {Mayya}, {Meier}, {De Looze}, {Herrera-Camus}, {Walter}, {Rela{\~n}o}, {Koziol}, {Marvil}, {Jim{\'e}nez-Donaire}, \& {Martini}}]{Bolatto2024}
{Bolatto}, A.~D., {Levy}, R.~C., {Tarantino}, E., {et~al.} 2024, \apj, 967, 63, \dodoi{10.3847/1538-4357/ad33c8}

\bibitem[{{Brinchmann} {et~al.}(2004){Brinchmann}, {Charlot}, {White}, {Tremonti}, {Kauffmann}, {Heckman}, \& {Brinkmann}}]{Brinchmann2004}
{Brinchmann}, J., {Charlot}, S., {White}, S.~D.~M., {et~al.} 2004, \mnras, 351, 1151, \dodoi{10.1111/j.1365-2966.2004.07881.x}

\bibitem[{{Cami} {et~al.}(2010){Cami}, {Bernard-Salas}, {Peeters}, \& {Malek}}]{Cami2010}
{Cami}, J., {Bernard-Salas}, J., {Peeters}, E., \& {Malek}, S.~E. 2010, Science, 329, 1180, \dodoi{10.1126/science.1192035}

\bibitem[{{Chastenet} {et~al.}(2023){Chastenet}, {Sutter}, {Sandstrom}, {Belfiore}, {Egorov}, {Larson}, {Leroy}, {Liu}, {Rosolowsky}, {Thilker}, {Watkins}, {Williams}, {Barnes}, {Bigiel}, {Boquien}, {Chevance}, {Dale}, {Kruijssen}, {Emsellem}, {Grasha}, {Groves}, {Hassani}, {Hughes}, {Kreckel}, {Meidt}, {Pan}, {Querejeta}, {Schinnerer}, \& {Whitcomb}}]{Chastenet2023}
{Chastenet}, J., {Sutter}, J., {Sandstrom}, K., {et~al.} 2023, \apjl, 944, L12, \dodoi{10.3847/2041-8213/acac94}

\bibitem[{{Chiappini} {et~al.}(2003){Chiappini}, {Romano}, \& {Matteucci}}]{Chiappini2003}
{Chiappini}, C., {Romano}, D., \& {Matteucci}, F. 2003, \mnras, 339, 63, \dodoi{10.1046/j.1365-8711.2003.06154.x}

\bibitem[{{Chown} {et~al.}(2024){Chown}, {Sidhu}, {Peeters}, {Tielens}, {Cami}, {Bern{\'e}}, {Habart}, {Alarc{\'o}n}, {Canin}, {Schroetter}, {Trahin}, {Van De Putte}, {Abergel}, {Bergin}, {Bernard-Salas}, {Boersma}, {Bron}, {Cuadrado}, {Dartois}, {Dicken}, {El-Yajouri}, {Fuente}, {Goicoechea}, {Gordon}, {Issa}, {Joblin}, {Kannavou}, {Khan}, {Lacinbala}, {Languignon}, {Le Gal}, {Maragkoudakis}, {Meshaka}, {Okada}, {Onaka}, {Pasquini}, {Pound}, {Robberto}, {R{\"o}llig}, {Schefter}, {Schirmer}, {Vicente}, {Wolfire}, {Zannese}, {Aleman}, {Allamandola}, {Auchettl}, {Baratta}, {Bejaoui}, {Bera}, {Black}, {Boulanger}, {Bouwman}, {Brandl}, {Brechignac}, {Br{\"u}nken}, {Buragohain}, {Burkhardt}, {Candian}, {Cazaux}, {Cernicharo}, {Chabot}, {Chakraborty}, {Champion}, {Colgan}, {Cooke}, {Coutens}, {Cox}, {Demyk}, {Meyer}, {Foschino}, {Garc{\'\i}a-Lario}, {Gavilan}, {Gerin}, {Gottlieb}, {Guillard}, {Gusdorf}, {Hartigan}, {He}, {Herbst}, {Hornekaer}, {J{\"a}ger}, {Janot-Pacheco}, {Kaufman}, {Kemper}, {Kendrew},
  {Kirsanova}, {Klaassen}, {Kwok}, {Labiano}, {Lai}, {Lee}, {Lefloch}, {Le Petit}, {Li}, {Linz}, {Mackie}, {Madden}, {Mascetti}, {McGuire}, {Merino}, {Micelotta}, {Misselt}, {Morse}, {Mulas}, {Neelamkodan}, {Ohsawa}, {Omont}, {Paladini}, {Palumbo}, {Pathak}, {Pendleton}, {Petrignani}, {Pino}, {Puga}, {Rangwala}, {Rapacioli}, {Ricca}, {Roman-Duval}, {Roser}, {Roueff}, {Rouill{\'e}}, {Salama}, {Sales}, {Sandstrom}, {Sarre}, {Sciamma-O'Brien}, {Sellgren}, {Shenoy}, {Teyssier}, {Thomas}, {Togi}, {Verstraete}, {Witt}, {Wootten}, {Zettergren}, {Zhang}, {Zhang}, \& {Zhen}}]{Chown2024}
{Chown}, R., {Sidhu}, A., {Peeters}, E., {et~al.} 2024, \aap, 685, A75, \dodoi{10.1051/0004-6361/202346662}

\bibitem[{{Cook} \& {Saykally}(1998)}]{Cook1998}
{Cook}, D.~J., \& {Saykally}, R.~J. 1998, \apj, 493, 793, \dodoi{10.1086/305156}

\bibitem[{{Croiset} {et~al.}(2016){Croiset}, {Candian}, {Bern{\'e}}, \& {Tielens}}]{Croiset2016}
{Croiset}, B.~A., {Candian}, A., {Bern{\'e}}, O., \& {Tielens}, A.~G.~G.~M. 2016, \aap, 590, A26, \dodoi{10.1051/0004-6361/201527714}

\bibitem[{{Daddi} {et~al.}(2007){Daddi}, {Dickinson}, {Morrison}, {Chary}, {Cimatti}, {Elbaz}, {Frayer}, {Renzini}, {Pope}, {Alexander}, {Bauer}, {Giavalisco}, {Huynh}, {Kurk}, \& {Mignoli}}]{Daddi2007}
{Daddi}, E., {Dickinson}, M., {Morrison}, G., {et~al.} 2007, \apj, 670, 156, \dodoi{10.1086/521818}

\bibitem[{{Donnan} {et~al.}(2024){Donnan}, {Garc{\'\i}a-Bernete}, {Rigopoulou}, {Pereira-Santaella}, {Roche}, \& {Alonso-Herrero}}]{Donnan2024}
{Donnan}, F.~R., {Garc{\'\i}a-Bernete}, I., {Rigopoulou}, D., {et~al.} 2024, \mnras, 529, 1386, \dodoi{10.1093/mnras/stae612}

\bibitem[{{Draine} {et~al.}(2021){Draine}, {Li}, {Hensley}, {Hunt}, {Sandstrom}, \& {Smith}}]{Draine2021}
{Draine}, B.~T., {Li}, A., {Hensley}, B.~S., {et~al.} 2021, \apj, 917, 3, \dodoi{10.3847/1538-4357/abff51}

\bibitem[{{Egorov} {et~al.}(2023){Egorov}, {Kreckel}, {Sandstrom}, {Leroy}, {Glover}, {Groves}, {Kruijssen}, {Barnes}, {Belfiore}, {Bigiel}, {Blanc}, {Boquien}, {Cao}, {Chastenet}, {Chevance}, {Congiu}, {Dale}, {Emsellem}, {Grasha}, {Klessen}, {Larson}, {Liu}, {Murphy}, {Pan}, {Pessa}, {Pety}, {Rosolowsky}, {Scheuermann}, {Schinnerer}, {Sutter}, {Thilker}, {Watkins}, \& {Williams}}]{Egorov2023}
{Egorov}, O.~V., {Kreckel}, K., {Sandstrom}, K.~M., {et~al.} 2023, \apjl, 944, L16, \dodoi{10.3847/2041-8213/acac92}

\bibitem[{{Elbaz} {et~al.}(2007){Elbaz}, {Daddi}, {Le Borgne}, {Dickinson}, {Alexander}, {Chary}, {Starck}, {Brandt}, {Kitzbichler}, {MacDonald}, {Nonino}, {Popesso}, {Stern}, \& {Vanzella}}]{Elbaz2007}
{Elbaz}, D., {Daddi}, E., {Le Borgne}, D., {et~al.} 2007, \aap, 468, 33, \dodoi{10.1051/0004-6361:20077525}

\bibitem[{Esposito {et~al.}(2024{\natexlab{a}})Esposito, Allamandola, Boersma, Bregman, Fortenberry, Maragkoudakis, \& Temi}]{esposito_anharmonic_2024}
Esposito, V.~J., Allamandola, L.~J., Boersma, C., {et~al.} 2024{\natexlab{a}}, Molecular Physics, 122, e2252936, \dodoi{10.1080/00268976.2023.2252936}

\bibitem[{Esposito {et~al.}(2024{\natexlab{b}})Esposito, Ferrari, Buma, Boersma, Mackie, Candian, Fortenberry, \& Tielens}]{esposito_anharmonicity_2024}
Esposito, V.~J., Ferrari, P., Buma, W.~J., {et~al.} 2024{\natexlab{b}}, Molecular Physics, 122, e2261570, \dodoi{10.1080/00268976.2023.2261570}

\bibitem[{Esposito {et~al.}(2024{\natexlab{c}})Esposito, Ferrari, Buma, Fortenberry, Boersma, Candian, \& Tielens}]{esposito_infrared_2024}
---. 2024{\natexlab{c}}, The Journal of Chemical Physics, 160, 114312, \dodoi{10.1063/5.0191404}

\bibitem[{{Galliano} {et~al.}(2008){Galliano}, {Madden}, {Tielens}, {Peeters}, \& {Jones}}]{Galliano2008}
{Galliano}, F., {Madden}, S.~C., {Tielens}, A.~G.~G.~M., {Peeters}, E., \& {Jones}, A.~P. 2008, \apj, 679, 310, \dodoi{10.1086/587051}

\bibitem[{{Garc{\'\i}a-Bernete} {et~al.}(2022){Garc{\'\i}a-Bernete}, {Rigopoulou}, {Alonso-Herrero}, {Donnan}, {Roche}, {Pereira-Santaella}, {Labiano}, {Peralta de Arriba}, {Izumi}, {Ramos Almeida}, {Shimizu}, {H{\"o}nig}, {Garc{\'\i}a-Burillo}, {Rosario}, {Ward}, {Bellocchi}, {Hicks}, {Fuller}, \& {Packham}}]{Garcia-Bernete2022}
{Garc{\'\i}a-Bernete}, I., {Rigopoulou}, D., {Alonso-Herrero}, A., {et~al.} 2022, \aap, 666, L5, \dodoi{10.1051/0004-6361/202244806}

\bibitem[{{Hony} {et~al.}(2001){Hony}, {Van Kerckhoven}, {Peeters}, {Tielens}, {Hudgins}, \& {Allamandola}}]{Hony2001}
{Hony}, S., {Van Kerckhoven}, C., {Peeters}, E., {et~al.} 2001, \aap, 370, 1030, \dodoi{10.1051/0004-6361:20010242}

\bibitem[{{Hudgins} \& {Allamandola}(1999)}]{Hudgins1999}
{Hudgins}, D.~M., \& {Allamandola}, L.~J. 1999, \apj, 513, L69, \dodoi{10.1086/311901}

\bibitem[{{Hudgins} {et~al.}(2005){Hudgins}, {Bauschlicher}, \& {Allamandola}}]{Hudgins2005}
{Hudgins}, D.~M., {Bauschlicher}, Charles~W., J., \& {Allamandola}, L.~J. 2005, \apj, 632, 316, \dodoi{10.1086/432495}

\bibitem[{{Kemper} {et~al.}(2004){Kemper}, {Vriend}, \& {Tielens}}]{Kemper2004}
{Kemper}, F., {Vriend}, W.~J., \& {Tielens}, A.~G.~G.~M. 2004, \apj, 609, 826, \dodoi{10.1086/421339}

\bibitem[{{Knight} {et~al.}(2021){Knight}, {Peeters}, {Stock}, {Vacca}, \& {Tielens}}]{Knight2021}
{Knight}, C., {Peeters}, E., {Stock}, D.~J., {Vacca}, W.~D., \& {Tielens}, A.~G.~G.~M. 2021, \apj, 918, 8, \dodoi{10.3847/1538-4357/ac02c6}

\bibitem[{{Lai} {et~al.}(2022){Lai}, {Armus}, {U}, {D{\'\i}az-Santos}, {Larson}, {Evans}, {Malkan}, {Appleton}, {Rich}, {M{\"u}ller-S{\'a}nchez}, {Inami}, {Bohn}, {McKinney}, {Finnerty}, {Law}, {Linden}, {Medling}, {Privon}, {Song}, {Stierwalt}, {van der Werf}, {Barcos-Mu{\~n}oz}, {Smith}, {Togi}, {Aalto}, {B{\"o}ker}, {Charmandaris}, {Howell}, {Iwasawa}, {Kemper}, {Mazzarella}, {Murphy}, {Brown}, {Hayward}, {Marshall}, {Sanders}, \& {Surace}}]{Lai2022}
{Lai}, T. S.~Y., {Armus}, L., {U}, V., {et~al.} 2022, \apjl, 941, L36, \dodoi{10.3847/2041-8213/ac9ebf}

\bibitem[{{Le Page} {et~al.}(2003){Le Page}, {Snow}, \& {Bierbaum}}]{LePage2003}
{Le Page}, V., {Snow}, T.~P., \& {Bierbaum}, V.~M. 2003, \apj, 584, 316, \dodoi{10.1086/345595}

\bibitem[{{Leger} \& {Puget}(1984)}]{Leger1984}
{Leger}, A., \& {Puget}, J.~L. 1984, \aap, 137, L5

\bibitem[{Mackie {et~al.}(2018)Mackie, Chen, Candian, Lee, \& Tielens}]{Mackie2018}
Mackie, C.~J., Chen, T., Candian, A., Lee, T.~J., \& Tielens, A. G. G.~M. 2018, The Journal of Chemical Physics, 149, 134302, \dodoi{10.1063/1.5038725}

\bibitem[{Mackie {et~al.}(2015)Mackie, Candian, Huang, Maltseva, Petrignani, Oomens, Buma, Lee, \& Tielens}]{mackieAnharmonicQuarticForce2015}
Mackie, C.~J., Candian, A., Huang, X., {et~al.} 2015, The Journal of Chemical Physics, 143, 224314, \dodoi{10.1063/1.4936779}

\bibitem[{{Mackie} {et~al.}(2016){Mackie}, {Candian}, {Huang}, {Maltseva}, {Petrignani}, {Oomens}, {Mattioda}, {Buma}, {Lee}, \& {Tielens}}]{Mackie2016}
{Mackie}, C.~J., {Candian}, A., {Huang}, X., {et~al.} 2016, Journal of Chemical Physics, 145, 084313, \dodoi{10.1063/1.4961438}

\bibitem[{Maltseva {et~al.}(2015)Maltseva, Petrignani, Candian, Mackie, Huang, Lee, Tielens, Oomens, \& Buma}]{maltseva_high-resolution_2015}
Maltseva, E., Petrignani, A., Candian, A., {et~al.} 2015, The Astrophysical Journal, 814, 23, \dodoi{10.1088/0004-637X/814/1/23}

\bibitem[{Maltseva {et~al.}(2018)Maltseva, Mackie, Candian, Petrignani, Huang, Lee, Tielens, Oomens, \& Buma}]{maltseva_high-resolution_2018}
Maltseva, E., Mackie, C.~J., Candian, A., {et~al.} 2018, Astronomy \& Astrophysics, 610, A65, \dodoi{10.1051/0004-6361/201732102}

\bibitem[{{Maragkoudakis} {et~al.}(2022){Maragkoudakis}, {Boersma}, {Temi}, {Bregman}, \& {Allamandola}}]{Maragkoudakis2022}
{Maragkoudakis}, A., {Boersma}, C., {Temi}, P., {Bregman}, J.~D., \& {Allamandola}, L.~J. 2022, \apj, 931, 38, \dodoi{10.3847/1538-4357/ac666f}

\bibitem[{{Maragkoudakis} {et~al.}(2018){Maragkoudakis}, {Ivkovich}, {Peeters}, {Stock}, {Hemachandra}, \& {Tielens}}]{Maragkoudakis2018a}
{Maragkoudakis}, A., {Ivkovich}, N., {Peeters}, E., {et~al.} 2018, \mnras, 481, 5370, \dodoi{10.1093/mnras/sty2658}

\bibitem[{{Maragkoudakis} {et~al.}(2020){Maragkoudakis}, {Peeters}, \& {Ricca}}]{Maragkoudakis2020}
{Maragkoudakis}, A., {Peeters}, E., \& {Ricca}, A. 2020, \mnras, 494, 642, \dodoi{10.1093/mnras/staa681}

\bibitem[{{Maragkoudakis} {et~al.}(2023{\natexlab{a}}){Maragkoudakis}, {Peeters}, \& {Ricca}}]{Maragkoudakis2023a}
---. 2023{\natexlab{a}}, \mnras, 520, 5354, \dodoi{10.1093/mnras/stad465}

\bibitem[{{Maragkoudakis} {et~al.}(2023{\natexlab{b}}){Maragkoudakis}, {Peeters}, {Ricca}, \& {Boersma}}]{Maragkoudakis2023b}
{Maragkoudakis}, A., {Peeters}, E., {Ricca}, A., \& {Boersma}, C. 2023{\natexlab{b}}, \mnras, 524, 3429, \dodoi{10.1093/mnras/stad2062}

\bibitem[{{Maragkoudakis} {et~al.}(2017){Maragkoudakis}, {Zezas}, {Ashby}, \& {Willner}}]{Maragkoudakis2017}
{Maragkoudakis}, A., {Zezas}, A., {Ashby}, M.~L.~N., \& {Willner}, S.~P. 2017, \mnras, 466, 1192, \dodoi{10.1093/mnras/stw3180}

\bibitem[{{Marshall} {et~al.}(2007){Marshall}, {Herter}, {Armus}, {Charmandaris}, {Spoon}, {Bernard-Salas}, \& {Houck}}]{Marshall2007}
{Marshall}, J.~A., {Herter}, T.~L., {Armus}, L., {et~al.} 2007, \apj, 670, 129, \dodoi{10.1086/521588}

\bibitem[{{Mattioda} {et~al.}(2020){Mattioda}, {Hudgins}, {Boersma}, {Bauschlicher}, {Ricca}, {Cami}, {Peeters}, {S{\'{a}}nchez de Armas}, {Puerta Saborido}, \& {Allamandola}}]{Mattioda2020}
{Mattioda}, A.~L., {Hudgins}, D.~M., {Boersma}, C., {et~al.} 2020, \apjs, 251, 22, \dodoi{10.3847/1538-4365/abc2c8}

\bibitem[{{McGuire} {et~al.}(2018){McGuire}, {Burkhardt}, {Kalenskii}, {Shingledecker}, {Remijan}, {Herbst}, \& {McCarthy}}]{McGuire2018}
{McGuire}, B.~A., {Burkhardt}, A.~M., {Kalenskii}, S., {et~al.} 2018, Science, 359, 202, \dodoi{10.1126/science.aao4890}

\bibitem[{{McGuire} {et~al.}(2021){McGuire}, {Loomis}, {Burkhardt}, {Lee}, {Shingledecker}, {Charnley}, {Cooke}, {Cordiner}, {Herbst}, {Kalenskii}, {Siebert}, {Willis}, {Xue}, {Remijan}, \& {McCarthy}}]{McGuire2021}
{McGuire}, B.~A., {Loomis}, R.~A., {Burkhardt}, A.~M., {et~al.} 2021, Science, 371, 1265, \dodoi{10.1126/science.abb7535}

\bibitem[{{Oliveira} {et~al.}(2024){Oliveira}, {Krabbe}, {Dors}, {Zinchenko}, {Hernandez-Jimenez}, {Cardaci}, {H{\"a}gele}, \& {Ilha}}]{Oliveira2024}
{Oliveira}, C.~B., {Krabbe}, A.~C., {Dors}, O.~L., {et~al.} 2024, \mnras, 531, 199, \dodoi{10.1093/mnras/stae1172}

\bibitem[{{Ossenkopf} {et~al.}(1992){Ossenkopf}, {Henning}, \& {Mathis}}]{Ossenkopf1992}
{Ossenkopf}, V., {Henning}, T., \& {Mathis}, J.~S. 1992, \aap, 261, 567

\bibitem[{{Pech} {et~al.}(2002){Pech}, {Joblin}, \& {Boissel}}]{Pech2002}
{Pech}, C., {Joblin}, C., \& {Boissel}, P. 2002, \aap, 388, 639, \dodoi{10.1051/0004-6361:20020416}

\bibitem[{{Peeters} {et~al.}(2017){Peeters}, {Bauschlicher}, {Allamandola}, {Tielens}, {Ricca}, \& {Wolfire}}]{Peeters2017}
{Peeters}, E., {Bauschlicher}, Jr., C.~W., {Allamandola}, L.~J., {et~al.} 2017, \apj, 836, 198, \dodoi{10.3847/1538-4357/836/2/198}

\bibitem[{{Peeters} {et~al.}(2002){Peeters}, {Hony}, {Van Kerckhoven}, {Tielens}, {Allamandola}, {Hudgins}, \& {Bauschlicher}}]{Peeters2002}
{Peeters}, E., {Hony}, S., {Van Kerckhoven}, C., {et~al.} 2002, \aap, 390, 1089, \dodoi{10.1051/0004-6361:20020773}

\bibitem[{{Peeters} {et~al.}(2004){Peeters}, {Mattioda}, {Hudgins}, \& {Allamandola}}]{Peeters2004b}
{Peeters}, E., {Mattioda}, A.~L., {Hudgins}, D.~M., \& {Allamandola}, L.~J. 2004, \apjl, 617, L65, \dodoi{10.1086/427186}

\bibitem[{{P{\'e}rez-Montero} {et~al.}(2013){P{\'e}rez-Montero}, {Contini}, {Lamareille}, {Maier}, {Carollo}, {Kneib}, {Le F{\`e}vre}, {Lilly}, {Mainieri}, {Renzini}, {Scodeggio}, {Zamorani}, {Bardelli}, {Bolzonella}, {Bongiorno}, {Caputi}, {Cucciati}, {de la Torre}, {de Ravel}, {Franzetti}, {Garilli}, {Iovino}, {Kampczyk}, {Knobel}, {Kova{\v{c}}}, {Le Borgne}, {Le Brun}, {Mignoli}, {Pell{\`o}}, {Peng}, {Presotto}, {Ricciardelli}, {Silverman}, {Tanaka}, {Tasca}, {Tresse}, {Vergani}, \& {Zucca}}]{Perez-Montero2013}
{P{\'e}rez-Montero}, E., {Contini}, T., {Lamareille}, F., {et~al.} 2013, \aap, 549, A25, \dodoi{10.1051/0004-6361/201220070}

\bibitem[{{Popesso} {et~al.}(2019){Popesso}, {Morselli}, {Concas}, {Schreiber}, {Rodighiero}, {Cresci}, {Belli}, {Ilbert}, {Erfanianfar}, {Mancini}, {Inami}, {Dickinson}, {Pannella}, \& {Elbaz}}]{Popesso2019}
{Popesso}, P., {Morselli}, L., {Concas}, A., {et~al.} 2019, \mnras, 490, 5285, \dodoi{10.1093/mnras/stz2635}

\bibitem[{{Ricca} {et~al.}(2012){Ricca}, {Bauschlicher}, {Boersma}, {Tielens}, \& {Allamandola}}]{Ricca2012}
{Ricca}, A., {Bauschlicher}, Charles~W., J., {Boersma}, C., {Tielens}, A. G.~G.~M., \& {Allamandola}, L.~J. 2012, \apj, 754, 75, \dodoi{10.1088/0004-637X/754/1/75}

\bibitem[{{Ricca} {et~al.}(2018){Ricca}, {Bauschlicher}, {Roser}, \& {Peeters}}]{Ricca2018}
{Ricca}, A., {Bauschlicher}, Charles~W., J., {Roser}, J.~E., \& {Peeters}, E. 2018, \apj, 854, 115, \dodoi{10.3847/1538-4357/aaa757}

\bibitem[{{Ricca} {et~al.}(2021){Ricca}, {Boersma}, \& {Peeters}}]{Ricca2021}
{Ricca}, A., {Boersma}, C., \& {Peeters}, E. 2021, \apj, 923, 202, \dodoi{10.3847/1538-4357/ac28fc}

\bibitem[{{Ricca} {et~al.}(2024){Ricca}, {Roser}, {Boersma}, {Peeters}, \& {Maragkoudakis}}]{Ricca2024}
{Ricca}, A., {Roser}, J.~E., {Boersma}, C., {Peeters}, E., \& {Maragkoudakis}, A. 2024, \apj, 968, 128, \dodoi{10.3847/1538-4357/ad4151}

\bibitem[{{Ricca} {et~al.}(2019){Ricca}, {Roser}, {Peeters}, \& {Boersma}}]{Ricca2019}
{Ricca}, A., {Roser}, J.~E., {Peeters}, E., \& {Boersma}, C. 2019, \apj, 882, 56, \dodoi{10.3847/1538-4357/ab3124}

\bibitem[{{Rigopoulou} {et~al.}(2024){Rigopoulou}, {Donnan}, {Garc{\'\i}a-Bernete}, {Pereira-Santaella}, {Alonso-Herrero}, {Davies}, {Hunt}, {Roche}, \& {Shimizu}}]{Rigopoulou2024}
{Rigopoulou}, D., {Donnan}, F.~R., {Garc{\'\i}a-Bernete}, I., {et~al.} 2024, arXiv e-prints, arXiv:2406.11415, \dodoi{10.48550/arXiv.2406.11415}

\bibitem[{{Sandstrom} {et~al.}(2012){Sandstrom}, {Bolatto}, {Bot}, {Draine}, {Ingalls}, {Israel}, {Jackson}, {Leroy}, {Li}, {Rubio}, {Simon}, {Smith}, {Stanimirovi{\'c}}, {Tielens}, \& {van Loon}}]{Sandstrom2012}
{Sandstrom}, K.~M., {Bolatto}, A.~D., {Bot}, C., {et~al.} 2012, \apj, 744, 20, \dodoi{10.1088/0004-637X/744/1/20}

\bibitem[{{Schutte} {et~al.}(1993){Schutte}, {Tielens}, \& {Allamandola}}]{Schutte1993}
{Schutte}, W.~A., {Tielens}, A.~G.~G.~M., \& {Allamandola}, L.~J. 1993, \apj, 415, 397, \dodoi{10.1086/173173}

\bibitem[{{Shannon} \& {Boersma}(2019)}]{Shannon2019}
{Shannon}, M.~J., \& {Boersma}, C. 2019, \apj, 871, 124, \dodoi{10.3847/1538-4357/aaf562}

\bibitem[{{Smith} {et~al.}(2007){Smith}, {Draine}, {Dale}, {Moustakas}, {Kennicutt}, {Helou}, {Armus}, {Roussel}, {Sheth}, {Bendo}, {Buckalew}, {Calzetti}, {Engelbracht}, {Gordon}, {Hollenbach}, {Li}, {Malhotra}, {Murphy}, \& {Walter}}]{Smith07b}
{Smith}, J.~D.~T., {Draine}, B.~T., {Dale}, D.~A., {et~al.} 2007, \apj, 656, 770, \dodoi{10.1086/510549}

\bibitem[{{Speagle} {et~al.}(2014){Speagle}, {Steinhardt}, {Capak}, \& {Silverman}}]{Speagle2014}
{Speagle}, J.~S., {Steinhardt}, C.~L., {Capak}, P.~L., \& {Silverman}, J.~D. 2014, \apjs, 214, 15, \dodoi{10.1088/0067-0049/214/2/15}

\bibitem[{{Tielens}(2008)}]{Tielens2008}
{Tielens}, A.~G.~G.~M. 2008, \araa, 46, 289, \dodoi{10.1146/annurev.astro.46.060407.145211}

\bibitem[{{Williams} \& {Leone}(1995)}]{Williams1995}
{Williams}, R.~M., \& {Leone}, S.~R. 1995, \apj, 443, 675, \dodoi{10.1086/175559}

\bibitem[{{Zang} {et~al.}(2022){Zang}, {Maragkoudakis}, \& {Peeters}}]{Zang2022}
{Zang}, R.~X., {Maragkoudakis}, A., \& {Peeters}, E. 2022, \mnras, 511, 5142, \dodoi{10.1093/mnras/stac214}

\end{thebibliography}
\bibliographystyle{aasjournal}

\end{document}